\documentclass[a4paper,11pt]{article}
\pdfoutput=1 % if your are submitting a pdflatex (i.e. if you have
\usepackage{Class/jcappub} % for details on the use of the package, please
\usepackage{lineno}
\usepackage[T1]{fontenc} % if needed
\usepackage{physics}
\usepackage{soul}
\usepackage{comment}
\usepackage{url}
\usepackage{amsmath}
\usepackage{amsfonts}
\usepackage{xspace}
\usepackage{bbm}
\usepackage{tabularx}
\usepackage{booktabs}
\usepackage{multirow}
\usepackage[dvipsnames]{xcolor}
\usepackage{cleveref}
\usepackage{orcidlink}

\newcommand{\LiteBIRD}{\textit{LiteBIRD}\xspace}
\newcommand{\LB}{\textit{\LiteBIRD}\xspace}
\newcommand{\Planck}{\textit{Planck}\xspace}

\newcommand{\WMAP}{\textit{WMAP}\xspace}
\newcommand{\EPIC}{\textit{EPIC}\xspace}
\newcommand{\PICO}{\textit{PICO}\xspace}

\newcommand{\SC}{\textit{standard configuration}\xspace}
\newcommand{\BC}{\textit{balanced configuration}\xspace}
\newcommand{\FC}{\textit{flipped configuration}\xspace}
\newcommand{\moire}{Moir\'e\xspace}
\newcommand{\etal}{et al.\xspace}

\newcommand{\Tbetalow}{$T_{\beta}^{\mathrm{lower}}$\xspace}
\newcommand{\tbl}{T_{\beta}^{\mathrm{lower}}}
\newcommand{\Nhits}{N_{\rm{hits}}}
\newcommand{\sigmahits}{\sigma_{\rm{hits}}}

\newcommand{\crosslinkmean}{$\langle \abs{{}_{n, m}\tilde{h}(\Omega)}^2 \rangle$\xspace}

\newcommand{\Nmod}{N_{\rm{mod}}}
\newcommand{\Nmargin}{N_{\rm{margin}}}
\newcommand{\spin}{\textit{spin}\xspace}

\newcommand{\Falcons}{\texttt{Falcons.jl}\xspace}

\newcommand{\Sd}[1][2]{{{}_{#1}\tilde{S}^d}}
\newcommand{\h}[1][2]{{{}_{#1}\tilde{h}}}
\newcommand{\St}[1][2]{{{}_{#1}\tilde{S}}}

\newcommand{\prd}{Physical Review D}
\newcommand{\apjl}{The Astrophysical Journal Letters}
\newcommand{\aaps}{Astrophysical Journal}
\newcommand{\aap}{A\&A}
\newcommand{\mnras}{Monthly Notices of the RAS}
\newcommand{\ptep}{Progress of Theoretical and Experimental Physics}
\newcommand{\jcap}{Journal of Cosmology and Astroparticle Physics}	
\newcommand{\apj}{Astrophysical Journal}

\crefname{figure}{figure}{figures} 
\Crefname{figure}{Figure}{Figures} 

\arxivnumber{2408.03040} % Only if you have one]

\title{Multi-dimensional optimisation of the scanning strategy for the \textit{LiteBIRD} space mission}

\collaborationImg{\includegraphics[width=0.15\columnwidth]{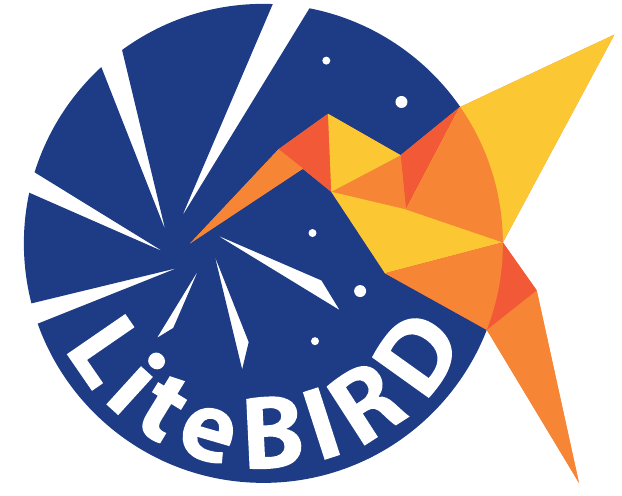}}
\author[1]{Y.\,Takase\orcidlink{0009-0002-5635-6009},}
\author[2]{L.\,Vacher\orcidlink{0000-0001-9551-1417},}
\author[1]{H.\,Ishino\orcidlink{0000-0002-8623-4080},}
\author[3,4,5,6]{G.\,Patanchon\orcidlink{0000-0002-9035-849X},}
\author[7]{L.\,Montier,}
\author[8,1,5]{S.\,L.\,Stever\orcidlink{0000-0002-3965-7080},}
\author[1]{K.\,Ishizaka,}
\author[1]{Y.\,Nagano\orcidlink{0009-0003-7744-8962},}
\author[6]{W.\,Wang\orcidlink{0000-0002-9226-7639},}
\author[7]{J.\,Aumont\orcidlink{0000-0001-6279-0691},}
% ...
\affiliation[1]{Okayama University, Department of Physics, Okayama 700-8530, Japan}
\affiliation[2]{International School for Advanced Studies (SISSA), Via Bonomea 265, 34136, Trieste, Italy}
\affiliation[3]{ILANCE, CNRS – University of Tokyo International Research Laboratory, Kashiwa, Chiba 277-8582, Japan}
\affiliation[4]{Université Paris Cité, F-75006 Paris, France}
\affiliation[5]{Kavli Institute for the Physics and Mathematics of the Universe (Kavli IPMU, WPI), UTIAS, The University of Tokyo, Kashiwa, Chiba 277-8583, Japan}
\affiliation[6]{Université Paris Cité, CNRS, Astroparticule et Cosmologie, F-75013 Paris, France}
\affiliation[7]{IRAP, Université de Toulouse, CNRS, CNES, UPS, Toulouse, France}
\affiliation[8]{Université Paris-Saclay, CNRS, Institut d’Astrophysique Spatiale, 91405, Orsay, France}
% ...

\emailAdd{takase\_y@s.okayama-u.ac.jp}

\abstract{
Large angular scale surveys in the absence of atmosphere are essential for measuring the primordial $B$-mode power spectrum of the Cosmic Microwave Background (CMB). Since this proposed measurement is about three to four orders of magnitude fainter than the temperature anisotropies of the CMB, in-flight calibration of the instruments and active suppression of systematic effects are crucial. We investigate the effect of changing the parameters of the scanning strategy on the in-flight calibration effectiveness, the suppression of the systematic effects themselves, and the ability to distinguish systematic effects by null-tests. 
Next-generation missions such as \textit{LiteBIRD}, modulated by a Half-Wave Plate (HWP), will be able to observe polarisation using a single detector, eliminating the need to combine several detectors to measure polarisation, as done in many previous experiments and hence avoiding the consequent systematic effects.
While the HWP is expected to suppress many systematic effects, some of them will remain. We use an analytical approach to comprehensively address the mitigation of these systematic effects and identify the characteristics of scanning strategies that are the most effective for implementing a variety of calibration strategies in the multi-dimensional space of common spacecraft scan parameters. 
We verify that \LiteBIRD's \textit{standard configuration} yields good performance on the metrics we studied.
We also present \texttt{Falcons.jl}, a fast spacecraft scanning simulator that we developed to investigate this scanning parameter space.
}

\begin{document}
\maketitle
\flushbottom

\section{Introduction} \label{sec:intro}

The quest for the primordial $B$ modes of the Cosmic Microwave Background (CMB) polarisation, which would be the smoking gun for the existence of Primordial Gravitational Waves~(PGWs), represents one of the biggest challenges of contemporary cosmology \cite{B-mode_Seljak,B-mode_Kamionkowski}. Their detection will require the mapping of the largest angular scales of the CMB, favoring the observation of the entire sky from out side Earth's atmosphere.

The next generation of space projects, including \LB \cite{PTEP2023}, are designed to accurately measure the primordial $B$ modes and test representative inflationary models by measuring the tensor-to-scalar ratio $r$ (corresponding to the amplitude of the PGW power spectrum), with an accuracy of $\delta r<0.001$.
The experiment must be able to distinguish primordial $B$ modes from the $E$ modes and temperature anisotropies, which are at least three to four orders of magnitude larger.

With the recent advent of highly-sensitive detector arrays, including Transition Edge Sensor (TES) bolometers that can reach the photon noise limit, the main obstacles to the $B$-mode observation are related to systematic uncertainties and the removal of Galactic foregrounds.
A first important consideration in order to suppress systematic effects in polarimetric observations is given by the careful choice of the sky scanning strategy, consisting of the combination of the spin and precession of the spacecraft. C.\,G.\,R.\,Wallis \etal\ have shown in ref.~\cite{OptimalScan} that an optimal scanning strategy, which allows each region of the sky to be observed such that the uniformity of the crossing angle distribution is maximised, is effective in suppressing systematic effects. 
However, since the role of scanning strategies is not only to suppress systematic effects, but also to effectively perform in-flight calibration of instruments and null-tests to reveal unknown systematic effects, the choices for the next generation of space missions should also carefully consider these aspects. Scanning strategies are characterised by a multi-dimensional parameter space, consisting of `geometric parameters', given by the angles of the spacecraft's spin and precession axes, and `kinetic parameters', consisting of the period/rate of rotations around these axes. 
The optimisation of such a parameter space has been previously performed for the missions \WMAP and \Planck \cite{Bennett2003,planck_scanning_strategy}. 
In particular, the data analysis method, and the coupling and removal of systematic effects (such as instrument drift and $1/f$ noise) with scanning strategies to achieve the high-resolution temperature anisotropy measurements targeted by these space missions were discussed in refs.~\cite{Wright1996,Janssen1996,Tegmark1997,Maino1999}. In addition, after the CMB polarisation was measured and the presence of $B$ modes predicted, similar analysis were also performed for polarisation in ref.~\cite{Hu2003}.

In this paper, we revisit the optimisation of the scanning strategy parameters in a multi-dimensional space with a fast scan simulator that we have developed, and provide clear metrics in order to assess the suppression of systematic effects and calibration for space missions.
We also account for a continuously rotating Half-Wave Plate (HWP) that modulates the polarisation of the incoming light. A HWP has been deployed in recent ground-based CMB experiments, such as \textit{POLARBEAR} and \textit{Simons Observatory} \cite{takakura_hwp,PB_HWP,SO-SAT_HWP}, and is also planned to be equipped on \LB.
By modulating the polarisation with a continuously rotating HWP, it is possible to effectively remove the $1/f$ noise that would otherwise degrade the sensitivity at the large angular scales where the primordial $B$-mode signal becomes prominent.
In addition, the pair-differencing between two orthogonal detectors, which is often used for polarisation observations, is no longer necessary with a HWP. Typical systematic effects due to pair-differencing, such as differential gain, differential pointing and differential beam ellipticity, discussed in ref.~\cite{BICEP_syst}, thus become irrelevant.
However, when installing a HWP on a spacecraft, it is necessary to consider other effects, such as whether the polarisation in a region of the sky to which the telescope's point spread function (i.e.\ main beam) is pointing would be modulated while it is in transit, that is, whether the HWP rotation is sufficiently faster than the sky scanning motion.

The paper is organised as follows: in \cref{sec:scanning_strategy}, we describe the relationship between the HWP rotational period and its subsequent constraint on the scanning angular velocity. In particular, we show that the lower limit on spin period depends on the requirement that the HWP is able to achieve at least one modulation during a pointing that transits the size of the main beam on the sky. In \cref{sec:metrics}, we introduce metrics to optimise the scanning strategy: visibility time of planets; hit-map uniformity; and `cross-link factor'.
The cross-link factor measures the uniformity of the crossing angles and is characterised by an integer number called \spin.\footnote{In this paper we distinguish the `spin' (normal font), which is the rotation around the maximum inertial axis of a spacecraft, and `\spin' (italic font), which is an integer characterising the cross-link factor and systematic effects.}
The cross-link factor metric for scanning strategy optimisation was proposed in ref.~\cite{OptimalScan} and it has been shown \cite{spin_characterisation, mapbased} that some systematic effects can be fully captured by this metric using the combination of specific \spin values. In \cref{sec:formalism}, we extend the formulation of ref.~\cite{mapbased} to include a continuously rotating HWP and introduce a framework for analytically assessing the coupling of systematic effects on the tensor-to-scalar ratio~$r$ and scanning strategies. \Cref{sec:Falcons} presents a scan simulator, \Falcons that enables fast computation of the scanning metrics and allows us to perform the optimisation of the multi-dimensional parameter space of a scanning strategy. The results for each metric thus obtained are then presented in \cref{sec:results}.
In \cref{sec:optimisation}, we consider the optimisation of the geometric parameters and kinetic parameters and evaluate the quality of the proposed parameters for the \LiteBIRD scanning strategy given in ref.~\cite{PTEP2023}.
\Cref{sec:implications} compares the scanning strategies considered in \Planck and the future \PICO missions \cite{PICO2019} with the effective scanning strategy for \LiteBIRD proposed in this work. We do so, by interpreting the capabilities of each scanning strategy in terms of null-tests and in-flight calibration, such as beam shape reconstruction and visit/revisit time of sky pixels and planets.
Finally, \cref{sec:conclusion} discusses the results obtained and presents our conclusions.
\section{Scanning strategies for spacecraft as CMB polarisation probes}\label{sec:scanning_strategy}

\subsection{The parameter space of a scanning strategy}
Past CMB space missions (\WMAP and \Planck) were conducted at the second Lagrange point~($\mathrm{L_2}$) in the Sun-Earth system. These two missions scanned the entire sky by combining the spin and precession of the spacecraft with its motion around the Sun.
As sketched in the left panel of \cref{fig:standard_config_and_T_beta}, 
the angle between the direction of observation of the spacecraft and its axis of maximal inertia---the spin axis---is called~$\beta$. The rotation of the spacecraft itself around this axis is thus called the spin. The angle between the spin axis and the vector from the Sun to the spacecraft is referred to as $\alpha$. While in orbit, the spacecraft witnesses a motion of precession of its spin axis around the Sun-spacecraft axis. The periods associated with these two rotations will hereafter be noted as $T_\beta$ and $T_\alpha$, respectively.

Some of the next generation of CMB experiments are also expected to be equipped with a continuous rotating HWP which modulates the polarised signal. Its angle around the optic axis is denoted $\phi$ and its period of rotation $T_{\phi}$.
Instead of the periods $T_j$ ($j \in \{\alpha,\beta,\phi \}$), one can equivalently consider the frequencies:
\begin{align}
    \omega_j &= 2\pi/T_j ~ \rm{[rad/s]}, \\
    f_j      &= 1/T_j    ~ \rm{[Hz]},    \\
    \nu_j    &= 60/T_j   ~ \rm{[rpm]}.
\end{align}
Depending on the context, we will switch between using $T_j,\omega_j,f_j$, or $\nu_j$ in the remainder of this work.
The detailed definition of the spacecraft's precession and spin motions are described in \cref{apd:scan_motion}.
Another crucial parameter defining the scanning strategy is the sampling rate $f_{\rm s}$, i.e., the number of data taken by the instrument per unit of time.
The angles $\alpha$ and $\beta$ are referred to as the `geometric parameters' of the scanning strategy, and the period/angular frequency/frequency/rotation rate, i.e., $T_j/\omega_j/f_j/\nu_j$ for rotation are referred to as the `kinematic parameters'.
The six-dimensional parameter space $\{\alpha, \beta, T_\alpha, T_\beta, \nu_\phi, f_{\rm s}\}$ characterises the scanning strategy of a spacecraft, and finding a reasonable compromise for the parameter set using various criteria represents a highly complex problem. However, combining the constraints that a scientific mission, in particular with a HWP demands, and using simplifying considerations, allows us to reduce the effective dimension of the problem and simplify its optimisation.

\subsection{Constraints on the parameter space}
\subsubsection{Constraints on geometric parameters \label{sec:angle-constraint}}
One commonly used constraint on the geometric parameters is to impose
\begin{align}
    \kappa= \alpha+\beta > 90^{\circ}.\label{eq:const_geometric}
\end{align}
As also noted in ref.~\cite{OptimalScan}, such a constraint is required in order to obtain full-sky coverage of the scanning. Since the effective $\beta$ is different from one detector to another, it is common to choose a higher value of $\kappa$, such as $\kappa \sim 95^{\circ}$ \citep{OptimalScan} in accordance with the specific Field of View~(FoV) of the experiment.
In so doing, one also optimises the configuration to avoid the exposure of the instruments to solar radiation.
This upper bound for $\kappa$ is related to the design of the sun-shield and other mechanisms to control heat input to the payload.
As examples, both \WMAP and \Planck had $\kappa=92.5^{\circ}$, although \EPIC \cite{Bock_epic} was allowed to have $\alpha=45^\circ, \beta=55^\circ$, i.e., $\kappa=100^\circ$ by using a relatively large sun-shield, which correspond to the largest $\kappa$ proposed for a CMB space mission. For a given experimental design, one can thus fix the value of $\kappa$ such that $\alpha$ and $\beta$ are no longer independent.

\subsubsection{Constraints on kinetic parameters}
Next, we consider the periods $T_j$.
The role of the HWP installed on the spacecraft is two-fold: first, it modulates the polarisation with a frequency sufficiently higher than the assumed $1/f$ noise knee frequency~$f_{\rm knee}$ of the instrument; second, it improves the uniformity of the effective crossing angles. The first point is not discussed here because it relates to the issue of optimising the HWP revolution frequency given a $1/f$ noise model.

Therefore, we assume a situation where the $1/f$ noise with a certain $f_{\rm knee}$ can be sufficiently suppressed by a HWP with revolution frequency $f_{\phi}$, and consider the conditions required for the spacecraft rotation periods.
Without a HWP an incoherent polarimetic receiver the polarisation modulation is solely provided by the spacecraft rotation.  Such a configuration requires a high spin rate to obtain high modulation efficiency.
The specific constraints in this case are discussed in ref.~\cite{OptimalScan}, but for a spacecraft with a HWP, the spacecraft's spin will no longer play a primary role in the modulation. On the other hand, we need to consider whether the HWP can modulate the polarisation in the sky pixel when the telescope's pointing transits over a sky pixel, i.e., whether the HWP rotation is fast enough for the sky scan motion. The maximum angular velocity at which the pointing sweeps the sky is given by
\begin{equation}
    \begin{split}
    \omega_{\rm max} &= \omega_\alpha \sin\kappa + \omega_\beta \sin{\beta}\\
    &= 2\pi \left( \frac{\sin\kappa}{T_\alpha} + \frac{\sin\beta}{T_\beta} \right).
    \end{split}
\end{equation}
For completeness, a detailed derivation of this expression of $\omega_{\rm max}$ is given in \cref{apd:sweeping_velocity}.\footnote{We assume here that the directions of rotation for precession and spin are the same. We further discuss the case where one of them is taken in the opposite direction in \cref{apd:rotation_direction}.}

As such, the time $\tau$ that the pointing spends in a region of the size of the FWHM (Full Width at Half Maximum) of the main beam $\Delta \theta$ is
\begin{align}
    \tau = \frac{\Delta \theta}{\omega_{\rm max}}.
\end{align}
When the HWP modulates the polarisation signal, the sampling rate must be determined such that the modulated signal can be demodulated correctly. A HWP revolving at a frequency $f_\phi$ modulates polarisation by $4f_\phi$, so its Nyquist frequency is $8f_\phi$. Introducing the variable $\Nmargin$ ($\Nmargin>1$) as the margin for sampling, the condition that the sampling rate $f_{\rm s}$ requires for demodulation is given by
\begin{align}
    f_{\rm s} > 8f_\phi\Nmargin. \label{eq:sampling_rate}
\end{align}
On the other hand, since the HWP must revolve $\Nmod$ ($\Nmod>1$) times while the pointing transits the angular distance of $\Delta \theta$ in order to modulate the polarisation in the sky pixel, it is required that
\begin{align}
    \Nmod T_\phi &< \tau. \label{eq:modulation_condition}
\end{align}

Combining \cref{eq:sampling_rate,eq:modulation_condition}, the following inequality must always be satisfied:
\begin{align}
    \frac{\Nmod}{\tau} < f_\phi < \frac{f_{\rm s}}{8\Nmargin}. \label{eq:total_condition}
\end{align}
Under this condition, both the proper modulation of the signal from an observed sky region within the FWHM $\Delta \theta$ and a sampling rate allowing for demodulation are satisfied. The constraint imposed on the spacecraft here is on $\tau$, which corresponds to making $\tau$ long enough to satisfy \cref{eq:modulation_condition}, i.e., this provides an upper bound on the maximum angular velocity $\omega_{\rm max}$ at which the instruments can sweep the sky.

In order to impose a constraint on the spacecraft's kinetic parameters, i.e., $T_\alpha$ and $T_\beta$, we can rewrite \cref{eq:modulation_condition} as follows
\begin{align}
    \frac{\Nmod}{f_\phi} &< \frac{\Delta \theta}{2\pi \left( \frac{\sin\kappa}{T_\alpha} + \frac{\sin\beta}{T_\beta} \right)}. \label{eq:req_for_HWP}
\end{align}
Solving this for $T_\beta$ yields to the lower bound
\begin{align}
    T_\beta^{\rm{lower}}\equiv\frac{2\pi \Nmod T_\alpha \sin\beta}{\Delta \theta f_\phi T_\alpha - 2\pi \Nmod \sin\kappa}. \label{eq:T_spin}
\end{align}
In ref.~\cite{OptimalScan}, the value of $T_\beta$ was also constrained by the time constant of the detectors. However, the situation is different in a scenario with a HWP modulation, as is considered here. Indeed, the rotation frequency at which the HWP modulates the signal is set according to the time constant of the detector in order to account for it. Therefore, the time constant becomes a parameter that limits the HWP rotation frequency and not the scanning strategy itself.

A lower bound on $T_\alpha$ can also be obtained by imposing the condition $T_\beta<T_\alpha$ in \cref{eq:T_spin}, to avoid complicating the inertial control of the spacecraft. One can thus define
\begin{align}
    T_\alpha^{\rm{lower}}\equiv\frac{2\pi \Nmod (\sin\beta + \sin\kappa)}{\Delta \theta f_\phi}. \label{eq:T_alpha}
\end{align}
Then, the relationship between the spin period and precession period can be summarised as follows
\begin{align}
    \tbl < T_\beta < T_\alpha.
\end{align}
For full-sky observations, there is also an upper limit on $T_\alpha$, which can be obtained only through simulations, but according to ref.~\cite{OptimalScan}, a precession period of more than 1\,year prevents full-sky observations.

Overall, combining the constraints on the angles and on the periods allows us to reduce the dimensionality of the problem from 6 to 3. Indeed, one can fix $\kappa$ from instrumental considerations, and consider the constraint $\tbl<T_\beta$, such that the parameter space becomes
\begin{align}
    \{\alpha, \beta, T_\alpha, T_\beta, \nu_\phi, f_{\rm s}\}\rightarrow\{\alpha, \kappa-\alpha, T_\alpha, \tbl(\alpha,T_\alpha)<T_\beta, \nu_\phi(f_{\rm knee}), f_{\rm s}(\nu_\phi)\}.
\end{align}
Keeping only the free parameters, the parameter space reduces to the triplet $\{\alpha, T_\alpha, \tbl<T_\beta\}$, where $\nu_\phi$ and $f_{\rm s}$ are scaled to fix a given $f_{\rm knee}$.

\subsection{The case of \LB}\label{sec:case_of_LB}

The \LB space telescope is a JAXA L-class mission selected for launch in 2032 in Japan, with a nominal mission duration of three years \cite{PTEP2023}. The instrument will be composed of three telescopes: LFT (Low Frequency Telescope), MFT (Medium Frequency Telescope) and HFT (High Frequency Telescope), observing respectively at low, medium and high electromagnetic frequencies in the GHz range. \LB's main scientific objective is the measurement of the faint primordial $B$ modes remaining from inflation, as imprinted on the CMB polarised signal anisotropies, to test representative inflationary models by measuring $r$ with an accuracy of $\delta r <0.001$. In order to achieve this, the polarisation is modulated by a continuously rotating HWP, and a sky scanning strategy must be carefully optimised to provide full-sky coverage with high-precision polarisation measurements. As such, \LB provides an ideal example for the present study.

The latest instrument model of the \LB uses the values:
$\alpha=45^{\circ}$, $\beta=50^{\circ}$, such that $\kappa = 95^{\circ}$, in agreement with our discussion in \cref{sec:angle-constraint}. 
Moreover, $T_\alpha= 3.2058$\,hours ($=192.348$\,min), $T_\beta = 20$\,min, $\nu_{\phi}= 46/39/61$\,rpm and $f_{\rm s}=19$\,Hz.\footnote{The exact value is given by $f_{\rm s} =20\,{\rm MHz}/2^{20}$ \cite{PTEP2023}.}
The instruments have FoV of $18^\circ\times8^\circ/14^\circ/14^\circ$ respectively for LFT, MFT, and HFT.\footnote{The MFT and HFT values represent the radius of the FoV.}

\begin{figure}
  \centering
  \includegraphics[width=0.49\columnwidth]{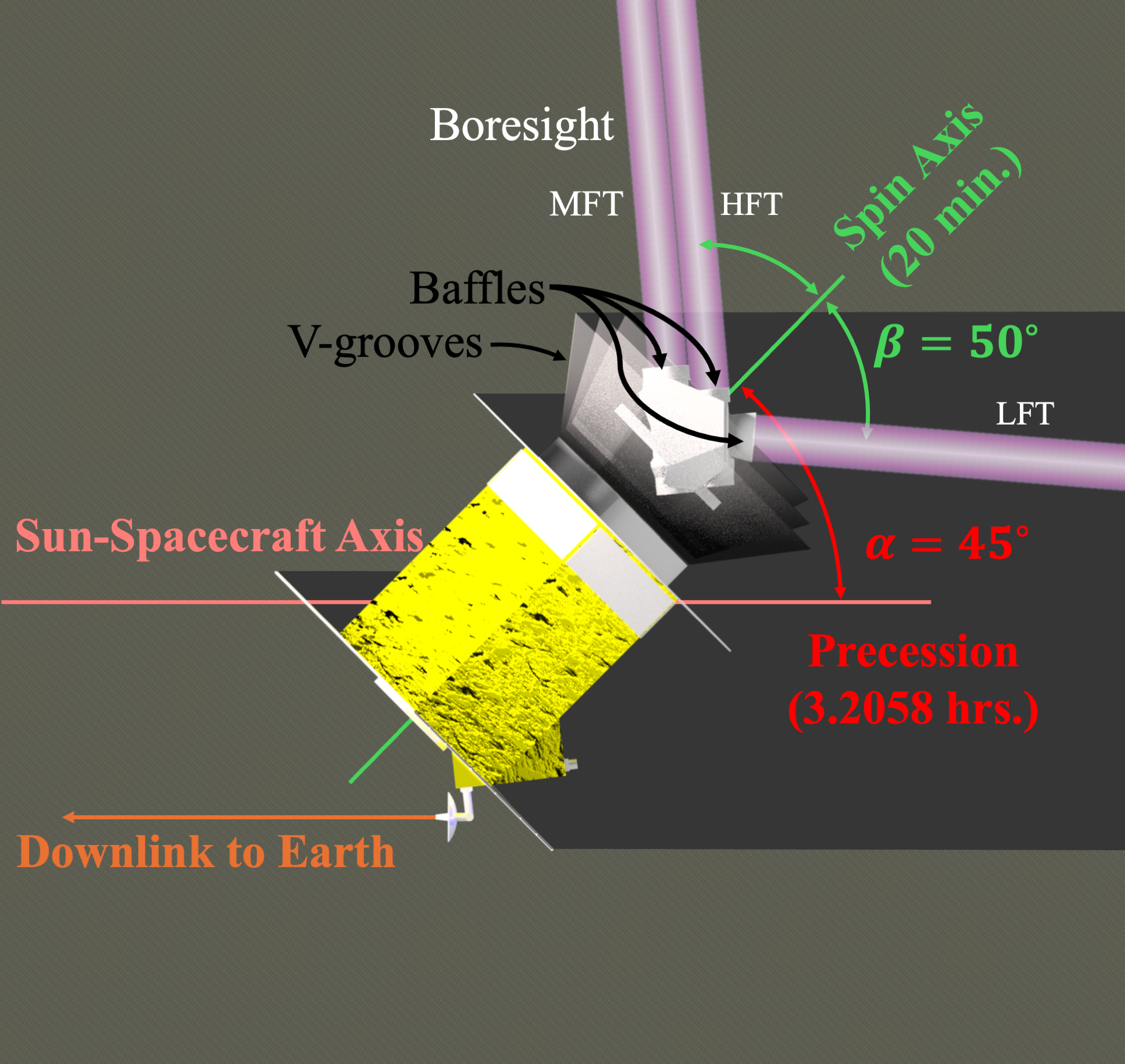}
  \includegraphics[width=0.49\columnwidth]{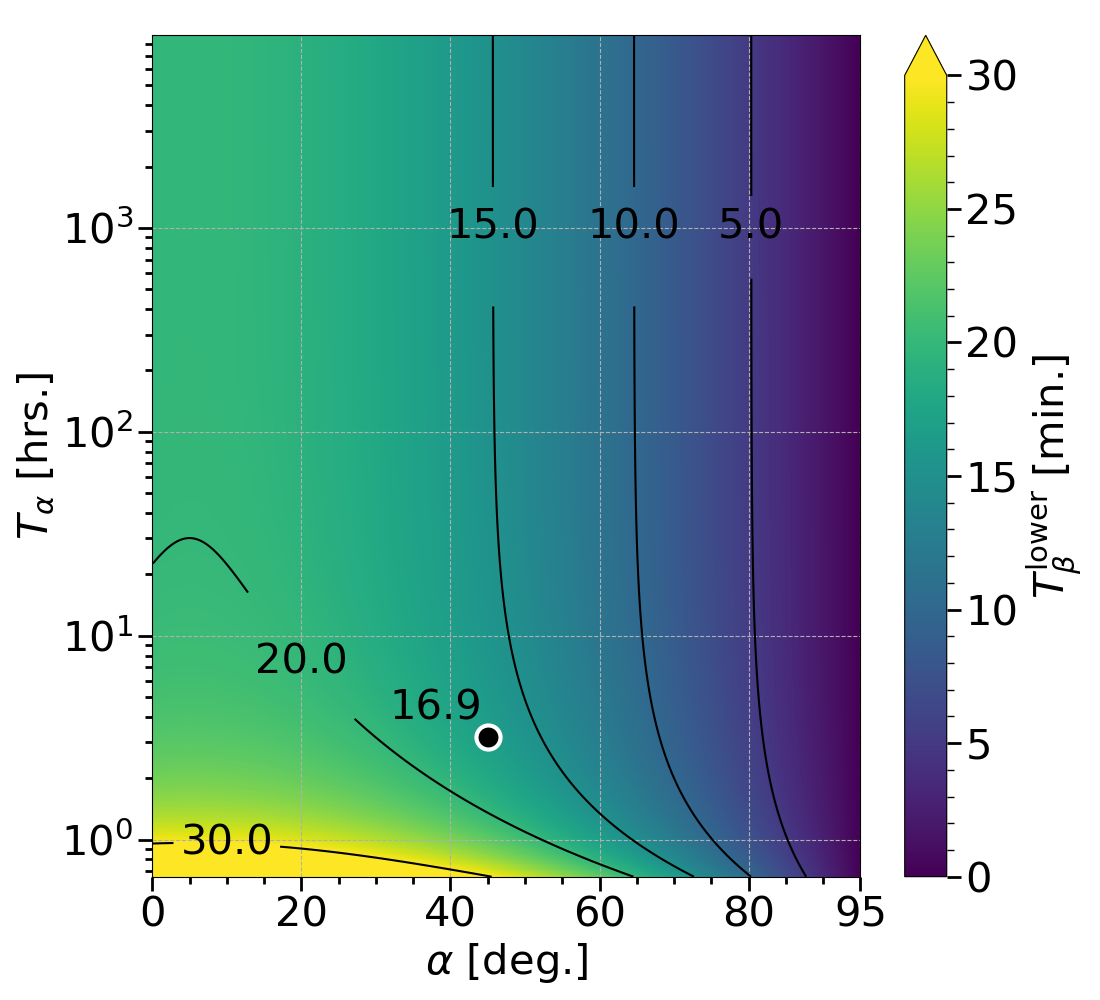}
  \caption{(left) The \SC of \LB's scanning strategy. (right) The lower limit of the spin period, $T_\beta^{\rm{lower}}$, which is determined by \cref{eq:T_spin} in case of the \SC when $\Nmod=1$. We consider an FWHM of the HFT with the highest center frequency of 402\,GHz and $\Delta \theta=17.9$\,arcmin. The black dot in this and subsequent similar figure represents the position which has $(\alpha, T_\alpha, T_\beta)=(45^\circ, 192.348\,\rm{min}, 16.9\,\rm{min})$. Of these, $\alpha$ and $T_\alpha$ are the values of \LiteBIRD's \SC.}
  \label{fig:standard_config_and_T_beta}
\end{figure}

\begin{figure}
  \centering
  \includegraphics[width=0.9\columnwidth]{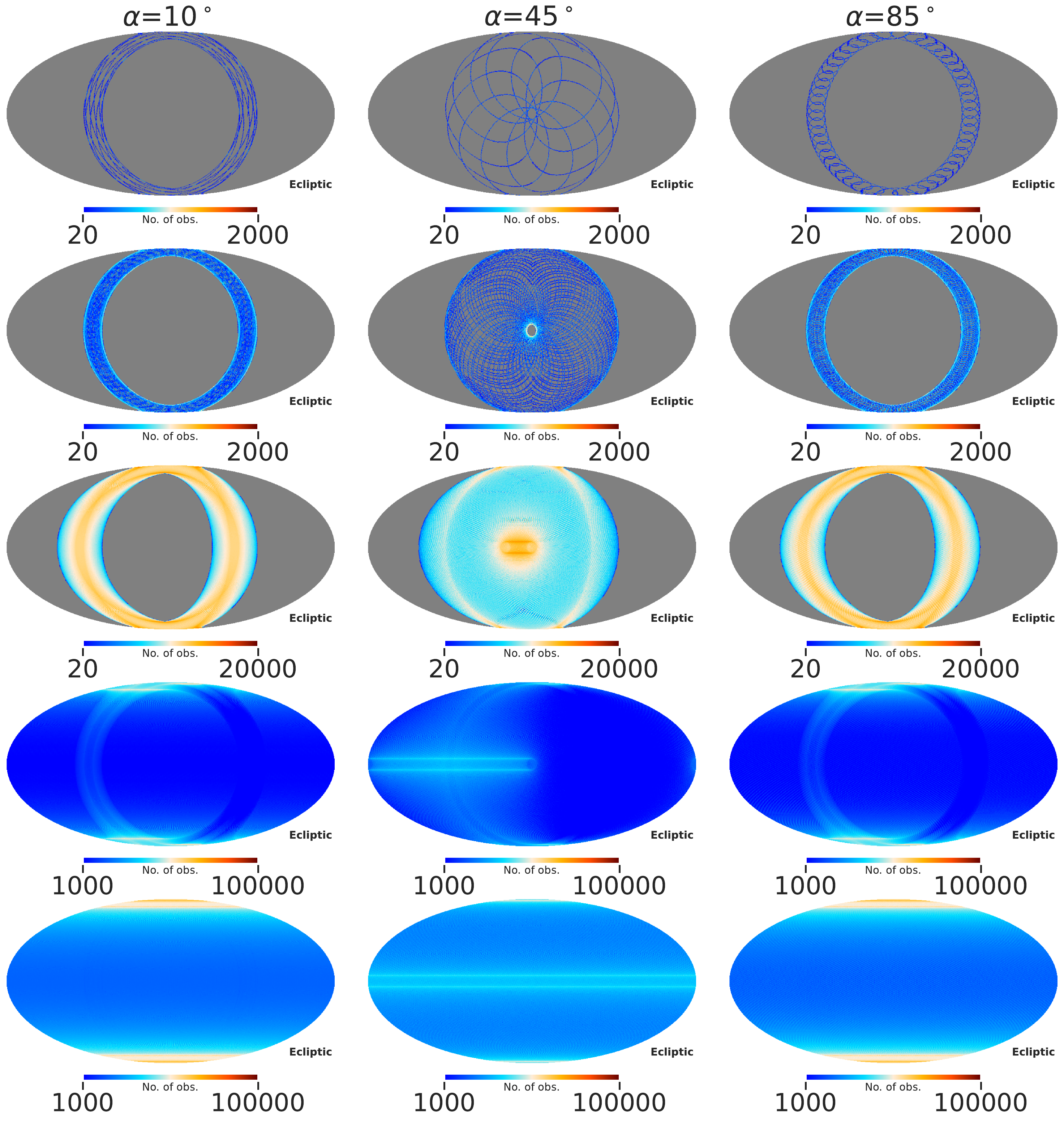}
  \caption{Hit-map which counts the number of observations, i.e., hits in each sky pixel and its time evolution for each geometric parameter. The middle column corresponds to the \SC, and in all columns $\kappa$ is fixed at $95^\circ$ and $T_\alpha$ at 192.348 min. In the left/right columns, $\alpha$ is $10^\circ/85^\circ$, and $T_\beta$ is 21.9/3.8 min, respectively, obtained by calculating \Tbetalow in \cref{eq:T_spin}. The top to bottom rows show the time evolution of the hit-map from a single precession period of the \SC, i.e., 192.348\,min, 1\,day, 1\,month, 6\,months, and up to 1\,year.
  Each hit-map is simulated for a single detector on the boresight with a sampling rate of 19\,Hz and $N_{\rm side}=128$, i.e., the size of a sky pixel is approximately $0.46^\circ$}
  \label{fig:hitmaps}
\end{figure}

We will refer to this setup as the \SC, illustrated in \cref{fig:standard_config_and_T_beta} (left) and summarised in \cref{tab:LB_standard_config}.
The hit-map, which counts the number of observations per sky pixel, is simulated with the \SC and its time evolution is shown in \cref{fig:hitmaps} (middle column).
The figure also shows the hit-maps simulated with $\alpha=10^\circ$ (left column) and $\alpha=85^\circ$ (right column), showing how the geometric parameters impact the scan trajectory.
In the two cases, $T_\alpha$ is the same as the \SC and $T_\beta$ is simulated by \Tbetalow calculated with the respective $\alpha$ and $T_\alpha$ pairs.

The goal of the present work is to explore and rigorously justify the choice of such a configuration with regard to in-flight calibration and suppression of systematic effects. To consider the constraints given by \cref{eq:T_spin}, we apply the HFT's HWP revolution rate which is the highest of the telescopes, with correspondingly the smallest FWHM ($\Delta \theta=17.9$\,arcmin) at 402\,GHz. We assume these parameters for all frequency bands, and we use $\Nmod=1$ and $\Nmargin=2$, as suggested in ref.~\cite{PTEP2023}. In the \LiteBIRD instrument model, as long as \cref{eq:T_spin} is satisfied for the HWP revolution rate and FWHM pair, it is also satisfied for the LFT and MFT cases.
Calculating \Tbetalow from these values, we can create the space shown in \cref{fig:standard_config_and_T_beta} (right), where $\tbl=16.9$\,min for $(\alpha,T_\alpha)=(45^\circ,192.348\,\rm{min})$.

\begin{table}
    \begin{tabular}{cccccccc}
        \hline
        $\alpha$     & $\beta$      & Precession period   & Spin period  & \multicolumn{3}{c}{HWP revolution rate}  & Sampling rate \\
                     &              &                     &              &  LFT         & MFT        & HFT          &               \\ 
        {[deg]}      & {[deg]}      & {[min]}             & {[min]}      & \multicolumn{3}{c}{[rpm]}                & {[Hz]}        \\ \hline
        45           & 50           & 192.348             & 20           & 46           & 39         & 61           & 19            \\ \hline
    \end{tabular}
    \caption{Parameters of the \SC for the \LB mission \cite{PTEP2023}.}
    \label{tab:LB_standard_config}
\end{table}

From the \LB instrument model and the discussion in the previous section, the effective free parameters of the scanning strategy to optimise are then given by
\begin{align}
    \{\alpha, \beta, T_\alpha, T_\beta, \nu_\phi, f_{\rm s}\}
    \rightarrow
    \{ \alpha, 95^{\circ}-\alpha, T_\alpha, 16.9\,\mathrm{min} < T_\beta, 61\,\rm{rpm}, 19\,\rm{Hz} \}. \label{eq:scan_parameter_space}
\end{align}

In the following analysis, all figures show results that are simulated considering a single boresight detector, i.e., a single detector located at the center of the telescope's focal plane.
We verified that performing the same study for detectors located at the edge of the FoV in the telescopes, does not change our conclusions regarding the optimisation choice discussed in \cref{sec:optimisation}. 
A detailed discussion of the detectors non located at the boresight can be found in \cref{apd:other_detector}.

\section{Metrics for optimisation} \label{sec:metrics}

\subsection{Visibility time of planets}

It is clear from the results of the \Planck experiment that compact sources such as solar planets and distant clusters of galaxies are valuable resources for in-flight beam and pointing calibration \cite{Planck_HFI_beam, Planck_LFI_beam}. 
One would then expect a scanning strategy to maximise observation time on these objects while still completing its primary science goals during the defined mission duration. The integrated time spent by the boresight on a compact source is a quantity that is expected to be proportional to the calibration accuracy, and we search for the parameters which can maximise the integrated visible time during the mission duration in the $\{\alpha, T_\alpha \}$ space.

\subsection{Hit-map uniformity}

Uniformity of the hit-map across the sky results in uniformity of the sensitivity across the sky. On the other hand, significant differences in sensitivity from one region of the sky to another would increase noise variance, which can degrade the accuracy of foreground removal.

We define the spherical coordinates $\Omega=(\theta,\varphi)$ and use the index $i$ to label the sky pixels as defined within the \texttt{HEALPix} framework \cite{healpix}.\footnote{\url{https://healpix.sourceforge.io/}} To quantify the uniformity of the sensitivity across the sky, we use the standard deviation of the hit-map, $\sigma_{\rm hits}$, which is defined as
\begin{align}
    \sigma_{\rm{hits}} = \sqrt{\frac{1}{N_{\rm pix}-1}\sum_{i=0}^{N_{\rm pix}-1}\left(\langle\Nhits\rangle-\Nhits(\Omega_i)\right)^2},
\end{align}
where $\Nhits$ represents total number of hits/observations and $\langle...\rangle$ denotes the entire sky average. A small value for $\sigma_{\rm hits}$ indicates that the hit-map is uniform and we can expect the sensitivity per sky pixel to be uniform as well. We examine the distribution of $\sigma_{\rm hits}$ in the $\{\alpha, T_\alpha\}$ space.

\subsection{Cross-link factor} \label{sec:spin_char}

According to ref.~\cite{OptimalScan}, a well-designed scanning strategy that has crossing angles for a sky pixel can suppress some systematic effects. In order to quantify the uniformity of the crossing angles, we define the following quantity which is referred to as the `cross-link factor':
\begin{equation}
    \begin{split}
        \abs{{}_{n, m}\tilde{h}(\Omega)}^2 &= \left(\frac{\sum_{j}^{\Nhits}\cos(n\psi_j+m\phi_j)}{\Nhits}\right)^2+\left(\frac{\sum_{j}^{\Nhits}\sin(n\psi_j+m\phi_j)}{\Nhits}\right)^2\\
        &= \langle \cos(n\psi_j+m\phi_j) \rangle^2 + \langle \sin(n\psi_j+m\phi_j) \rangle^2,
    \end{split}\label{eq:crosslink}
\end{equation}
where $\psi$ is the crossing angle, which is defined as the angle between the scan direction of the detector and the meridian, and $\phi$ is the HWP angle.\footnote{The HWP angle $\phi$ is defined such that the direction of increasing crossing angle $\psi$ and rotation angle obey either the IAU or COSMO (\texttt{HEALPix}) convention (\url{https://lambda.gsfc.nasa.gov/product/about/pol_convention.html}), allowing us to define polarisation consistently on the sphere.} In this paper we assume that the crossing angle and the detector polarisation angle are always parallel. The subscript $j$ represents the $j^{\rm{th}}$ measurement at the sky pixel and the pair of integers $(n,m)$ is called the \spin. 
A perfectly uniform distribution of the crossing leads to $\abs{{}_{n, m}\tilde{h}(\Omega)}^2=0$ for all pairs $(n,m)$. As such, the crosslink factor is a quantity to be minimised and we will generally say that a scanning strategy has a `good cross-linking' when the cross-link factor is relatively small.
It has been shown in refs.~\cite{OptimalScan, mapbased} that a scanning strategy minimising the cross-link factors across the sky can efficiently suppress the systematic effects. 

In this paper, we include the continuous rotating HWP in \cref{eq:crosslink} and examine the distribution of the cross-link factors per \spin-$(n,m)$ in the $\{\alpha, T_\alpha\}$ space. As the final metric, we consider the entire sky average of the cross-link factor, \crosslinkmean as suggested in ref.~\cite{OptimalScan}. 
In order to demonstrate how the cross-link factor can contribute to the suppression of specific systematic effects, we introduce in \cref{sec:formalism} a formalism in which some systematic effects are expressed in term of the cross-link factors in the \spin space.

\section{Formalism and method} \label{sec:formalism}

\subsection{Contribution of cross-link factor for systematic effects suppression}

Following the formalism introduced in refs.~\cite{spin_characterisation, mapbased}, we can express the signal detected by a CMB experiment within a pixel of spherical coordinates $\Omega$ as a function of the orientation angles as
\begin{equation}
    S^d(\Omega,\psi,\phi)=h(\Omega,\psi,\phi)S(\Omega,\psi,\phi),
\end{equation}
where $S$ describes the signal field, which expresses the signal received by the instrument at each visit of the sky pixel, and the real space scan field, $h$, describes the observation by the detector modulated by the HWP in each sky pixel under a specific scanning strategy. As discussed in ref.~\cite{wallis_new_map-making}, $h$ can be defined as
\begin{equation}
  h(\Omega,\psi, \phi)= \frac{4 \pi^2}{N_{\rm hits}(\Omega)}\sum_j\delta(\psi-\psi_j)\delta(\phi-\phi_j),
\end{equation}
where $\delta$ is the Dirac delta function, so that the scan field $h$ acts as a window function on the signal fields.
It is then possible to decompose the signal in a Fourier series in the discrete \spin space $(\psi,\phi) \to (n,m)$ as
\begin{align}
    S^d(\Omega,\psi,\phi)       &= \sum_{n,m} \Sd[n,m](\Omega)e^{i n\psi}e^{i m\phi}, \\
    \Sd[n,m](\Omega) &= \sum_{n'=-\infty}^{\infty} \sum_{m'=-\infty}^{\infty} \h[\Delta n, \Delta m](\Omega) \St[n',m'](\Omega), \label{eq:kSd}
\end{align}
where we here introduce $\Delta n = n-n'$ and $\Delta m = m-m'$, and define the orientation function, $\h[\Delta n,\Delta m]$ by Fourier transform of the real space scan field as
\begin{equation}
    \begin{split}
        {}_{n,m}\tilde{h}(\Omega) &= \frac{1}{4\pi^2}\int d\psi \int d\phi h(\Omega,\psi,\phi)e^{-i n\psi}e^{-i m\phi} \\
        &= \frac{1}{N_{\rm hits}}\sum_{j}e^{-i(n\psi_j + m \phi_j)}.\label{eq:exp_crosslink}
    \end{split}
\end{equation}
By taking the norm of the orientation function, we obtain \cref{eq:crosslink}, i.e., the cross-link factor.
The ${}_{\Delta n,\Delta m}\tilde{h}$ term in \cref{eq:kSd}, which corresponds to the cross-link factor, couples the \spin-$(n-n',m-m')$ signal to the \spin-$(n,m)$ quantity of interest, or, in the case where $(n',m')=(n,m)$, behaves as the window function of the observation. In general, one should consider the infinite series of \spin-$(n',m')$ in \cref{eq:kSd}, but it can be shown that specific values of the pair $(n,m)$ can be associated with specific systematic effects, such that a finite series of \spin terms can give an accurate representation of the signal under simplifying assumptions.
(See \cref{apd:pointing_offset,apd:HWP_sys} and refs.~\cite{spin_characterisation,mapbased} for details).

In addition, since the maps of \cref{eq:exp_crosslink} can be pre-computed for a particular scanning strategy, we can efficiently evaluate the impact of the systematic effect associated with each spin term by using a map-based approach without running Time Ordered Data (TOD) simulations each time.

\subsection{Map-based simulation using the \spin formalism}\label{sec:map-based-sim}
According to ref.~\cite{mapbased}, we can assess some systematic effects without time-dependency by map-based approach by using the \spin. In the previous section, we included the HWP revolution to the orientation function, then we can construct similar map-based approach while taking into account the HWP revolution. 
We define the bolometric equation which gives us a $j^{\rm th}$ TOD as following
\begin{align}
    d_j = I + \frac{1}{2}Pe^{-i(4\phi_j-2\psi_j)} + \frac{1}{2}P^*e^{i(4\phi_j-2\psi_j)},
\end{align}
where $I$ represents the intensity of Stokes parameter, and $P$ is a complex number which is defined as $P=Q+iU$ and its complex conjugate, $P^*$ with the Stokes parameter of $Q$ and $U$.
To estimate the Stokes parameters from the signal, a linear regression analysis can be performed, which in \spin space is denoted as
\begin{align}
    \mqty(\hat{I} \\ \hat{P} \\ \hat{P}^*)
    &=
    \mqty(
    1                   & \frac{1}{2}\h[-2,4] & \frac{1}{2}\h[2,-4] \\
    \frac{1}{2}\h[2,-4] & \frac{1}{4}         & \frac{1}{4}\h[4,-8] \\
    \frac{1}{2}\h[-2,4] & \frac{1}{4}\h[-4,8] & \frac{1}{4}
    )^{-1}
    \mqty(
    \Sd[0,0]  \\ \frac{1}{2}\Sd[2,-4] \\ \frac{1}{2}\Sd[-2,4]
    ).\label{eq:map-making_spin}
\end{align}
%\yusukeadd{
A detailed derivation is given in \cref{apd:map-making}. This formalism has applications beyond the one presented in ref.~\cite{mapbased} and can be generalised to allow for a HWP to be taken into account.
We verify explicitly in \cref{apd:TOD_comparison} that this new method gives results which are consistent with TOD-based binning map-making simulations. 
Also, here the demodulation is done in the vector part of the right side of \cref{eq:map-making_spin} as we can see in \cref{eq:map-making_TOD} which is describing TOD-based binning map-maker.
This method is different from the method used to obtain independent TOD of $I$, $Q$ and $U$ with the lock-in technique using HWP, and its implementation into the map-based simulation is left for future study.\footnote{It has not yet been determined what demodulation method will be used in the case of \LiteBIRD.}
However, the time-independent systematic effects discussed in the next section can be accounted for by cross-link factors, so the conclusions of this paper based on them are not dependent on the demodulation method used in map-making.
%}

In this formalism, if the map of $\h[n,m]$ is computed in advance under a certain scanning strategy, all calculations in \cref{eq:map-making_spin} are just quadratic operations of the maps, and this is a faster approach than the computationally expensive TOD-based simulations. However, it is difficult for this method to handle time-dependent systematic effects, such as $1/f$ noise or scan-synchronised signals, and TOD simulations are still required in order to correctly assess these.

\subsection{Instance of systematic effects}\label{sec:formalism_syst}

We introduce a formalism to characterise systematic effects in \spin space. This formalism can be used to express some systematic effects without having to consider any time-dependencies. 

As an instance of a systematic effect, we consider the pointing offset and the instrumental polarisation due to the HWP, whose formalism in \spin space is described in \cref{apd:pointing_offset,apd:HWP_sys}. These are just a couple of examples of possible systematic effects impacting the next generation of CMB polarimetric space missions, but they are good choices for the purpose of demonstrating the methods and to investigate how the value of the cross-link factor maps can be propagated to the bias on $r$.

\subsection{Scanning observation simulator: \Falcons}\label{sec:Falcons}

For the purpose of this study, we developed the public package \Falcons, allowing for fast simulation of spacecraft observations in the \texttt{Julia} programming language \cite{Julia-2017}.\footnote{\url{https://github.com/yusuke-takase/Falcons.jl}} It computes the spacecraft pointing and cross-link factors for a given observation time from a set of input scanning strategy parameters.

Internally, the code uses the \texttt{HEALPix} framework through the \texttt{Healpix.jl} module \cite{healpix.jl}.\footnote{\url{https://github.com/ziotom78/Healpix.jl}} The loop computation for spacecraft pointing is thread-parallel, which is expected to be faster on multi-core workstations and clusters. 
In benchmarks using a modern laptop, we can compute the pointings of a single detector case in about 3\,seconds of run-time during a 1\,year mission duration with a 1\,Hz sampling rate. The benchmark was performed with a single thread and it is expected that the computation time will scale with the number of threads on the CPU.\footnote{A Macbook Pro 2020 (RAM-16\,GB) with Apple M1 processor was used for the benchmarks.}
\section{Results}\label{sec:results}

In this section, we display the results given by each optimisation metric: visibility time of planets; hit-map uniformity; and cross-link factor. 
For display purposes, we consider only the special case $T_\beta=\tbl$, so that every metrics is evaluated in the $\{\alpha,T_\alpha\}$ space, as in \cref{fig:standard_config_and_T_beta} (right). 
It is possible to show that a different choice of the value of $T_\beta>\tbl$ only scales the distribution of the metrics in the $\{\alpha, T_\alpha\}$ space, such that the different optimum values remain unchanged. This assumption is further justified in \cref{apd:T_beta_scaled}.

\subsection{Visibility time of planets}\label{sec:comp_source_obs}

We consider that the boresight has observed a planet when it approaches at an angular distance less than $0.5^\circ$ away from it. When this condition is met, the observation is accounted for in the visibility time. The assumptions made for the planetary motion are as follows.
\begin{itemize}
    \item Using \texttt{Astropy} \cite{astropy}, the position of the planet at 2032-04-01T00:00:00 is taken as the initial condition in Barycentric Dynamical Time.\footnote{\url{https://www.astropy.org/}}
    \item The position of the planet is updated every second and is assumed not to move for that second.
    \item Mars, Jupiter, Saturn, Uranus and Neptune are all considered as calibration sources.
\end{itemize}

\Cref{fig:planet_and_sigma_hit} (left) shows the distribution in $\{\alpha, T_\alpha\}$ space of the integrated visibility time over the 3\,years of the mission duration, which considers all of the observations of the boresight for all of the planets considered. 
There is no significant difference in the shape of the distribution, including the position of the peaks, among the planets considered. The peak heights are as follows: Mars, 0.77\,hours; Jupiter, 1.00\,hour; Saturn, 0.87\,hours; and Neptune, 0.83\,hours.
It can be seen that the value of $T_\alpha$ has a small impact on the integrated visibility time in the range $T_\alpha\lesssim100$\,hours.
On the other hand, when $\alpha$ varies, the integrated visibility time reaches its maximum value for $\alpha=\beta=47.5^\circ$, and the distribution presents a symmetrical shape centered on this value.
This can be explained by the difference in trajectories for each $\alpha$, shown in \cref{fig:hitmaps}. This is because the two parallel lines symmetrically formed across the equator by the shift of the tangent line of the scan due to orbital rotation overlap at the equator when $\alpha=\beta=47.5^\circ$.
On the other hand, while the planet is traversing the inner gap, that we refer to as the `scan pupil', the planet cannot be observed. The pupil can be seen at the center of the hit-map simulated for a 1-day duration in \cref{fig:hitmaps}. Therefore, scanning strategies with a value of $\alpha$ significantly different from $47.5^\circ$ are associated with a lower integrated visibility time.

\subsection{Hit-map uniformity}\label{sec:hitmap_dist}

\begin{figure}[htbp]
  \centering
  \includegraphics[width=0.49\columnwidth]{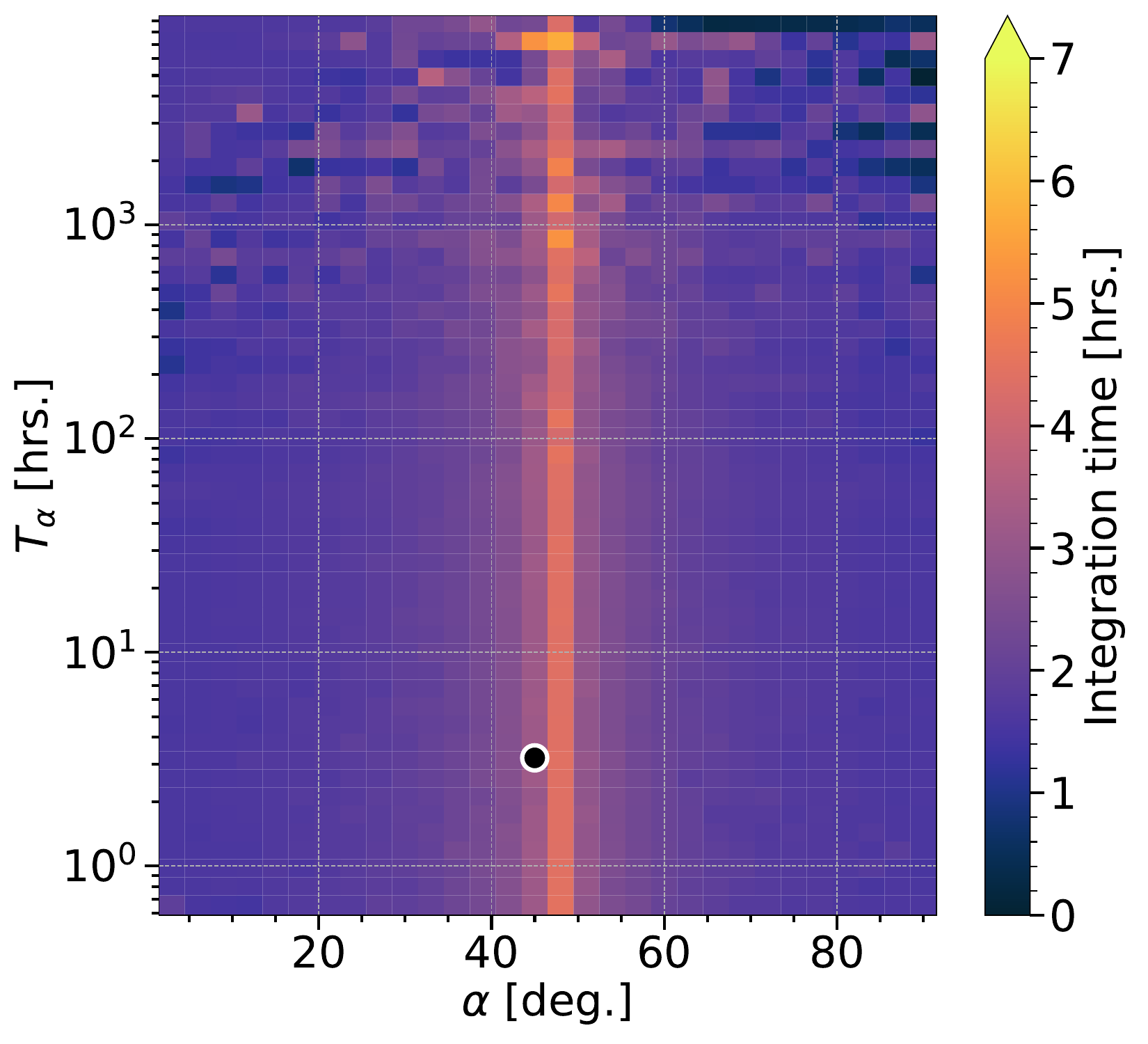}
  \includegraphics[width=0.49\columnwidth]{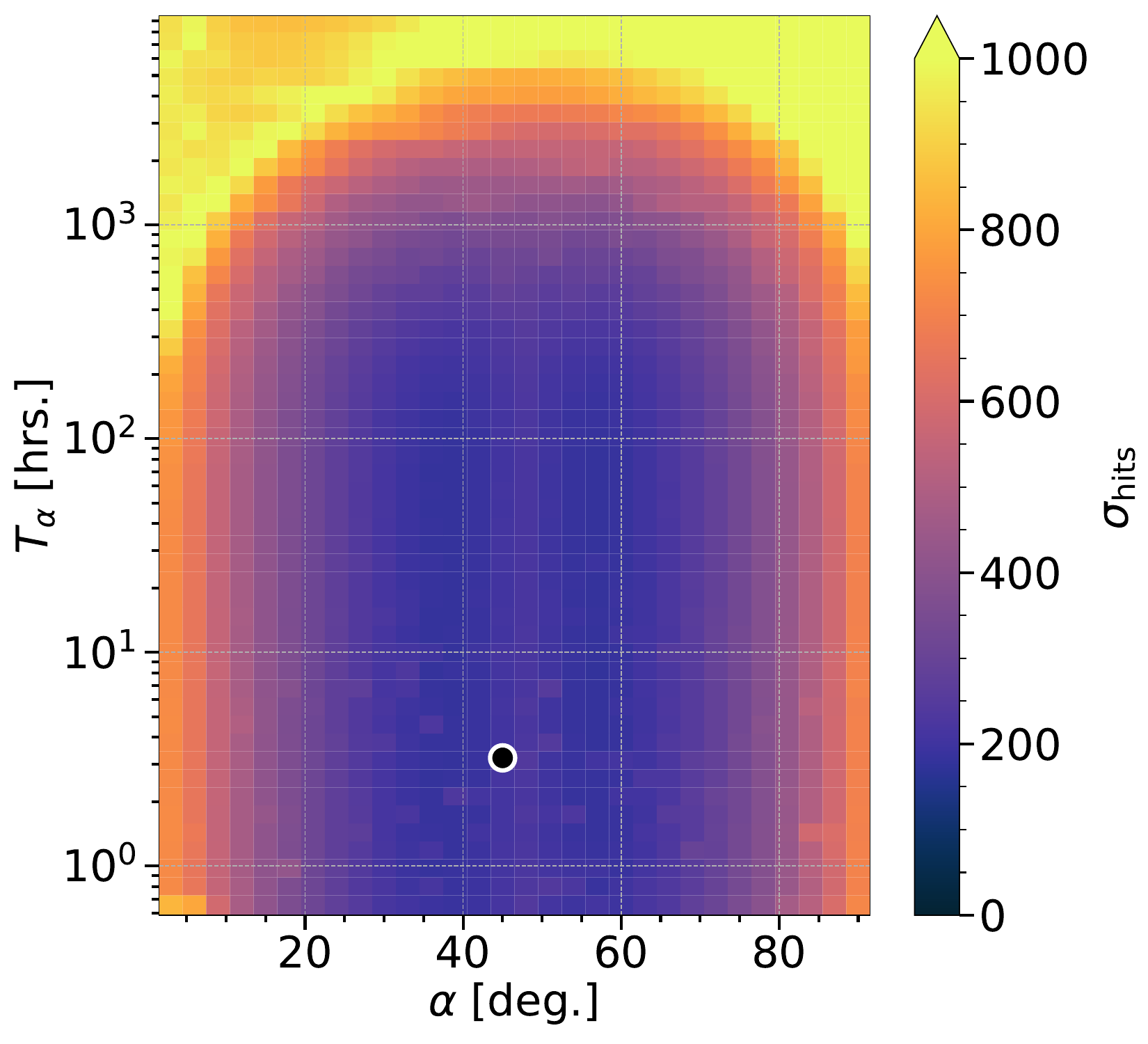}
  \caption{(left) The distribution of integrated visible time which is accumulated by considered planets.
  (right) Standard deviation of the hit-map which has $N_{\rm side}=256$. In the cases where $\alpha$ is extremely large or small, the $\sigmahits$ is larger because the number of hits on the poles is higher and the number of hits on the tropics region is lower.
  }
  \label{fig:planet_and_sigma_hit}
\end{figure}

\Cref{fig:planet_and_sigma_hit} (right) shows the distribution of $\sigma_{\rm hits}$. The results suggest that scanning strategies with extremely large or small values for $\alpha$, such as ones shown in \cref{fig:hitmaps} (left/right), can be associated with a large value of $\sigma_{\rm{hits}}$.
The hit-map after 1\,year of observation shows that such scanning strategies realise many observations near the poles, while the regions around the equator are not observed as often, leading to an increase in $\sigma_{\rm{hits}}$.
In addition, scanning strategies with $T_\alpha>100$\,hours allow
only for a few degrees of orbital rotation during the precession period. This prevents multiple precession cycles in the same region of the sky, leading to the loss of azimuthal symmetry (seen in \cref{fig:hitmaps}) and an increase in $\sigma_{\rm{hits}}$.

It can be seen that $\sigmahits$ is almost invariant in the range $25^\circ \lesssim \alpha \lesssim 75^\circ$ and $T_\alpha\lesssim100$\,hours, with a slight increase for the special case $\alpha=\beta$.
The reason for this increase is that, in this special case, the trajectories intersect on the equator, i.e., the scan pupil closes, such that the poles and the equator are scanned intensively, while the number of observations in the equatorial region is reduced, resulting in a larger variance.

From the discussion of the previous section, we can thus note the existence of a trade-off to be found between the minimisation of $\sigmahits$ and the maximisation of the visibility time of the planet dictated by the size of the pupil.

In \cref{fig:planet_and_sigma_hit} (right), one can also notice the presence of outlier pixels in the $T_\alpha\lesssim10$\,hours region, for which the value of $\sigmahits$ deviate slightly from the trend of neighboring pixels. This effect can be explained by the existence of a resonance between the spin and the precession periods, and will be discussed in more detail in \cref{sec:Opt_kinetic}. 

\subsection{Cross-link factor} \label{sec:result_crosslink}

\begin{figure}
  \centering
  \includegraphics[width=0.32\columnwidth]{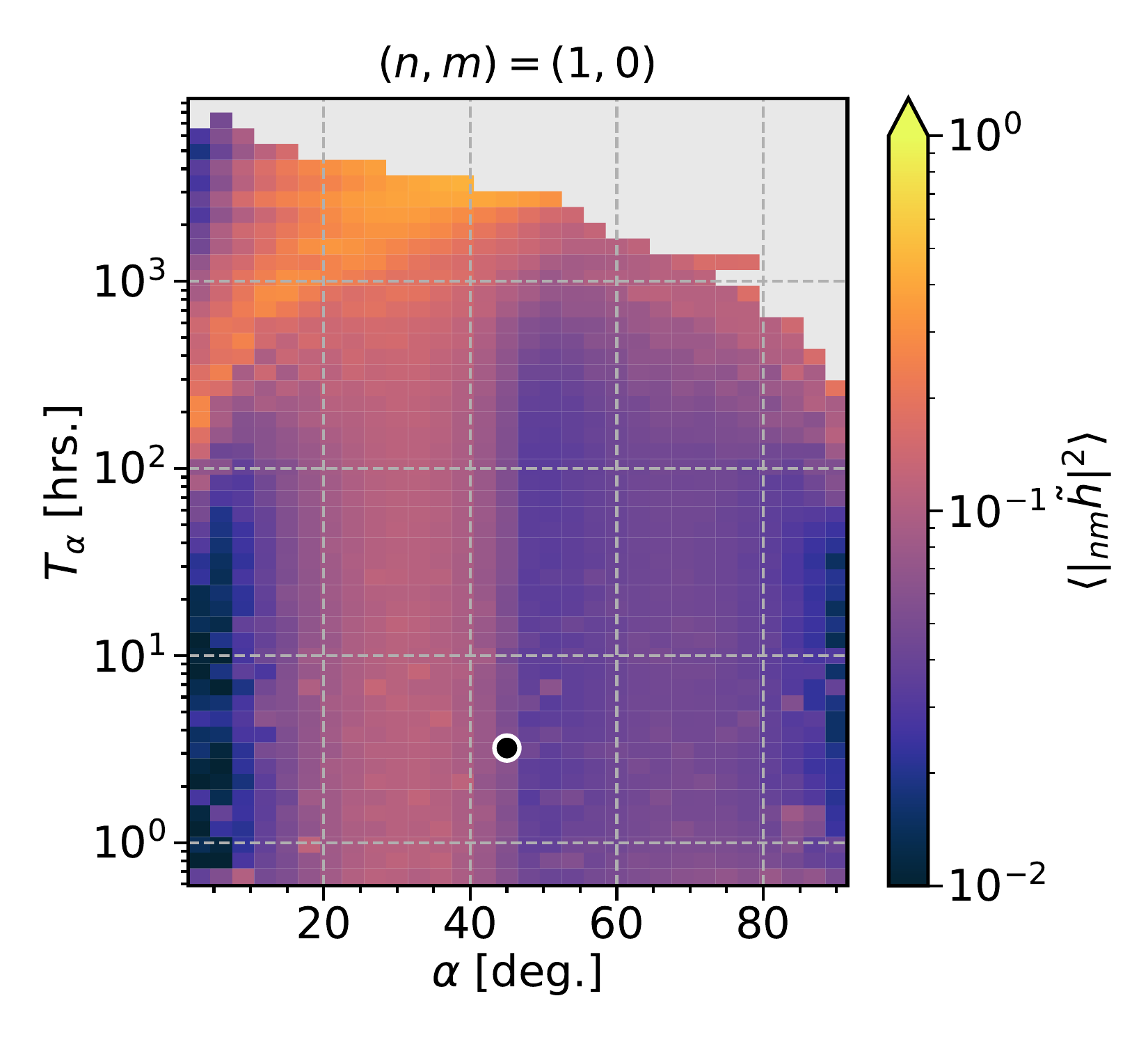}
  \includegraphics[width=0.32\columnwidth]{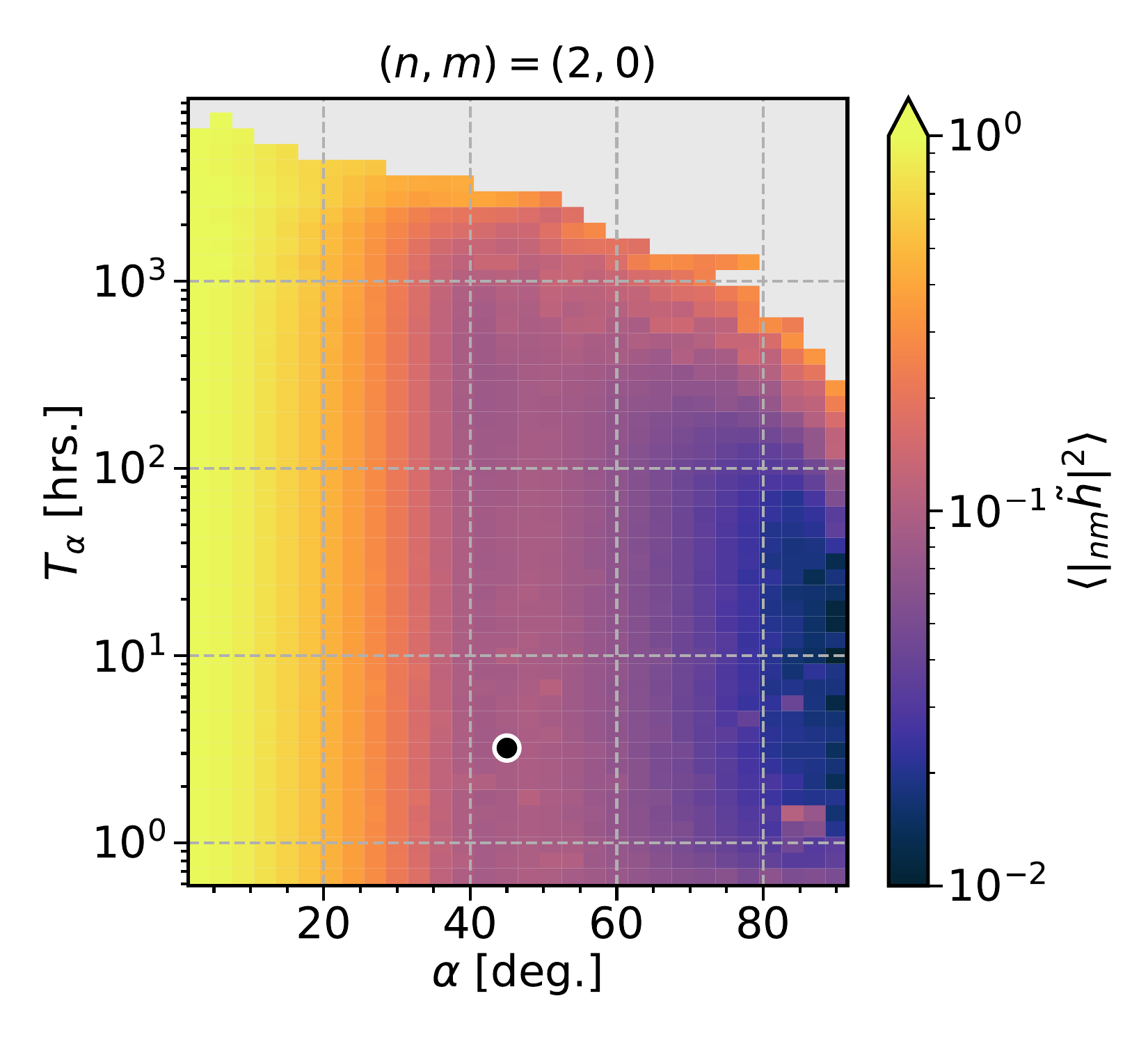}
  \includegraphics[width=0.32\columnwidth]{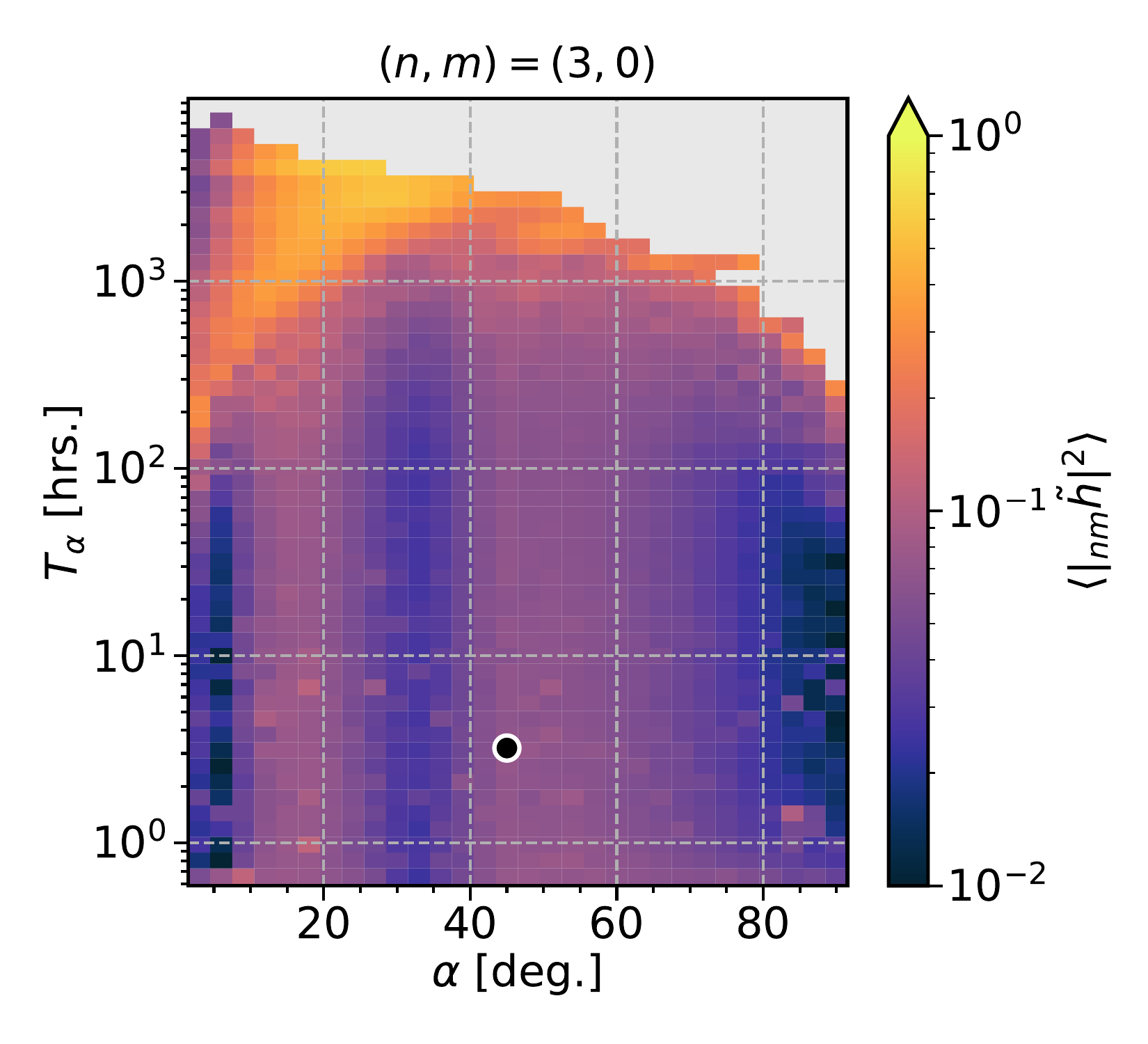}\\
  \includegraphics[width=0.32\columnwidth]{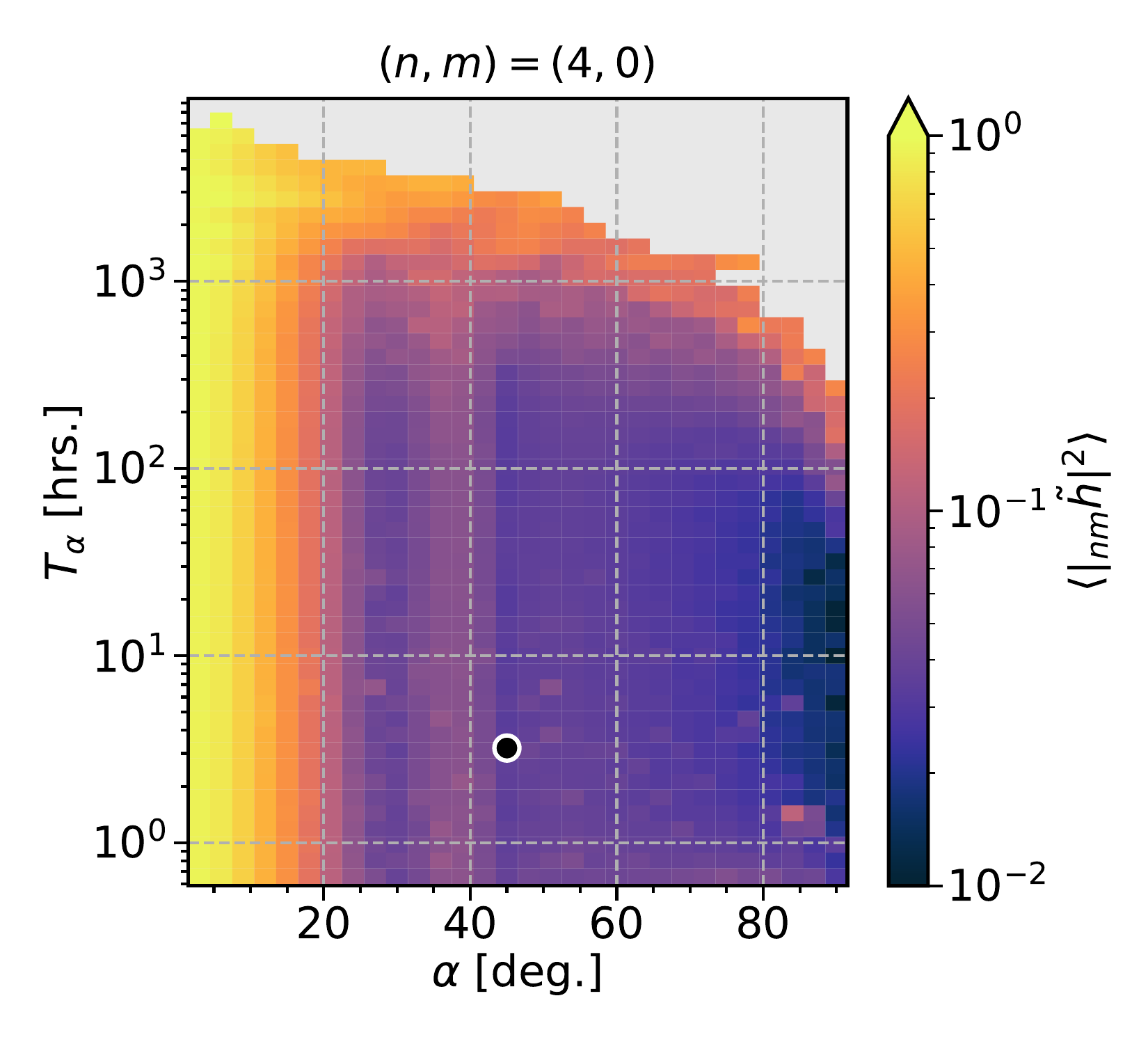}
  \includegraphics[width=0.32\columnwidth]{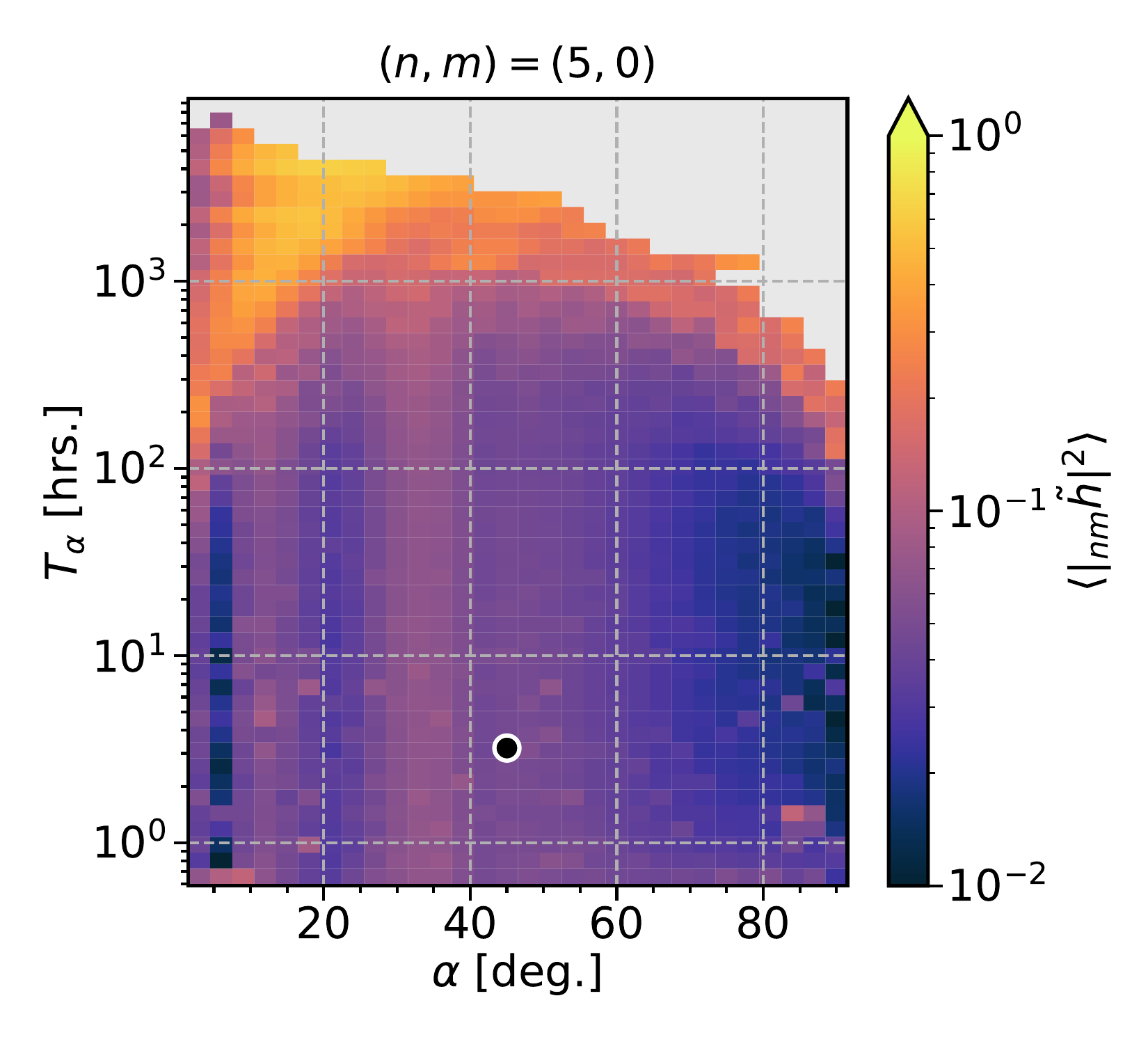}
  \includegraphics[width=0.32\columnwidth]{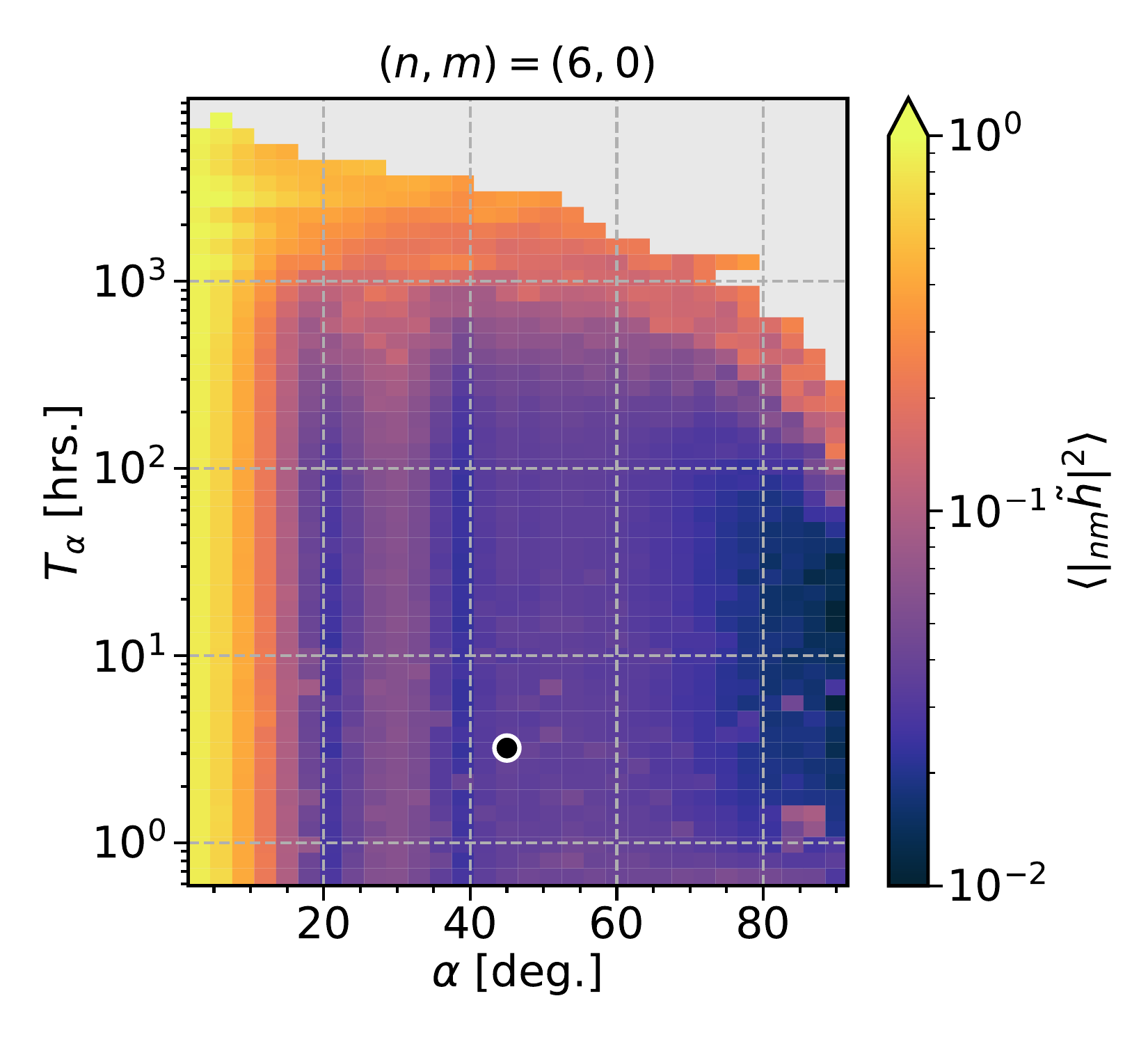}\\
  \includegraphics[width=0.32\columnwidth]{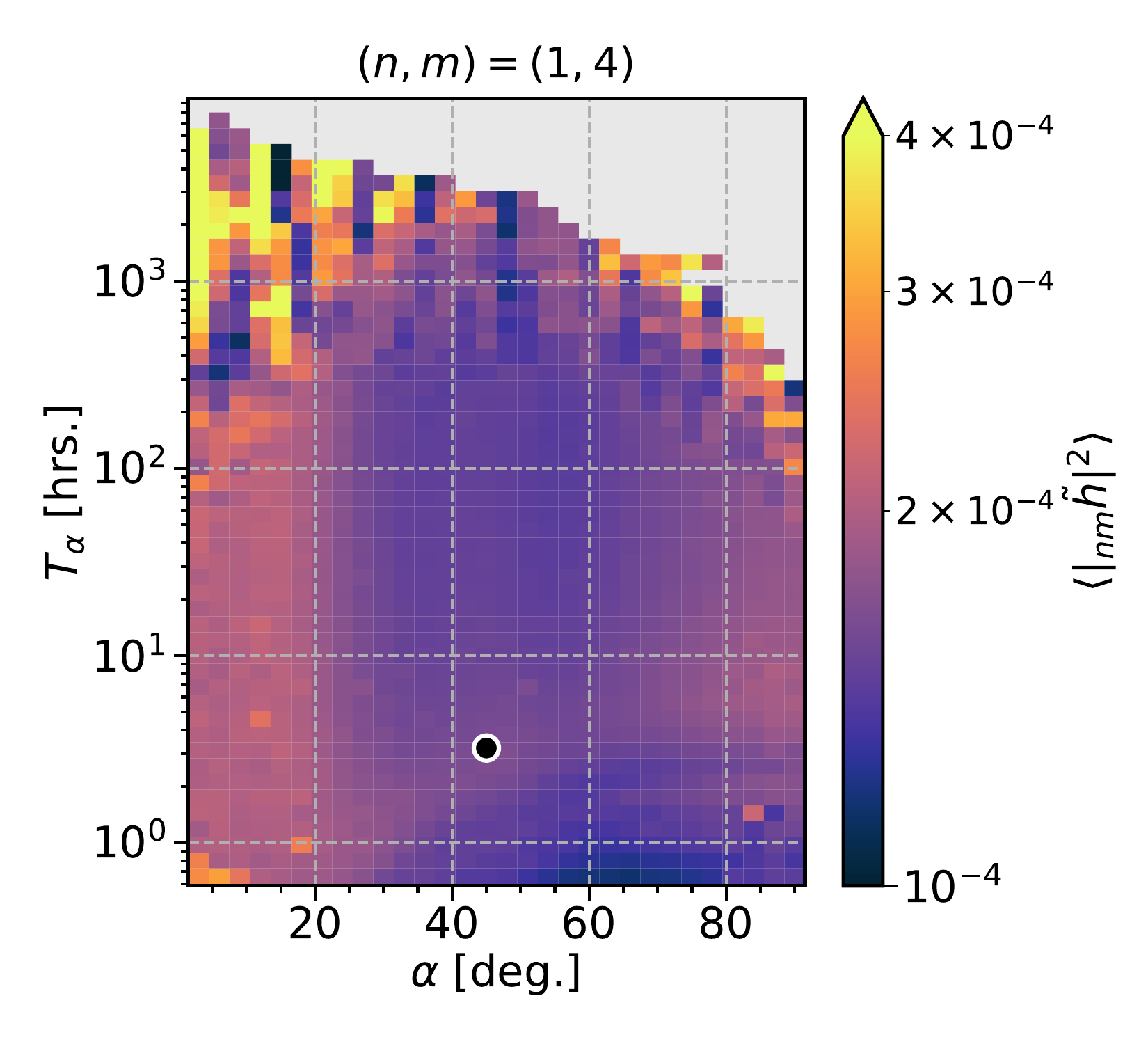}
  \includegraphics[width=0.32\columnwidth]{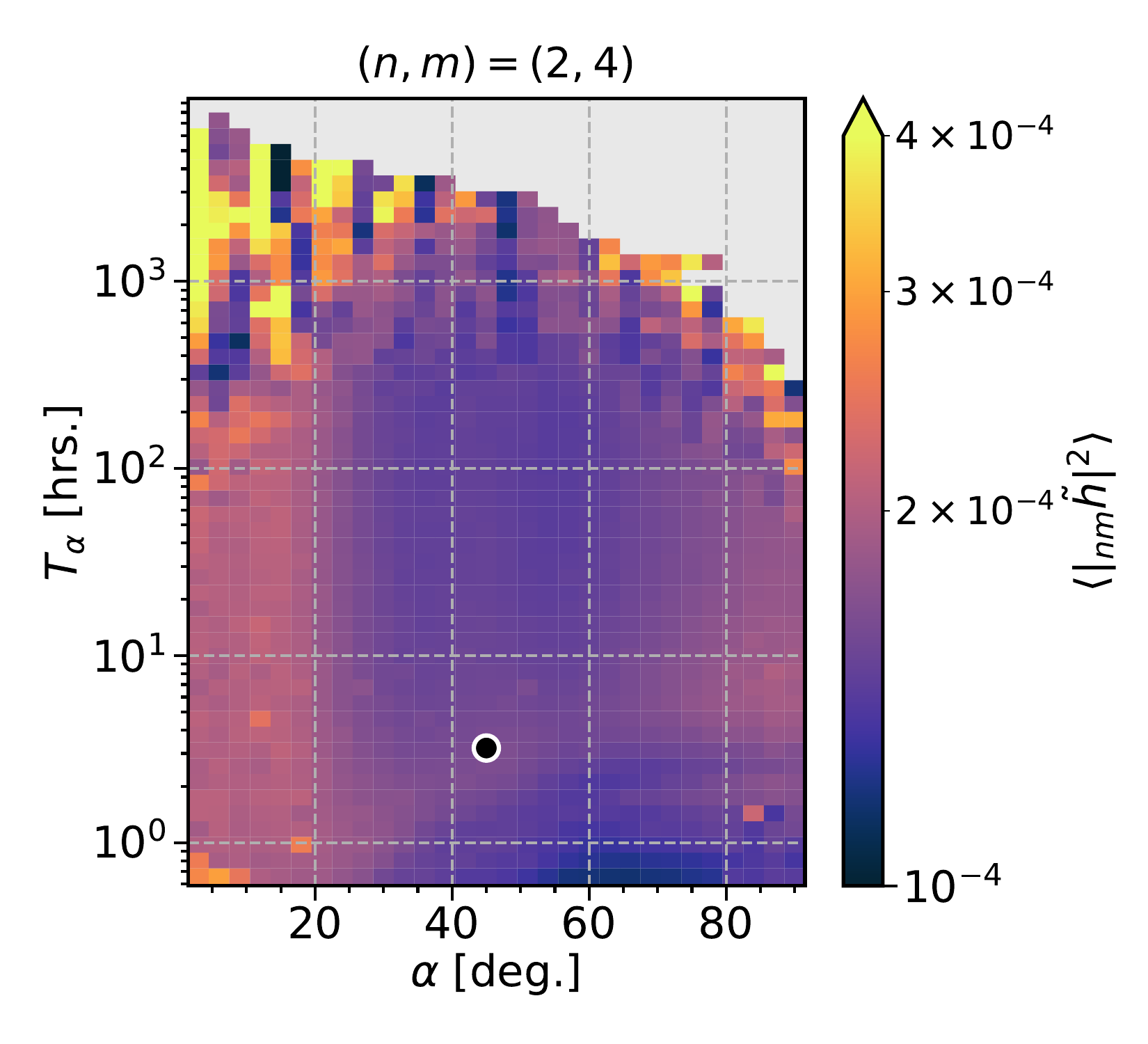}
  \includegraphics[width=0.32\columnwidth]{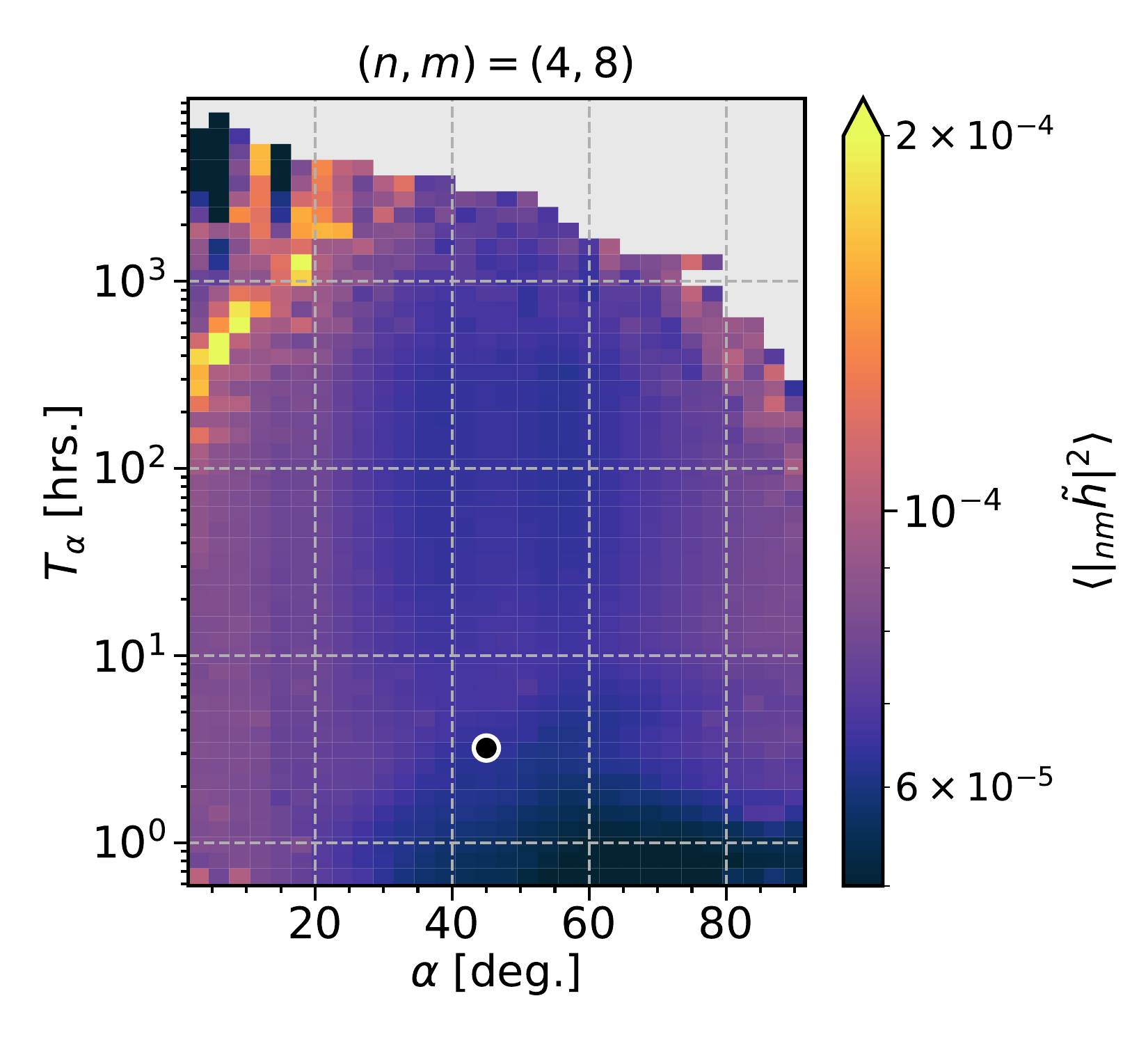}
  \caption{The entire sky average of cross-link factors per \spin-$(n,m)$. The top two panels represent \spin-$(n,0)$ cross-link factors with no HWP contribution, while the third panel represents cross-link factors with \spin-$(1,4),(2,4)$ and $(4,8)$ from left to right and with HWP contribution. The \spin-$(n,0)$ cross-link factors are almost independent of $T_\alpha$ and are smaller for larger $\alpha$. It can be seen that the \spin-$(n,m)|_{m=4,8}$ cross-link factors are kept constant over a wide region by the constraints of \cref{eq:T_spin}. In addition, since the HWP rotates independently of the crossing angle, there is almost no difference for each \spin-$n$ with finite \spin-$m$ in the cross-link factors.}
  \label{fig:cross-links}
\end{figure}

\Cref{fig:cross-links} shows the distribution of the cross-link factors values for the most relevant values of \spin-$(n,m)$ in the $\{\alpha, T_\alpha \}$ space. The top two rows are \spin-$(n,m)|_{m=0}$ cross-link factors with no HWP contribution and the last row contains \spin-$(n,m)|_{m=4,8}$ cross-link factors with HWP contribution.
We note that the obtained \spin-$(n,0)$ cross-link factors distributions have a very similar structure to the ones of ref.~\cite{OptimalScan}, with differences that can be explained by the different choices of the parameter spaces regions that are sampled (the authors of ref.~\cite{OptimalScan} explore faster spin rates regions in order to suppress the $1/f$ noise using the spacecraft's rotation). Regarding \spin-$(n,0)$ cross-link factors, scanning strategies with large $\alpha$ and small $\beta$ tend to be associated with smaller cross-link factors. This is because the ring drawn by the spin is smaller in this region, as we can see in \cref{fig:hitmaps}, making the crossing angles more uniform per sky pixel.

The distribution of the \spin-$(n,m)|_{m=4,8}$ cross-link factors associated to the HWP tends to be mostly flat across the $\{\alpha, T_\alpha\}$ space.
The reason is that, according to constraint \cref{eq:T_spin}, the HWP makes more than one revolution during the transit of the sky pixel, and uniform crossing angles between $0$ and $2\pi$ are obtained in one observation.
All the values of the \spin-$(n,m)|_{m=4,8}$ cross-link factors ($1\leq n \leq 6$) can be similarly computed, displaying an identically flat behavior in the $\{\alpha,T_\alpha\}$ space for each \spin-$(n,4)|_{1\leq n \leq 6}$ cross-link factors, and \spin-$(n,8)|_{1\leq n \leq 6}$ cross-link factors as displayed for the \spin-$(1,4)$ and \spin-$(2,4)$ cross-link factors in \cref{fig:cross-links}.

\subsection{Propagation of cross-link factor to bias} \label{sec:Propagation}

We use the map-based method using the \spin formalism presented in \cref{sec:formalism} to estimate the measurement bias on $r$, i.e., $\Delta r$ induced by the pointing offset and the HWP non-ideality, and to demonstrate how the value of the cross-link factor can be directly linked to $\Delta r$. A detailed description about the way we estimate $\Delta r$ is given in \cref{apd:delta_r}.

The CMB map used for the demonstration is generated by \texttt{CAMB} \cite{CAMB}.\footnote{\url{https://camb.readthedocs.io/en/latest/}}
A 6-parameter $\Lambda\rm{CDM}$ model is chosen based on the best fit of the \Planck 2018 results as follows:
Hubble constant, $H_0=67.32$, baryon density, $\Omega_b=0.0494$, dark matter density, $\Omega_{cdm}=0.265$, optical depth to reionisation, $\tau=0.0543$, scalar spectral index, $n_s=0.966$, and amplitude of scalar perturbations, $A_s = 2.10\times10^{-9}$ \cite{Planck2018}. The tensor-to-scalar ratio is set to $r=0$, assuming no primordial $B$ modes.

For the pointing offset simulation, we impose the offset parameter $(\rho,\chi)=(1,0)$\,arcmin in \cref{eq:pointing_offset_field} and use CMB only map as input. We smooth the CMB map with a symmetric Gaussian beam of $\rm{FWHM}=17.9$\,arcmin, simulating the smallest instrumental beam of \LB, as done in \cref{sec:case_of_LB}. For the simulation of the instrumental polarisation due to the HWP, we impose $\epsilon_1=1.0\times10^{-5}$ and $\phi_{QI}=0$ in \cref{eq:HWP_IP_field}, and use the CMB solar dipole map as input because it is one of the dominant components of temperature as is discussed in ref.~\cite{guillaume_HWPIP}.
Indeed, CMB and foreground anisotropies can contribute as leakage signals, but CMB anisotropies are sufficiently negligible compared to the dipole. 
In general, the complexity of the frequency-dependent foregrounds is assessed using specific models and addressed through masking and component separation. 
This discussion forms a whole topic on itself and we will not consider it here in the optimisation of the scanning strategy.
Note that the magnitude of the systematic effect, i.e., $\rho,\chi$ and $\epsilon_1$ themselves are not treated here, but only the relative penalty of each scanning strategy is identified. In order to deal with the impact of these systematic effects on the scientific goals, calibration as in ref.~\cite{planck_pointing_cal} and mitigation techniques as in ref.~\cite{guillaume_HWPIP} also need to be considered.
Notably, the map-based simulations using the \spin formalism shown in \cref{sec:map-based-sim,apd:spin_space_analysis} is about $10^4$ times faster than the TOD-based simulation.

\begin{figure}
  \centering
  \includegraphics[width=0.49\columnwidth]{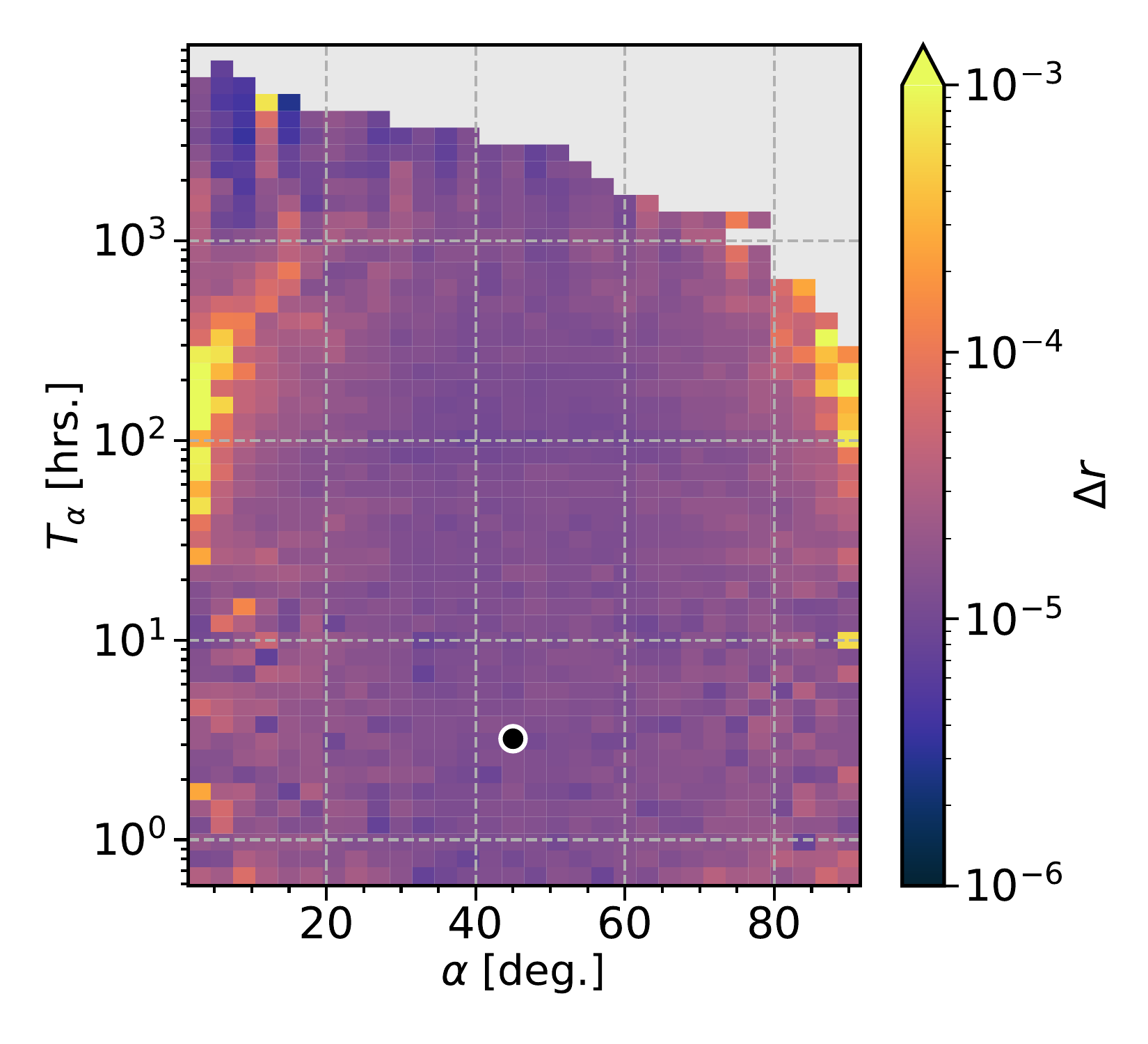}
  \includegraphics[width=0.49\columnwidth]{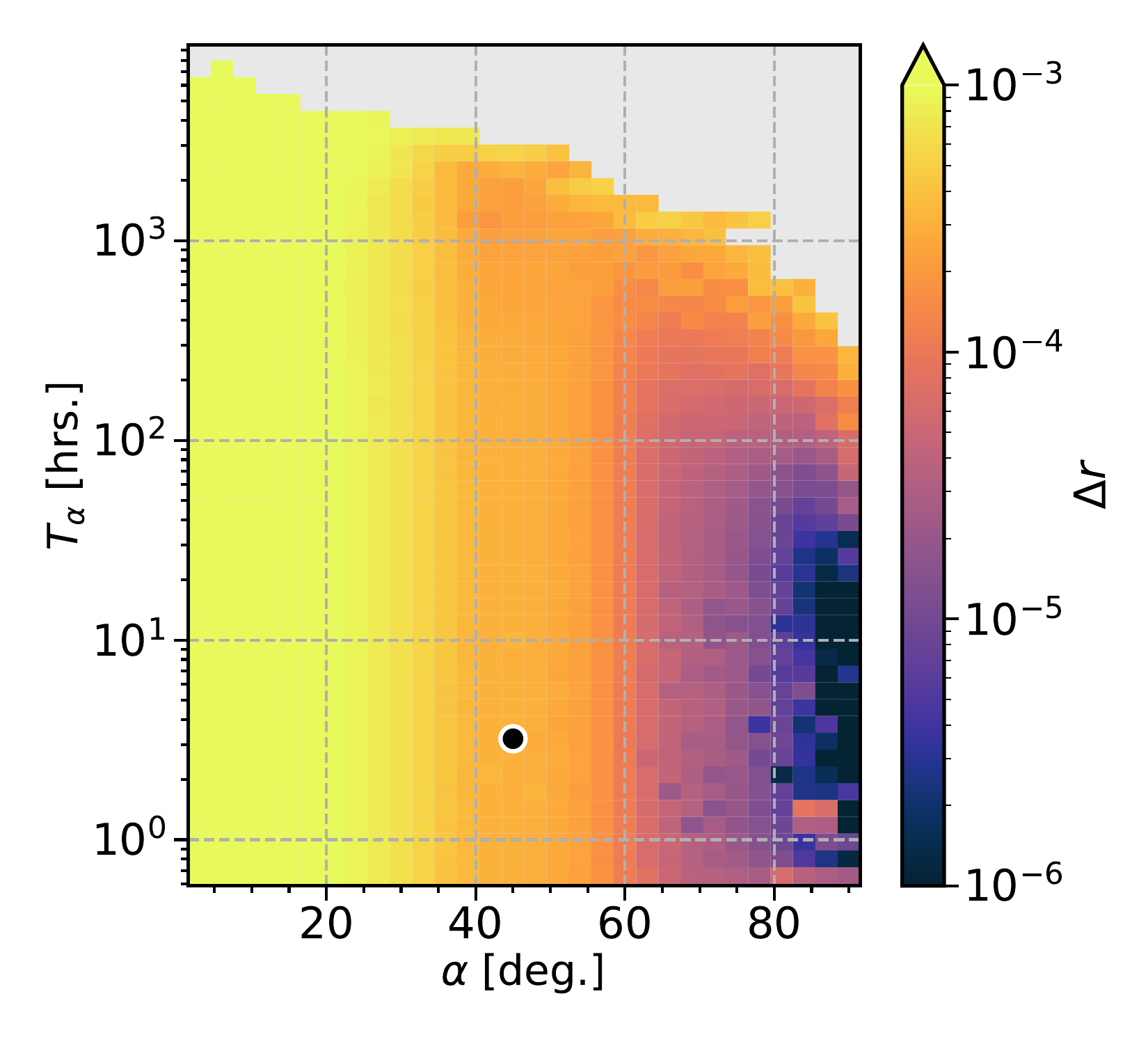}
  \caption{
  (left) The distribution of $\Delta r$ due to the pointing offset that has $(\rho,\chi)=(1,0)$\,arcmin as an offset parameter. The input map here is only the CMB, which is smoothed by the Gaussian beam with $\rm{FWHM}=17.9$\,arcmin.
  (right) The distribution of $\Delta r$ due to the instrumental polarisation with the HWP. We impose $\epsilon_1=1.0\times10^{-5}$ and $\phi_{QI}=0$. The input map here is only the CMB solar dipole.}
  \label{fig:delta_r}
\end{figure}

\Cref{fig:delta_r} (left) shows the distribution of $\Delta r$ due to the pointing offset. As expected, the distribution of $\Delta r$ has an overall flat distribution, inherited from the \spin-$(n,4)$ cross-link factors, indicating that the HWP effectively suppresses leakage from temperature due to the pointing offset.
This makes it clear that the full-sky average of cross-link factors considering the HWP expressed in \cref{eq:crosslink} is an appropriate indicator to represent the penalty of the scanning strategy for a particular systematic effect that couples with the HWP modulation. 

\Cref{fig:delta_r} (right) shows the distribution of $\Delta r$ due to the instrumental polarisation with the HWP. The distribution has a similar structure to the \spin-$(2,0)$ cross-link factor in \cref{fig:cross-links}, as expected. This implies that this systematic effect is not suppressed by the reduction of cross-link factors by the HWP rotation, meaning that only the \spin-$(2,0)$ cross-link factor acts as a suppression factor. 
The justification of the cross-link factor dependency of the HWP non-ideality can be found in \cref{apd:HWP_sys}.
These results indicate that the systematic effect coupled with the \spin-$(n,4)$ cross-link factor is suppressed independently of the scanning strategy and that the choice of scanning strategy to reduce \spin-$(n,0)$ is important even with the use of HWPs.

\section{Optimisation} \label{sec:optimisation}

In the previous section, we studied how the metrics of interest behave in the $\{\alpha,T_\alpha\}$ space and we confirmed that an increase of $T_\beta$ from $\tbl$ does not lead to a different optimised configuration in the cross-link factors (\cref{apd:T_beta_scaled}). 
We will now discuss how to optimise these parameters in regards of the multiple metrics considered.
Many metrics are found to be depend mainly on $\alpha$ and independent of $T_\alpha$ when it is less than $100$\,hours. As such, we propose to start by optimising the choice of the geometric parameters, assuming that this choice can be made independently of the kinetic parameters. 
In this section, we first list the suggested values of the effective choice of the geometric parameters for the next generation of CMB space missions and compare them in regards to the various metrics considered.
After discussing the optimisation of geometric parameters, we fix them and discuss the optimisation of kinematic parameters, allowing us to propose a reasonable compromise for the scanning strategy parameters.

\subsection{Optimisation of the geometric parameters}\label{sec:Opt_geometric}

We now discuss the optimisation of the geometric parameters $\alpha$ and $\beta$. Note that, determining $\alpha$ automatically determines $\beta$, since $\beta$ is a function of $\alpha$, depending on the constraints of \cref{eq:const_geometric}. What we are looking for in a scanning strategy is the ability to have long visibility time on the 
planets and to minimise both $\sigmahits$ and the cross-link factors.

In this respect, the $\alpha$ with the longest visibility time on the planets is given by the case $\alpha=\beta=47.5^\circ$. As shown on \cref{fig:planet_and_sigma_hit} (left), the distribution is symmetrical around this point, which we refer to as the \BC, associated with 4.4\,hours of integration time for planets. However, from a hit-map uniformity point of view such a configuration is slightly disfavored, as $\sigmahits$ displays an excess at this point. 
The \SC has 3.2\,hours of integration time, which means that despite losing about a factor of 0.7 of planet integration time during the mission duration compared to the \BC, it provides a hit-map that is 1.1 times more uniform.

On the other hand, a scanning strategy with $\alpha=50^\circ$ can also guarantee a similar planet integration time and $\sigmahits$ as the \SC, with smaller cross-link factor values. We thus consider the \FC, in which the values of $\alpha$ and $\beta$ are reversed compared to the \SC. 
Cross-link factor is smaller in the \FC than in the \SC due to the smaller value of $\beta$, but the difference is less than 5\%, such that the ability to suppress systematic effects is still expected to be similar.

In that context, we now seek for a reasonable compromise for the scanning strategy applied to the specific case of \LB. To begin with, we reject the \BC in order to preserve the uniformity of the hit-map. 
As discussed further in \cref{sec:implications}, the adoption of the \SC results in a loss of integrated visibility time of planets compared to the \BC, but the \SC still has a relatively long integrated visibility time of planets compared to past (\Planck) and future (\PICO) space missions.

The metrics we have examined do not show significant differences between the \SC and the \FC, but there are significant differences in terms of heat input from the Sun and telescope design.
Firstly, in terms of heat input, the \FC tilts the entire spacecraft $5^\circ$ more towards the Sun than the \SC, increasing the heat input by $\sin50^\circ/\cos45^\circ=1.08$. Conversely, the solar panels are tilted in a direction parallel to the Sun radiation, so the power generation efficiency is reduced by 0.92 through the reciprocal of this factor.

Even allowing for these degradations, we are forced to optimise a thermal protection mechanism for the payload modules such as the V-grooves (\cref{fig:standard_config_and_T_beta}) in \LB. However, considering \LB as a model, this optimisation takes place in the form of reducing the allowable volume of the telescopes.
On the other hand, trying to keep the V-groove design of the \SC would imply a shrinkage of the baffle (\cref{fig:standard_config_and_T_beta}) in front of the aperture of the telescopes to mitigate the effect of the Moon's shadow.
This would generate a larger sidelobe leading to significant systematic effects, and it is clear that the 5\% better cross-link factors would not represent a significant enough benefit.

The $60^\circ\lesssim \alpha \lesssim70^\circ$ case seems attractive from the point of view of suppressing systematic effects, since the cross-link factors and the systematic error are significantly reduced in \cref{fig:cross-links} (\spin-$(n,0)$ cross-link factors) and \cref{fig:delta_r} (right). However, such a configuration would increase the heat input even more than in the \FC case, and the power generation efficiency of the solar panels would correspondingly be greatly reduced.

In addition, a smaller $\beta$ (associated with the larger $\alpha$) would lead to a reduction in the amplitude of the CMB solar dipole signal used for gain calibration, which is not favored from both the engineering and scientific points of view.
In this respect, the \SC has been well studied by refs.~\cite{PTEP2023,Odagiri_SPIE} to ensure that systematic effects and heat inputs meet the scientific objectives of \LiteBIRD.
We therefore conclude that the combination $(\alpha,\beta)=(45^\circ,50^\circ)$ is an effective choice for the \LB mission.

\subsection{Optimisation of the kinetic parameters}\label{sec:Opt_kinetic}

The geometric parameters adopted in the \SC $(\alpha,\beta)=(45^\circ,50^\circ)$, which determine the essential properties of the scanning strategy, are found to be favorable in terms of scientific objectives and spacecraft design for the \LiteBIRD mission. In this section, we fix the values of $\alpha$ and $\beta$ to these values and further optimise $T_\alpha$ and $T_\beta$.

\subsubsection{Global survey of the kinetic parameter space}

In the simulations so far, $T_\beta$ was fixed at the value of $T_\beta^{\rm lower}(\alpha,T_\alpha)=16.9$\,min instead of the proposed 20\,min of the \SC.
This value is the lower limit of the spin period of the spacecraft and can also be converted into an upper limit on the spin rate, $\nu_\beta^{\rm upper}$ that allows the spacecraft to spin as fast as the constraint determined by \cref{eq:req_for_HWP} permits.
Without the HWP, the spacecraft has to maintain a high spin rate to suppress the $1/f$ noise. Since the HWP plays this role, we do not have to maintain $\nu_\beta^{\rm upper}$, and a slower spin rate is permitted for easier attitude control.
However, as the spin rate becomes slower, the number of spins per precession becomes smaller, and the uniformity of the crossing angles becomes worse, resulting in a larger \spin-$(n,0)$ cross-link factor. Therefore, we examine the trade-off between the spin rate and the cross-link factors.

\begin{figure}
  \centering
  \includegraphics[width=0.32\columnwidth]{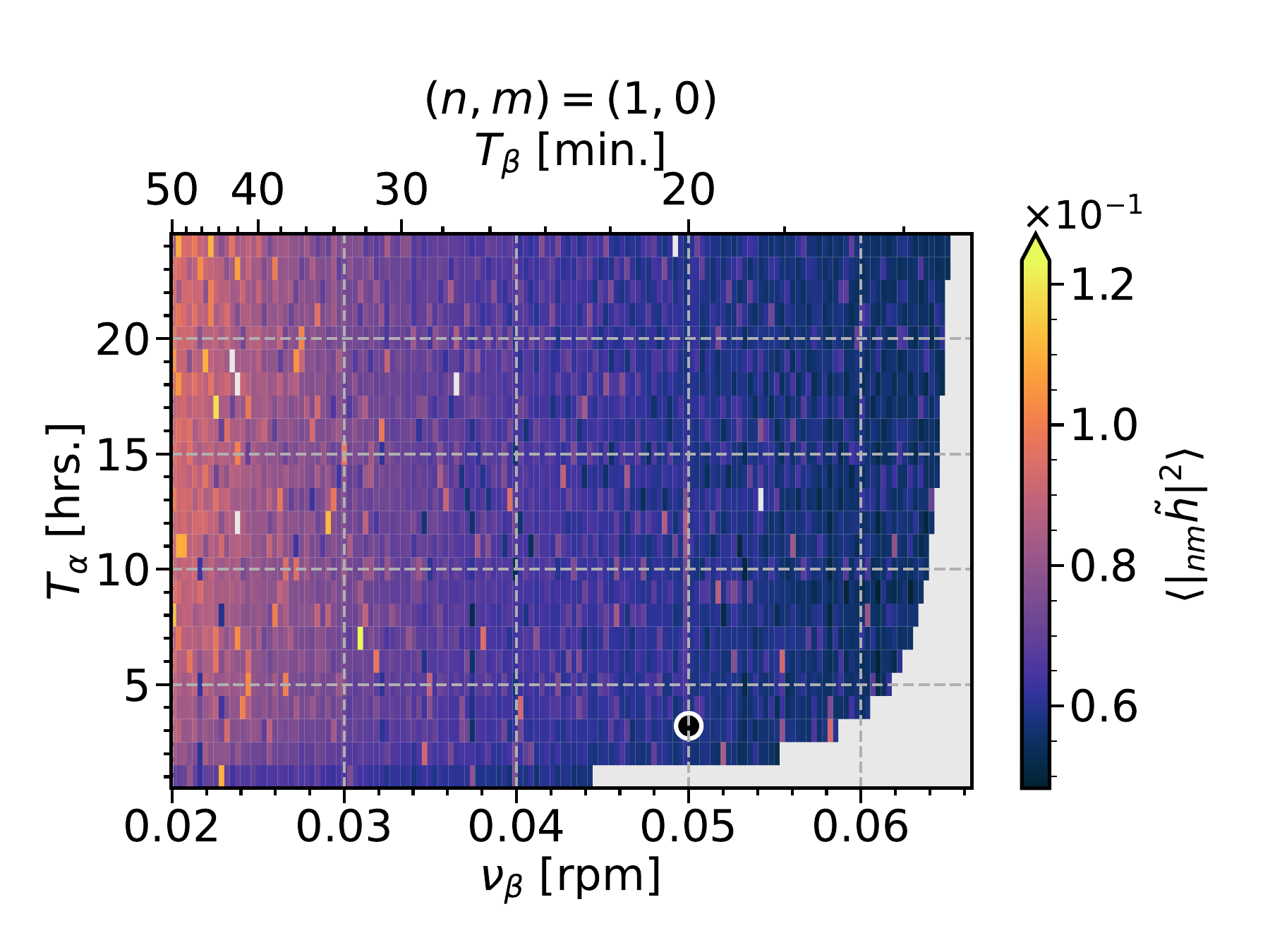}
  \includegraphics[width=0.32\columnwidth]{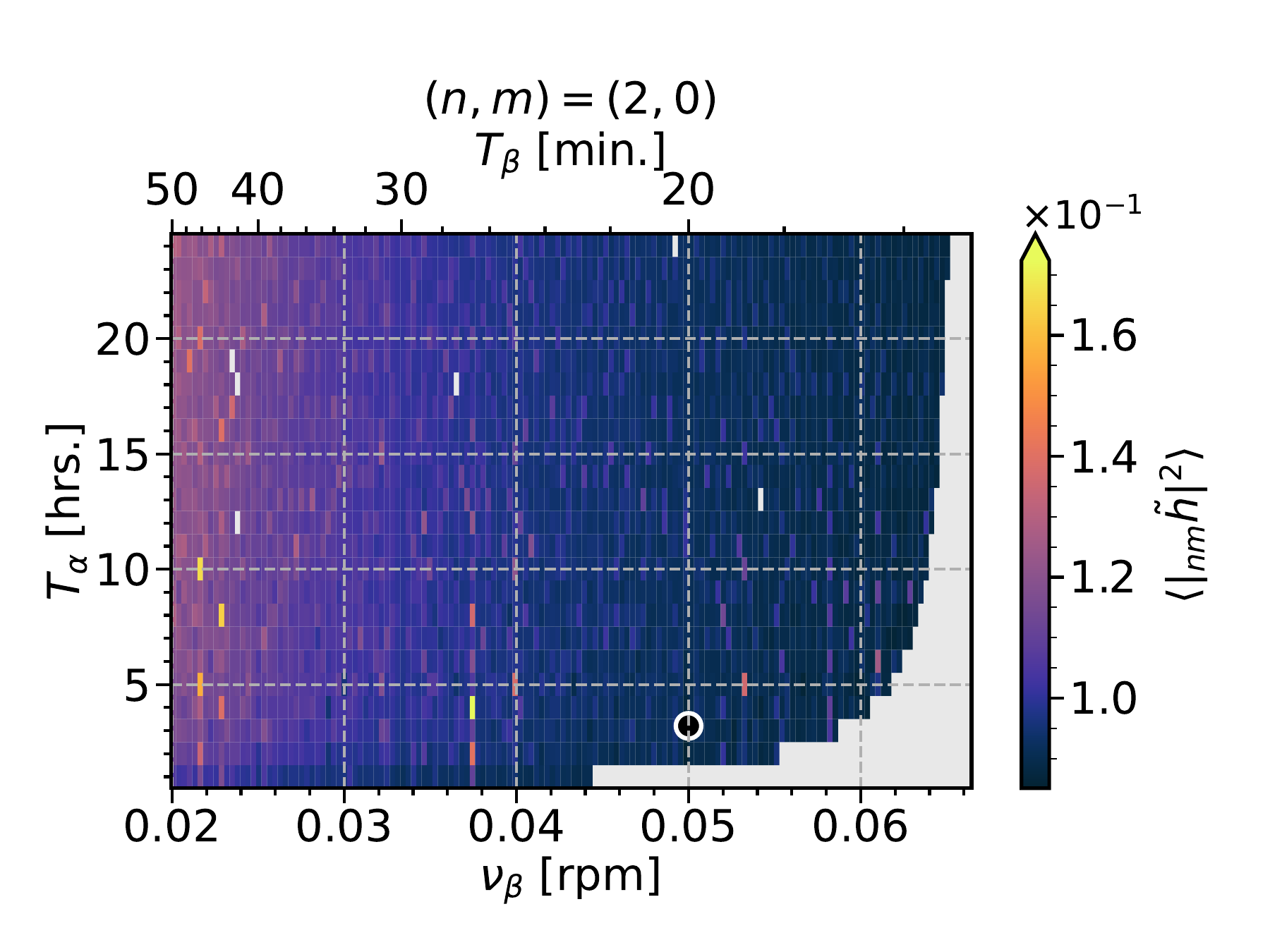}
  \includegraphics[width=0.32\columnwidth]{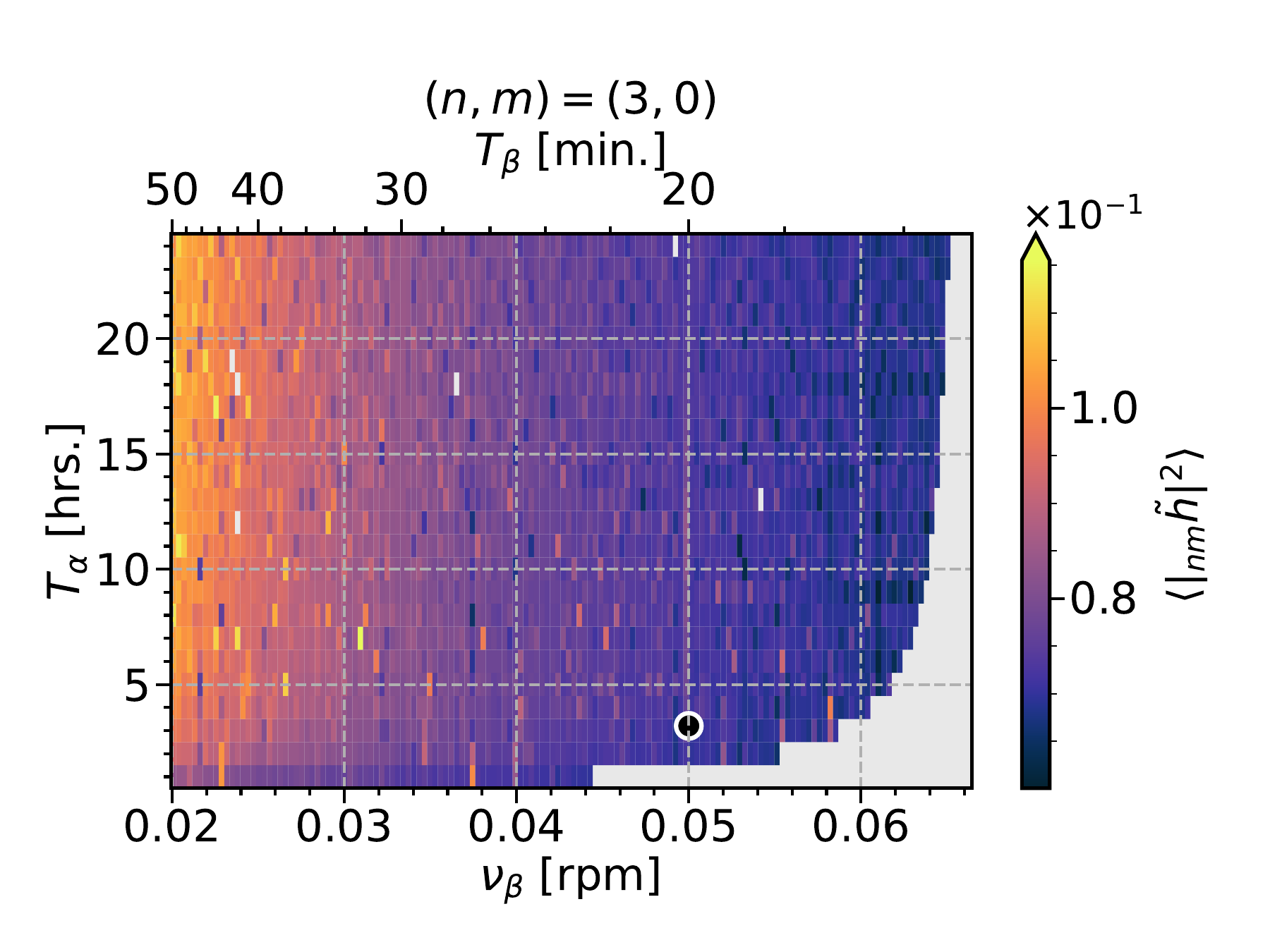}\\
  \includegraphics[width=0.32\columnwidth]{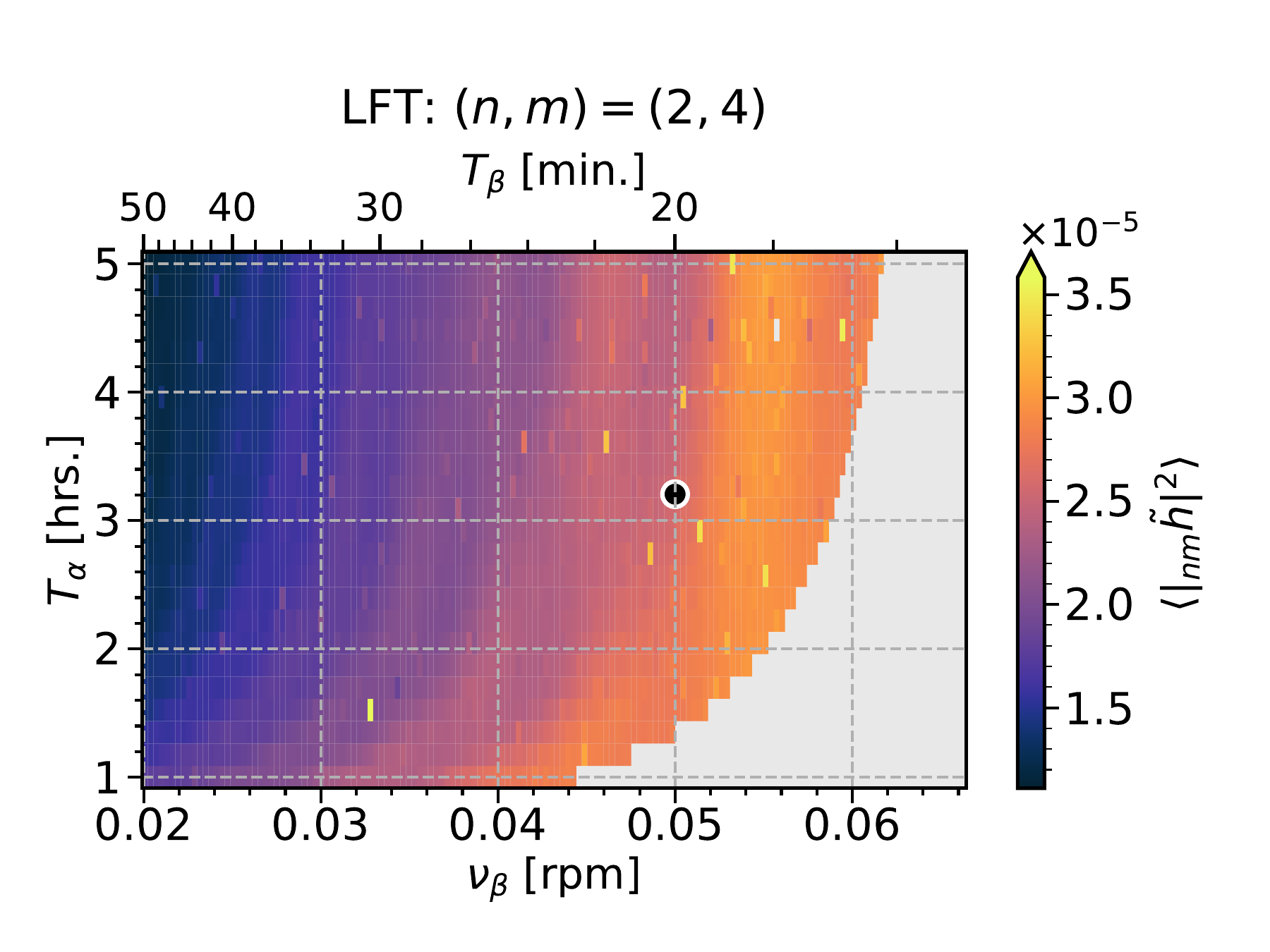}
  \includegraphics[width=0.32\columnwidth]{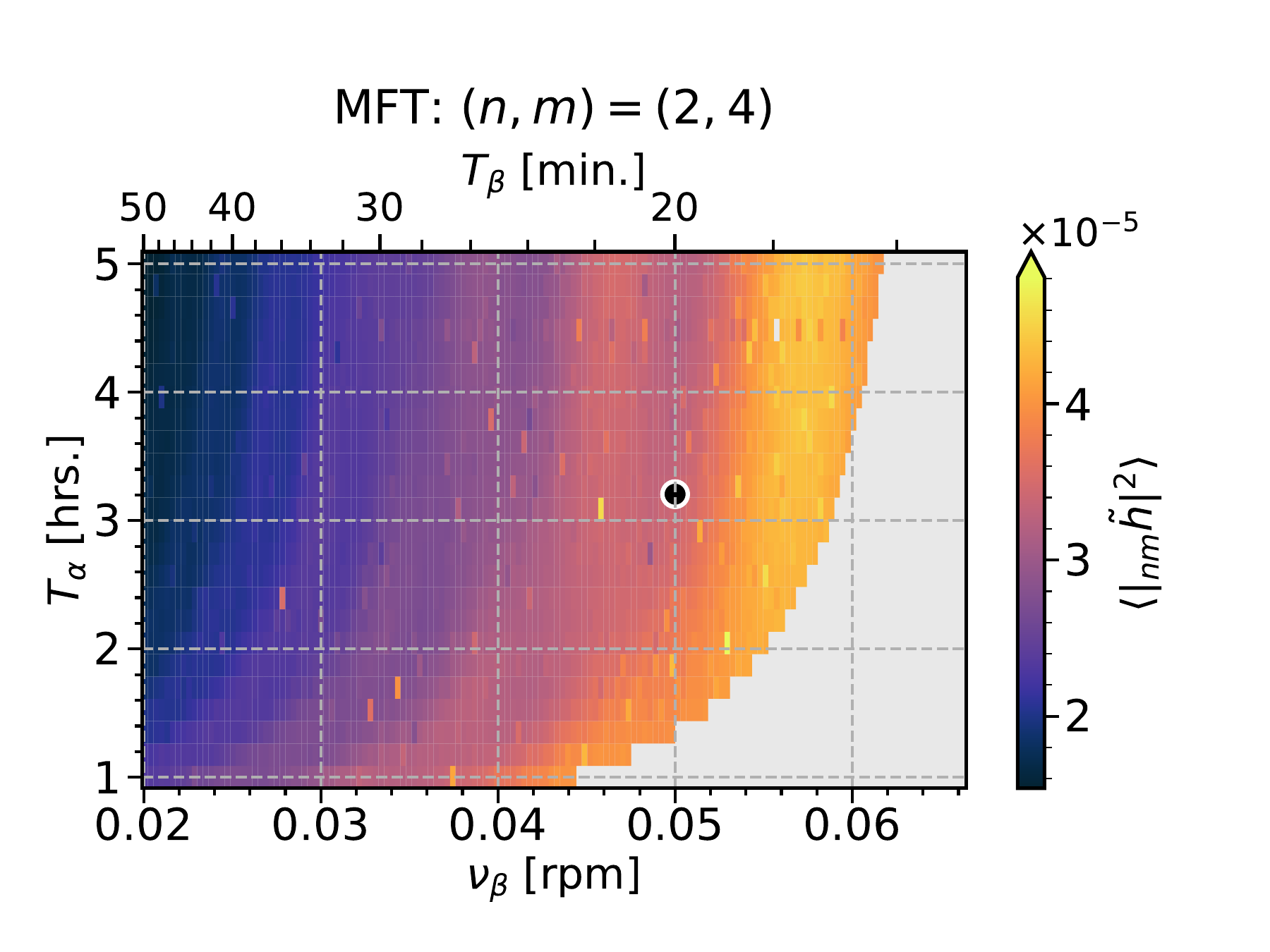}
  \includegraphics[width=0.32\columnwidth]{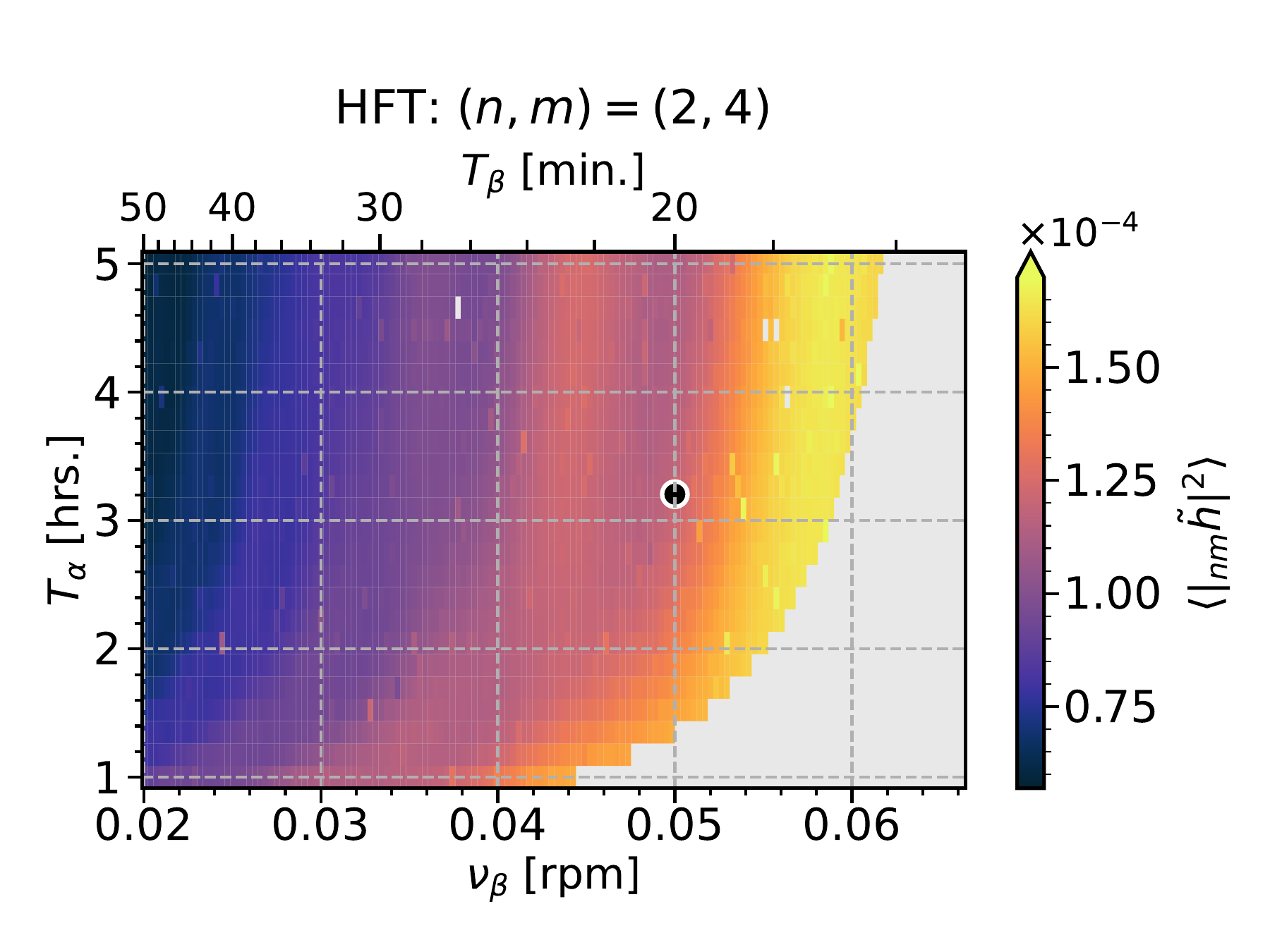}
  \caption{(top panels) The cross-link factors are shown for each \spin-$(n,0)$ cross-link factor. 
  (bottom panels) The \spin-$(2,4)$ cross-link factors are shown for each telescope: LFT (left), MFT (middle) and HFT (right). The gray area in higher $\nu_\beta$ region indicates where the condition $\nu_\beta<\nu_\beta^{\rm{upper}}$ (\cref{eq:T_spin}) is violated. Some points are not associated with a value even though the condition is not violated. This is due to the resonance between spin and precession, which will be discussed in \cref{sec:fine-tuned}. In practice, the scanning strategy for these points also covers most of the sky, but reflects the fact that a few sky pixels may not be observed due to the redundant trajectories created by the synchronisation of the spin and precession.}
  \label{fig:rot_period_opt}
\end{figure}

\Cref{fig:rot_period_opt} (top panels) shows the $T_\alpha$ and $T_\beta$ dependence of the \spin-$(n,0)$ cross-link factors.
\Cref{fig:rot_period_opt} (bottom panels) shows the \spin-$(2,4)$ cross-link factors calculated for each of the three different telescopes. Note that we zoom into a narrower parameter space region at higher resolution than the upper panels where $T_\alpha$ ranges up to 5\,hours.
For LFT/MFT/HFT, the HWP revolution rates are set to 46/39/61\,rpm and $N_{\rm side}$ is set to 128/128/256 because the narrowest FWHM of beams are 23.7/28.0/17.9 arcmin, respectively.\footnote{The size of a \texttt{HEALPix} pixel is $\sqrt{4\pi/N_{\rm pix}}$. When $N_{\rm side}$ is 128/256, the pixel size is 27.5/13.7\,arcmin.}
Parameters other than the spin and precession period are fixed to the values of the \SC. Note that the use of the spin rate $\nu_\beta$ (instead of $T_\beta$), which is often used in spacecraft operations, is given linearly on the horizontal axis of the figures. 

In the top panels of \cref{fig:rot_period_opt}, there is a tendency for \spin-$(n,0)$ cross-link factors to become smaller when $\nu_\beta$ is larger in the figures.\footnote{We have verified that this tendency remains unchanged for higher \spin-$(n,0)$ cross-link factors, e.g., \spin-$(n,0)|_{n=4,5,6}$. }
When $T_\alpha$ is longer than 2\,hours, the cross-link factors are almost constant, regardless of the value of $T_\alpha$.\footnote{We examined the limit of the $T_\alpha\leq100$\,hours, which keeps the metrics that we considered in a favorable state, but the trend of cross-link factor independence on $T_\alpha$ does not change.} 
For pixels with large values in the figures, such as outliers, the map has a regular geometric pattern called a \moire pattern, due to resonance between spin and precession, as described in ref.~\cite{hoang2017bandpass}.
Mathematically, it is known that the \moire pattern disappears when the ratio, $T_\alpha/T_\beta$ becomes an irrational number; therefore as we will discuss in the next section, a fine-tuning around the desired $T_\alpha$ and $T_\beta$ is always possible. 

For all \spin-$(n,0)$ cross-link factors, slowing down $\nu_\beta$ does not significantly increase the cross-link factors up to 0.05\,rpm, but a 5\% increase is observed between 0.04 and 0.05\,rpm. On the other hand, the cross-link factor with a HWP contribution increases proportionally to $\nu_\beta$, but never monotonically, because it presents an oscillating behavior due to the synchronisation of spin and the revolution of the HWP. 
The value $\nu_\beta=0.05$\,rpm is the first local minimum that appears when $\nu_\beta$ is reduced from $\nu_\beta^{\rm upper}$.
Therefore, we conclude that $\nu_\beta=0.05$\,rpm, i.e., $T_\beta=20$\,min is a compromise point where the spin rate can be slowed down to favor attitude control without dramatically degrading the ability of the scanning strategy to suppress the systematic effects.

We now discuss the optimisation of the precession period. From the viewpoint of gain calibration using the CMB solar dipole, the precession rate as well as the spin rate should be faster to avoid  contamination by $1/f$ noise. However, when $\nu_\beta$ is 0.05\,rpm, the \spin-$(n,4)$ cross-link factor is observed to decrease by 15\% from $T_\alpha^{\rm lower}$ to $T_\alpha=3.2$\,hours in every telescope. On the other hand, the cross-link factor does not change significantly when $T_\alpha$ is longer than 3.2\,hours (192\,minutes), and we conclude that this value is a good trade-off between fast precession period and cross-link factors.

\subsubsection{Fine-tuned study of the precession period}\label{sec:fine-tuned}

Now that our geometric parameters are determined, we can perform some fine tuning of the rotation periods in order to remove the resonances. We define the ratio $\eta$, characterising the resonance between spin and precession:
\begin{align}
    \eta = \frac{T_\alpha}{T_\beta}.
\end{align}
So far, we considered $\eta=192~[\rm{min}]/20~[\rm{min}]=9.6$, which is a rational number. The spin thus synchronises with the precession in exactly 9.6 cycles, such that the intersections of the trajectory do not drift and are dragged along by the revolution around the Sun. This symmetry leads to an undesired straight line pattern on the hit-map in the azimuthal direction.

In ref.~\cite{hoang2017bandpass}, the authors suggested to replace the decimal part of this ratio with the decimal part of an irrational number. Since the actual irrational numbers have an infinite number of decimals, they cannot be handled numerically and it is necessary to approximate them by rational numbers with a high accuracy. It is known from the Diophantine approximation that this can be achieved by terminating the continued-fraction expansion of the irrational numbers in the middle.

The adoption of the golden ratio, which has the slowest convergence, was the choice of ref.~\cite{hoang2017bandpass}. This value is unique and provides a logical choice in order to avoid resonance between oscillations. A similar argument is discussed in ref.~\cite{berry1978} from the perspective of dynamical systems. However, in actual spacecraft operation, the accuracy of the rotation control is limited and the spacecraft motion will be unavoidably subject to unpredictable disturbances. 
To account for this fact, we perform a focused analysis of $\sigmahits$ in the range of $T_\alpha \in [192, 193]$\,min with a resolution of 0.1\%. In doing so, we hope to identify stable regions of the parameter space where the resonances can be avoided in a way that could be robust against disturbances induced to the precession.

\begin{figure}
  \centering
  \includegraphics[width=1\columnwidth]{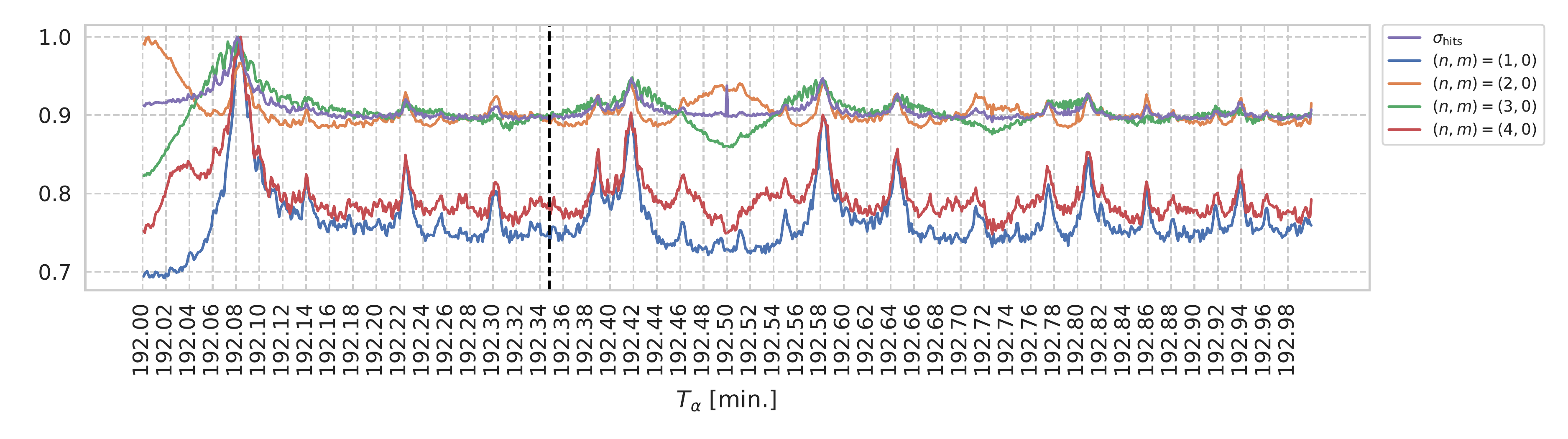 }
  \caption{$T_\alpha$ dependence of normalised $\sigma_{\rm hits}$ and \spin-$(n,0)$ cross-link factors. The dashed black line shows the $T_\alpha=192.348$\,min that the value we adopt for the \SC. The peaks are associated with values for which the spin and precession motion enter in resonance, resulting in \moire patterns appearing on the maps as shown in \cref{fig:prec_tuning_maps}.}
  \label{fig:prec_tuning}
\end{figure}

\begin{figure}[ht]
  \centering
  \includegraphics[width=0.32\columnwidth]{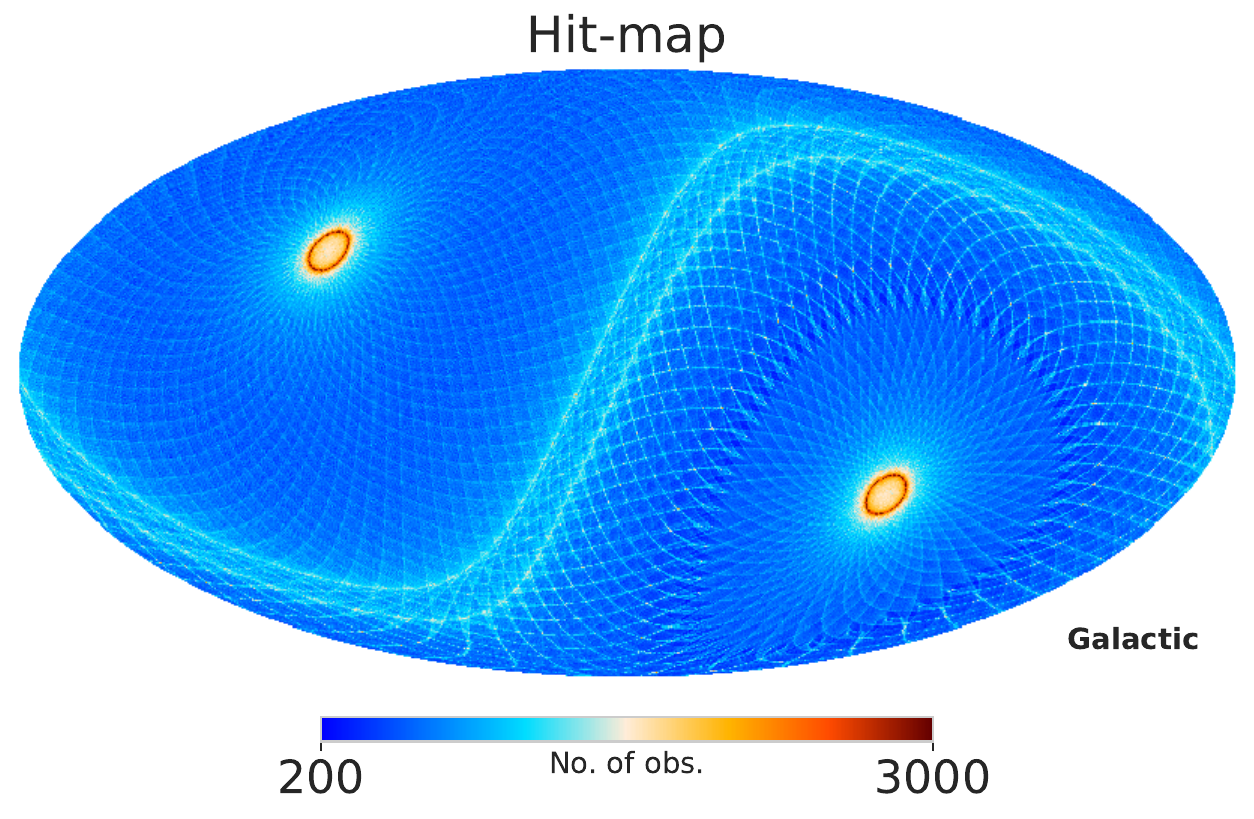}
  \includegraphics[width=0.32\columnwidth]{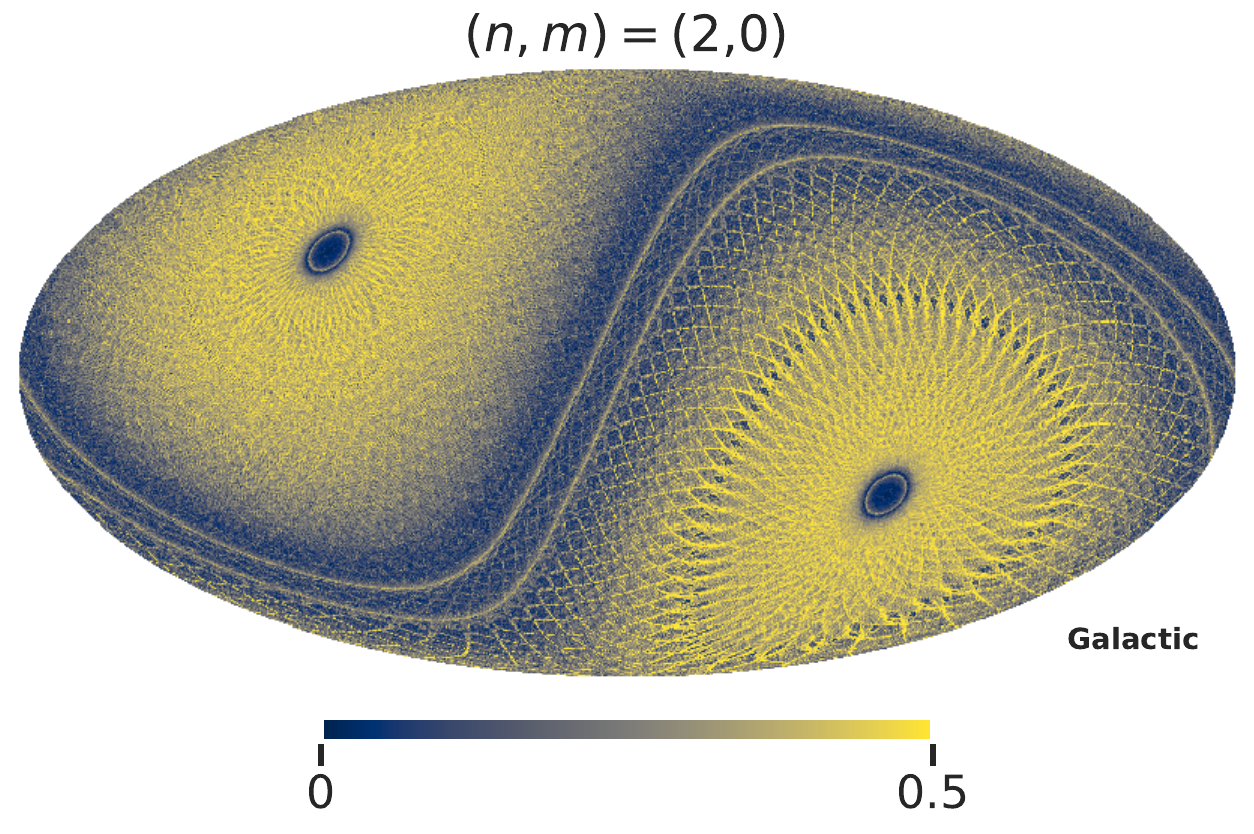}
  \includegraphics[width=0.32\columnwidth]{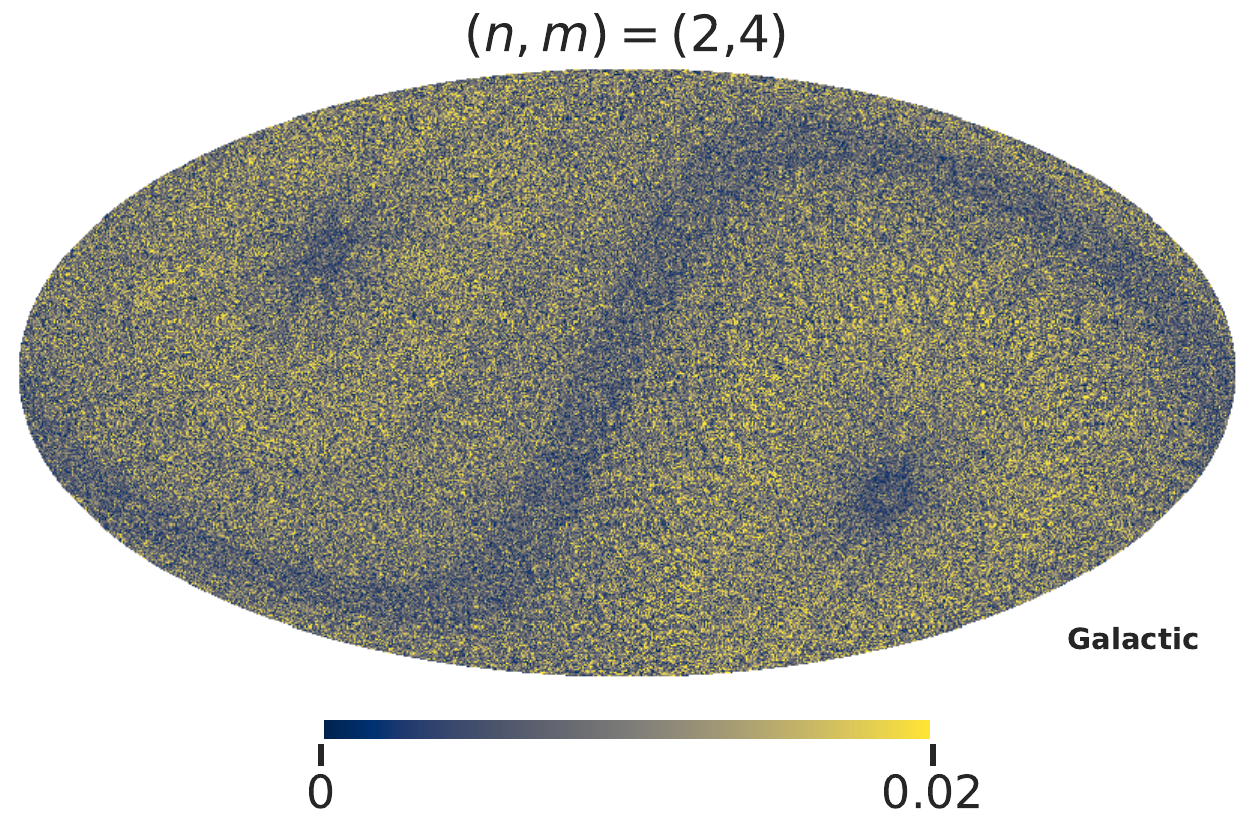}\\
  \includegraphics[width=0.32\columnwidth]{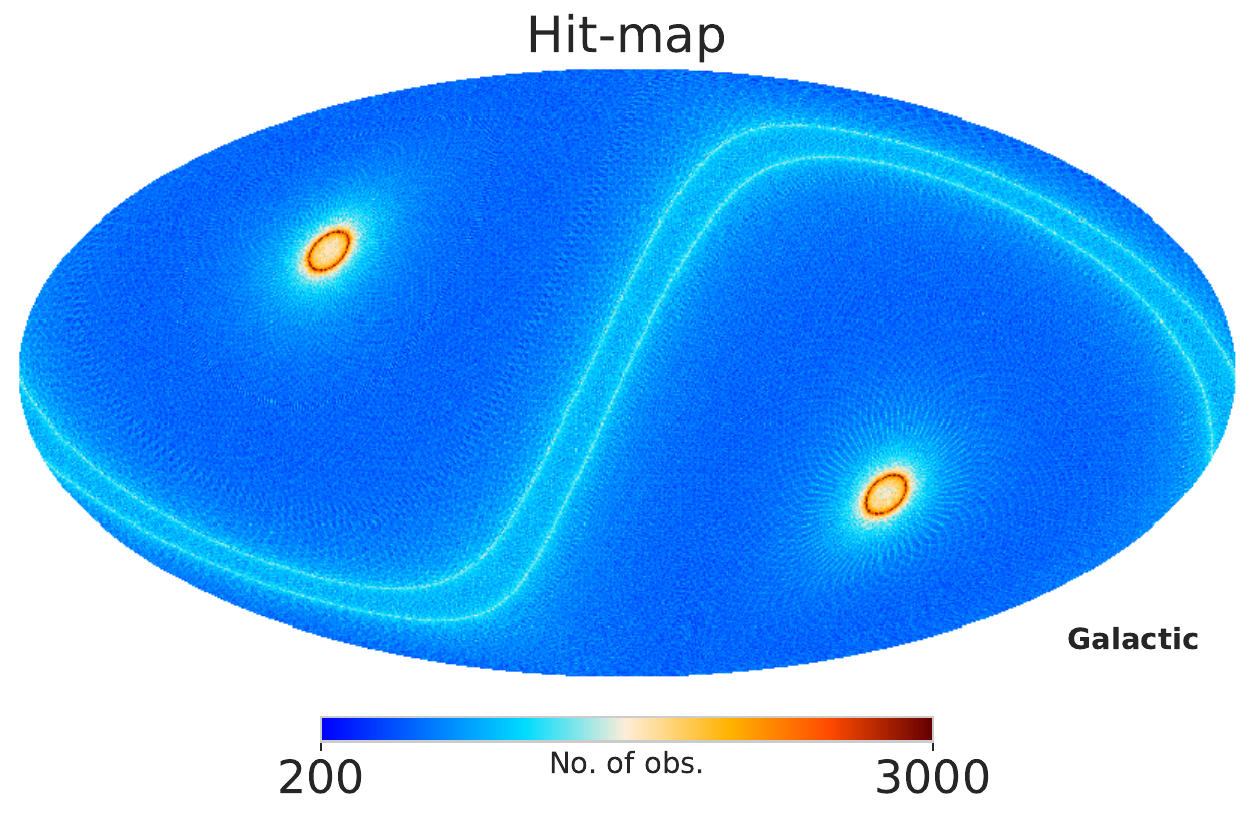}
  \includegraphics[width=0.32\columnwidth]{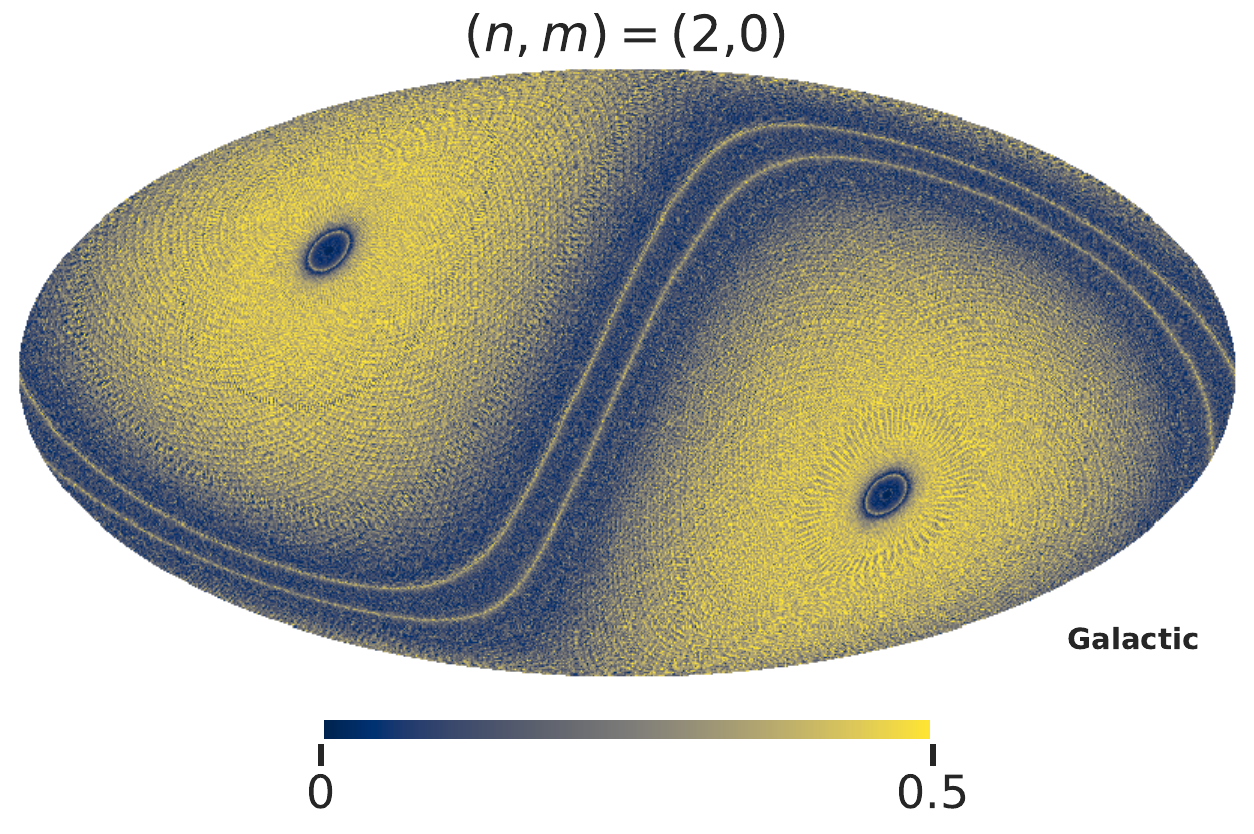}
  \includegraphics[width=0.32\columnwidth]{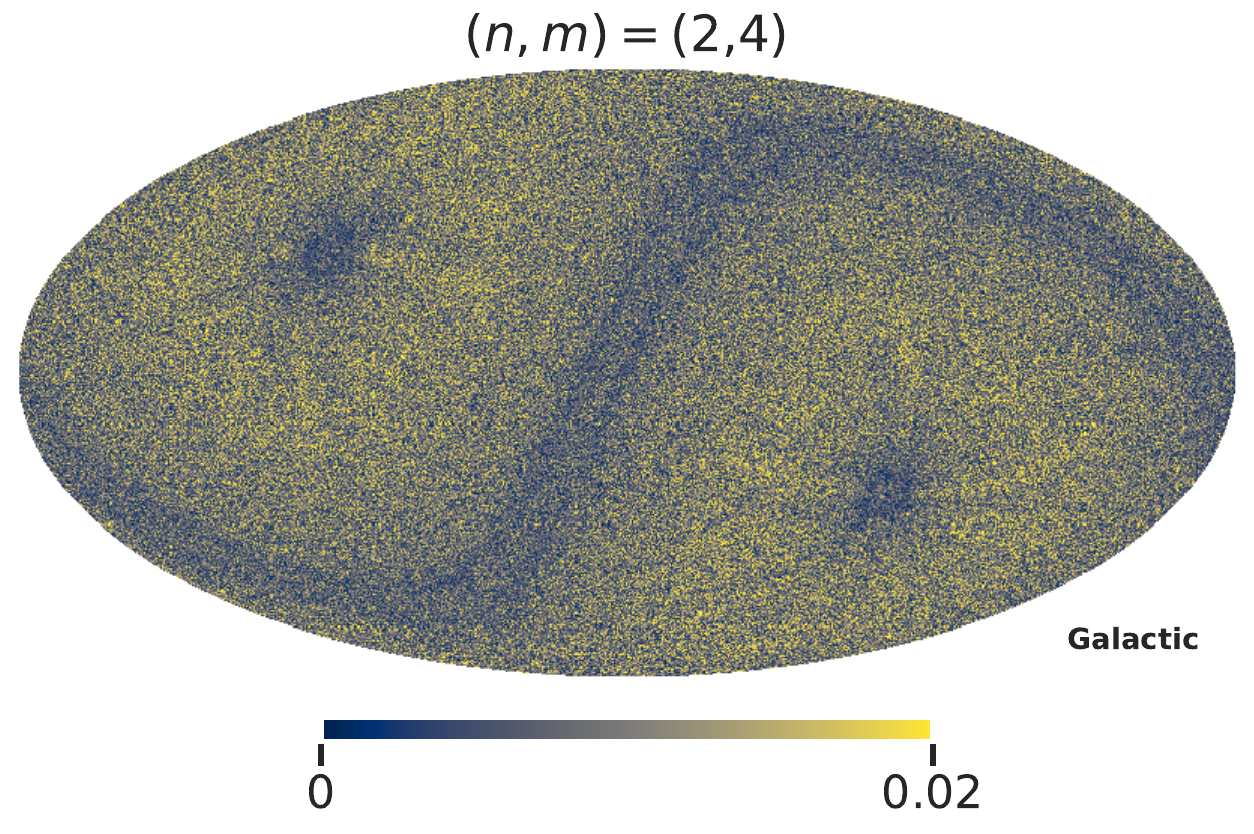}
  \caption{(top panels) Hit-map (left), \spin-$(2,0)$ (middle) and \spin-$(2,4)$ (right) cross-link factors in Galactic coordinates for $T_\alpha=192.08$\,min. (bottom panels) Identical to the top panels with $T_\alpha=192.348$\,min used in the \SC. As suggested in \cref{fig:prec_tuning}, the resonance between spin and precession is large for the top panels, leading to the appearance of \moire patterns. These patterns disappear in the bottom panels, where a fine-tuned precession period is used.}
  \label{fig:prec_tuning_maps}
\end{figure}

\Cref{fig:prec_tuning} shows the results of this analysis, displaying the dependence of the normalised $\sigma_{\rm hits}$ and \spin-$(n,0)$ cross-link factors with respect to $T_\alpha$ in the domain of interest. We identify $192.320 \,\mathrm{min}\leq T_\alpha \leq192.370\,\mathrm{min}$\, as an interval in which all the metrics are close to minimal and do not vary significantly in a close neighborhood.\footnote{The same analysis can be performed considering the cross-link factors of \spin-$(n,m)|_{m=4,8}$, but since the revolution rate of the HWP is relatively faster than the spacecraft's spin and precession rate, resonances are highly damped and much less significant for these quantities than for the \spin-$(n,0)$ cross-link factors.}
We thus propose the value of $T_\alpha=192.348$\,min as the point with the smallest cross-link factors located in this region.
For example, at $T_\alpha=192.08$\,min, when the resonance is expected to be the largest, a \moire pattern appears on the large angular scales in the hit-maps and cross-link factor maps as shown in \cref{fig:prec_tuning_maps} (top panels), but tuning the value to $T_\alpha=192.348$\,min (bottom panels) makes this pattern vanish. 

In the remainder of this work, we will keep the assumption of perfect stability under rotations for the instrument. As mentioned above, such perfect stability cannot be achieved in practice and further studies modeling the inertia of spacecraft and its possible in-flight stabilisation are required to assess the accuracy to which $T_\alpha$ can be fixed in practice, but this is a future work. 

\section{Implications}\label{sec:implications}

In this section we discuss other important items to consider when designing scanning strategies, beyond the metrics discussed previously.
First, we discuss in \cref{sec:beam_reconstruction} the impact of the choice of scanning strategy on beam shape reconstruction through the concept of scanning beam angle.
In \cref{sec:skypix}, we then compare CMB missions with different scanning strategies in regard to visit/revisit times of sky pixels, which are important in null-tests. Finally, we discuss in \cref{sec:visit} the duration and frequency at which calibrations can be performed during the mission duration by considering observations of planets.

\begin{figure}[ht]
  \centering
  \includegraphics[width=1\columnwidth]{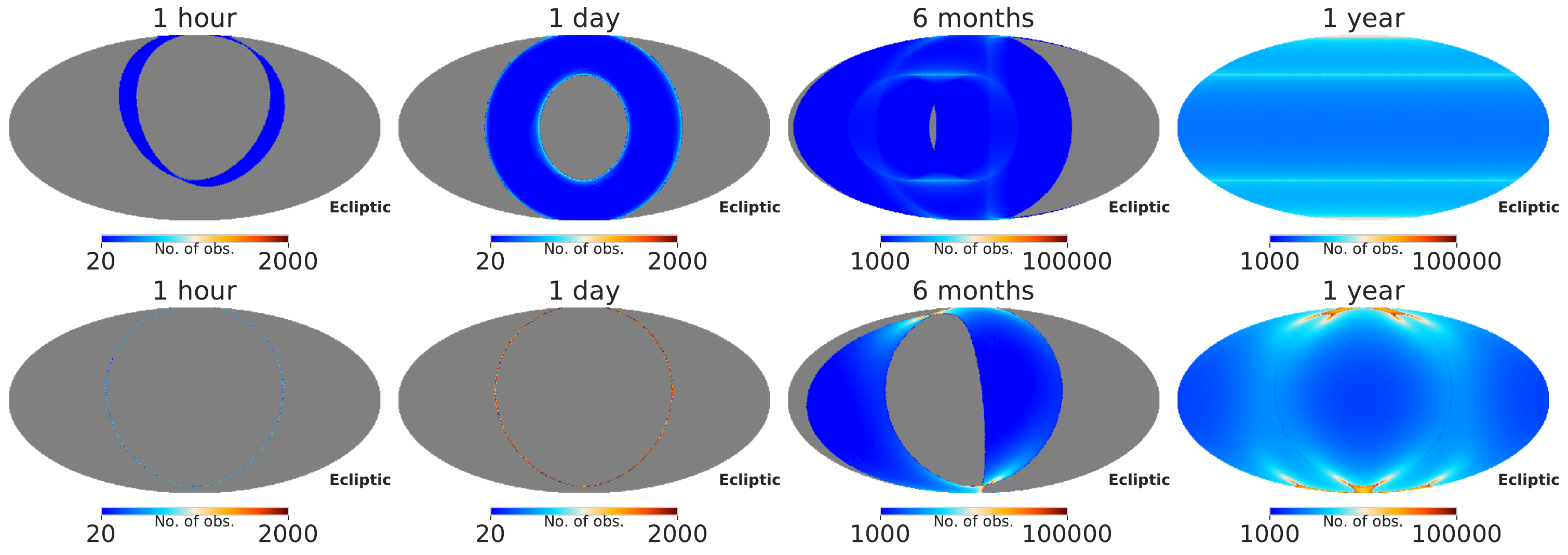}
  \caption{Time evolution of the hit-map simulated by \PICO's scanning strategy (top panels), and \Planck's scanning strategy (bottom panels). The times used in the simulations are, from left to right, 1\,hour, 1\,day, 6\,months and 1\,year. The sampling rate is set to 19\,Hz and the simulations are performed with $N_{\rm side}=128$. Both scanning strategies show continuous sky mapping with overlapping rings, with the rings described by spin shifting in slow precession, which is a significant difference from the \LiteBIRD's \SC shown in \cref{fig:hitmaps}.}
  \label{fig:hitmap_pico_planck}
\end{figure}

The \Planck and the planned \PICO missions are considered for comparison. \Cref{fig:hitmap_pico_planck} shows the time evolution of the hit-maps simulated for the scanning strategies of the two spacecraft, detailed in \cref{tab:pico_and_planck}.

\begin{table}[t]
\centering
\begin{tabular}{lllll}
\hline
{}              & $\alpha$      & $\beta$      & $T_\alpha$   & $T_\beta$ \\ \hline
\LiteBIRD       & $45^\circ$    & $50^\circ$   & 3.2058\,hr  & 20\,min   \\ \hline
\PICO           & $26^\circ$    & $69^\circ$   & 10\,hr      & 1\,min    \\ \hline
\Planck         & $7.5^\circ$   & $85^\circ$   & 6\,month     & 1\,min    \\ \hline
\end{tabular}
\caption{Geometric/kinetic parameters of \LiteBIRD, \PICO and \Planck \cite{PICO2019, planck_prelaunch2010}.}
\label{tab:pico_and_planck}
\end{table}

\subsection{Beam reconstruction systematics}\label{sec:beam_reconstruction}

From the TOD of planet observations, it is possible to reconstruct the beam shape. This process is therefore extremely important in order to mitigate systematic effects and the choice of scanning strategy plays a key role in this regard \cite{far_sidelobe}.
As discussed in ref.~\cite{Planck_HFI_beam}, the degeneracies between systematic effects in the direction of the scan (such as detector time constant and pointing systematics) and the shapes of sidelobes can negatively impact the reconstruction capabilities. 

Consequently, although a degeneracy between the beam shape and the detector time constant was observed in \Planck, it is interesting to note that the accuracy of the beam shape reconstruction with Jupiter was improved by using a `deep scan mode', which slows down the spin axis shift angle once Jupiter is in the period when it can be observed \cite{Planck_LFI_beam}. The choice of scanning strategy is helpful in this respect, as \Planck devised a scanning mode to measure the beam shape.

\begin{figure}[h]
  \centering
  \includegraphics[width=0.45\columnwidth]{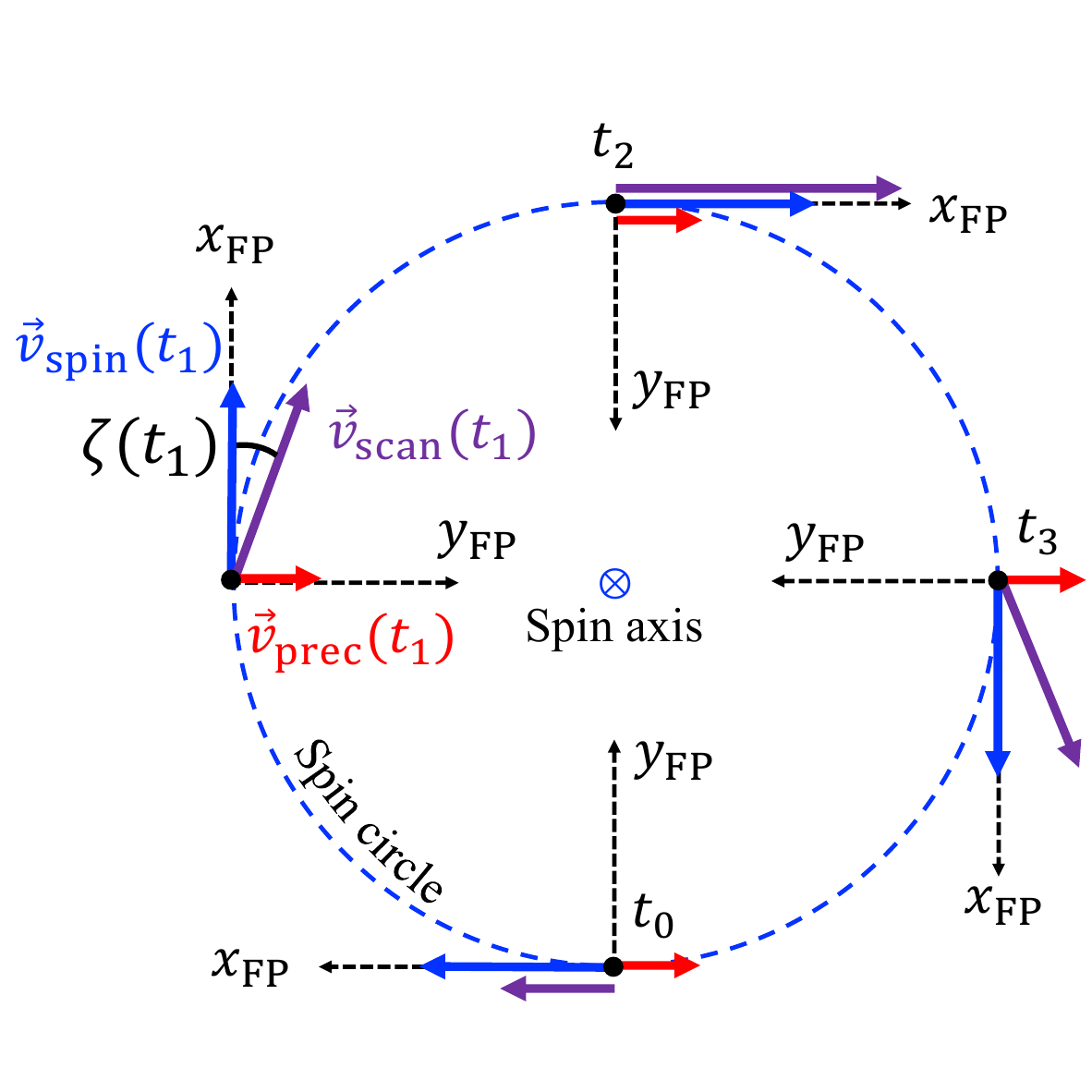}
  \includegraphics[width=0.45\columnwidth]{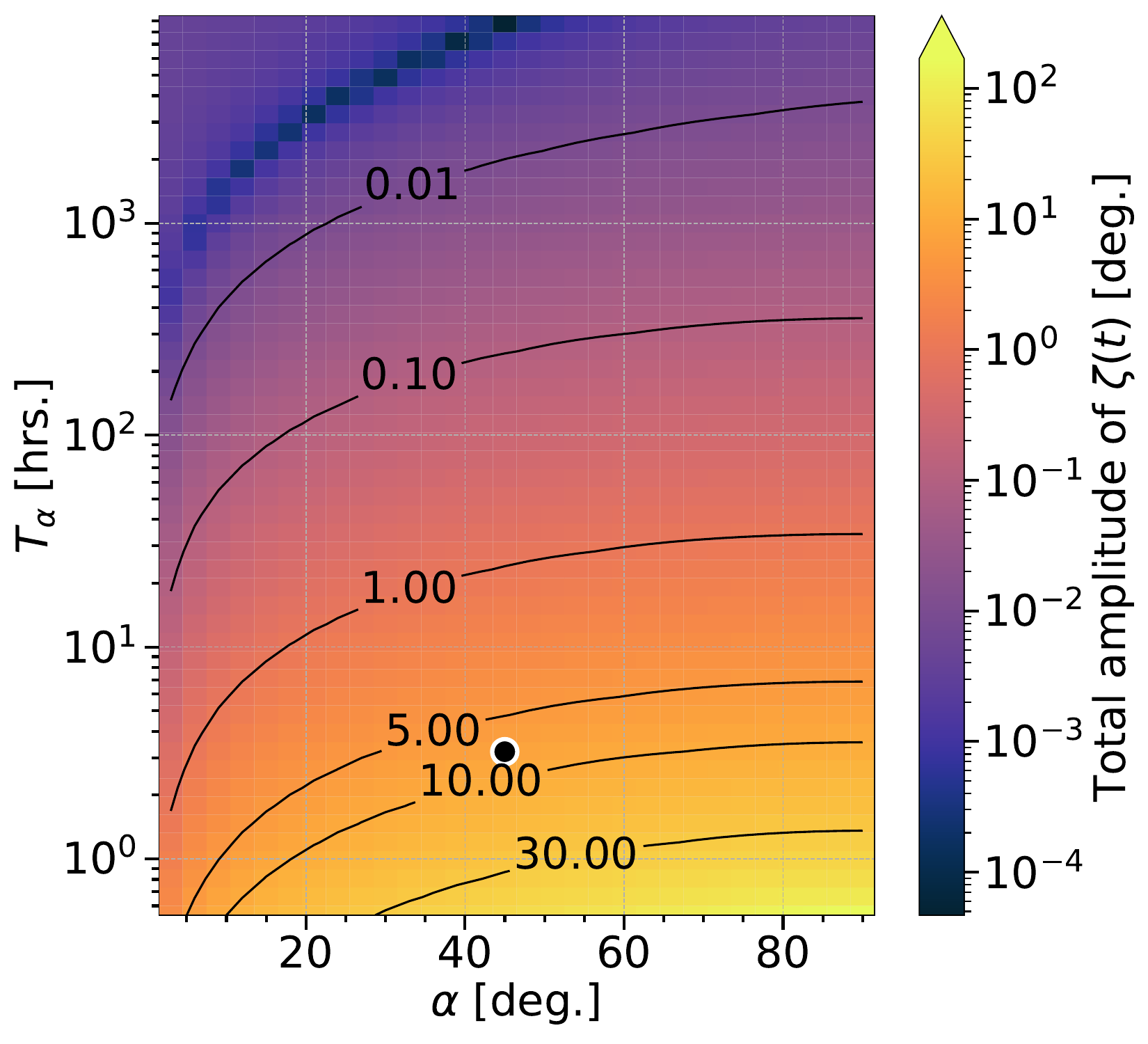}
  \caption{(left) Illustration of velocity vectors during one spin cycle looking down from the spin axis. Here $x_{\rm FP}$ and $y_{\rm FP}$ represent a focal plane coordinate which indicates the detector reference frame, while $\vec{v}_{\rm prec}$ and $\vec{v}_{\rm spin}$ show the velocity vectors of precession and spin, respectively.
  The precession motion acts to push the spin axis in the same direction (to the right in the figure) as seen from the spin axis, and the scan velocity vector $\vec{v}_{\rm scan}$, a combination of $\vec{v}_{\rm prec}$ and $\vec{v}_{\rm spin}$, changes the direction across the $x_{\rm FP}$-axis.
  The angle $\zeta$, referred to as the scanning beam angle, is given by the angle between $\vec{v}_{\rm scan}$ and the $x_{\rm FP}$-axis.
  (right) Value of the total amplitude of $\zeta(t)$ during one spin cycle in the $\{\alpha,T_\alpha\}$ space. }
\label{fig:scanbeam}
\end{figure}

\Cref{fig:scanbeam} (left) shows a sketch of the velocity vectors during one spin cycle. We introduce here the scanning beam angle $\zeta$, which is the angle between the $x$-axis of the focal plane coordinate ($x_{\rm FP}$) and the scan velocity vector $\vec{v}_{\rm scan}$, which can itself be expressed as the combination of spin and precession velocity vectors, $\vec{v}_{\rm spin}$ and $\vec{v}_{\rm prec}$ respectively. During the scanning,
$\vec{v}_{\rm scan}$ changes direction across the $x_{\rm FP}$-axis, such that $\zeta(t)$ oscillates as a function of time.

\Cref{fig:scanbeam} (right), displays the total amplitude of $\zeta(t)$, defined as the range of variation of the angle made by the direction of scanning in the detector's frame during one spin cycle. The greater the amplitude, the easier it becomes to solve the degeneracies between unknown systematic effects related to the direction of $\vec{v}_{\rm scan}$. A greater amplitude also facilitate beam-shape reconstruction and characterisation of the sidelobes.

The small value of $T_\alpha$ for \LiteBIRD is associated with an total amplitude of $\zeta$ of approximately $5^\circ$, which is a much larger value compared to a mission such as \Planck, which had a total amplitude of $\zeta$ of 0.5\,arcsec.
As such, sidelobe reconstruction must be easier with missions like \LiteBIRD due to its faster precession. Because of this much larger total amplitude of $\zeta(t)$, the presence of unanticipated systematics -- such as longer detector time constant or transfer function effects -- would be noticeable during the beam reconstruction, leaving some flexibility in order to account for them.
However, it is true that the \Planck and \PICO scanning strategies, with larger $T_\alpha$, have the advantage of being able to observe compact sources more frequently over shorter periods.

\subsection{Sky pixel visit/revisit times} \label{sec:skypix}

In order to detect unknown time-dependent systematic effects, data sets are usually divided into splits associated with different time periods, so that null-tests may be performed by differencing these splits. The distribution of the time at which a sky pixel is observed, i.e., the numbers of iterations at which the pointing visited a pixel, is one indicator for considering how effectively null-tests can be performed. For example, if the same pixel has been observed uniformly over the entire mission duration, it is effective for detecting systematic effects that vary over long time scales, such as gain drift.
The revisit time separating the re-observation of a sky pixel is also an important indicator and is defined as
\begin{align}
    t^{\rm re}_j = t_{j+1} - t_j,
\end{align}
where $j$ represents the $j^{\rm th}$ measurement on the sky pixel and the revisit time $t^{\rm re}_j$ can be defined by subtracting the $j^{\rm th}$ from the $(j+1)^{\rm th}$ observation time.
If the distribution composed of $t^{\rm re}_j$ is uniform and the number of revisit events at each time scale is abundant, a data set can be created for any time scale required.

We focus on the following three characteristic points in the \texttt{HEALPix} map in ecliptic coordinates: $(\theta,\varphi)=(0^\circ,0^\circ),(45^\circ,180^\circ)$, and $(90^\circ,180^\circ)$,
and calculate the visit time of these pixels. In all spacecraft cases, we used the same sampling rate as in the \SC, 19\,Hz, and discretise the pixels of the map with $N_{\rm side}=64$.

\begin{figure}[h]
  \centering
  \includegraphics[width=0.95\columnwidth]{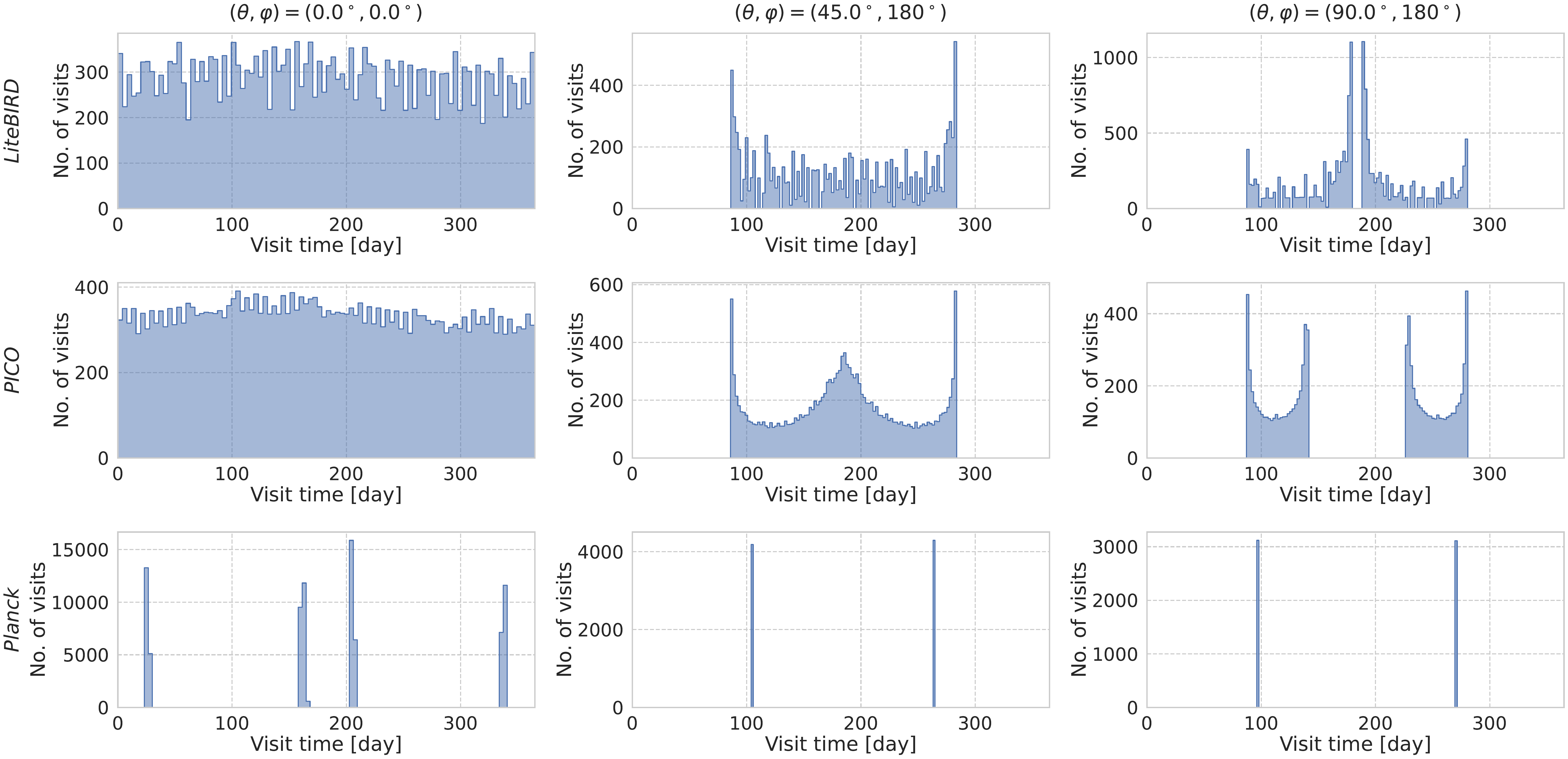}
  \caption{Distribution of visit times per spacecraft per sky position. From the top row, in the order of \LiteBIRD, \PICO and \Planck, the distributions are $(\theta,\varphi)$=$(0^\circ,0^\circ), (45^\circ,180^\circ)$, and $(90^\circ,180^\circ)$ from  left to right, i.e., at the North pole, between the North pole and the equator, and at sky pixels located at the equator in ecliptic coordinates. The pixels are given in a discretised \texttt{HEALPix} map with $N_{\rm side}=64$ and the sampling rate is 19\,Hz for all simulations. Note that the vector from the Sun to $\rm{L_2}$ at the start of the simulation was chosen to point to $(\theta,\varphi)=(90^\circ,0^\circ)$.}
  \label{fig:visit_hist}
\end{figure}

\Cref{fig:visit_hist} shows the distribution of visit times in different sky positions for each mission. It can be seen that for pixels with $(\theta,\varphi)=(0^\circ,0^\circ)$, \LiteBIRD and \PICO have abundant visits throughout the year, while for \Planck there is a gap in the observation period. This is because the $\alpha$ of \Planck's scanning strategy describes large rings over the spin cycle due to the large $\beta$, as we can see in \cref{fig:hitmap_pico_planck}, so that many visits occur to the same pixel in the sky on short timescales, but once the ring of the scan drifts away from the region due to orbital rotation, the next visit will not happen until about 6\,months later.
Also, large gaps in visit time exist for all pixels considered, making it difficult to detect some effects, for example, gain drifts that occur on longer timescales.

On the other hand, \LiteBIRD and \PICO have a larger $\alpha$ and shorter precession period than \Planck. This reduces the size of the ring described by the spin and allows the pointing to visit many pixels on the sky many occasions for long periods of time due to its relatively fast precession.
A characteristic difference between the two spacecraft is the width of the gap found in the center of the visit time distribution of equatorial pixels. 
This gap is created by the fact that the equatorial sky pixels are contained within the scan pupil mentioned in \cref{sec:comp_source_obs}.
Note that the size of the scan pupil is determined by $2|\alpha-\beta|$, so there is an annual unobservable time determined by the scan pupil size and the orbital velocity in regions that are $|\alpha-\beta|$ away from the equator.

\begin{figure}[h]
  \centering
  \includegraphics[width=0.95\columnwidth]{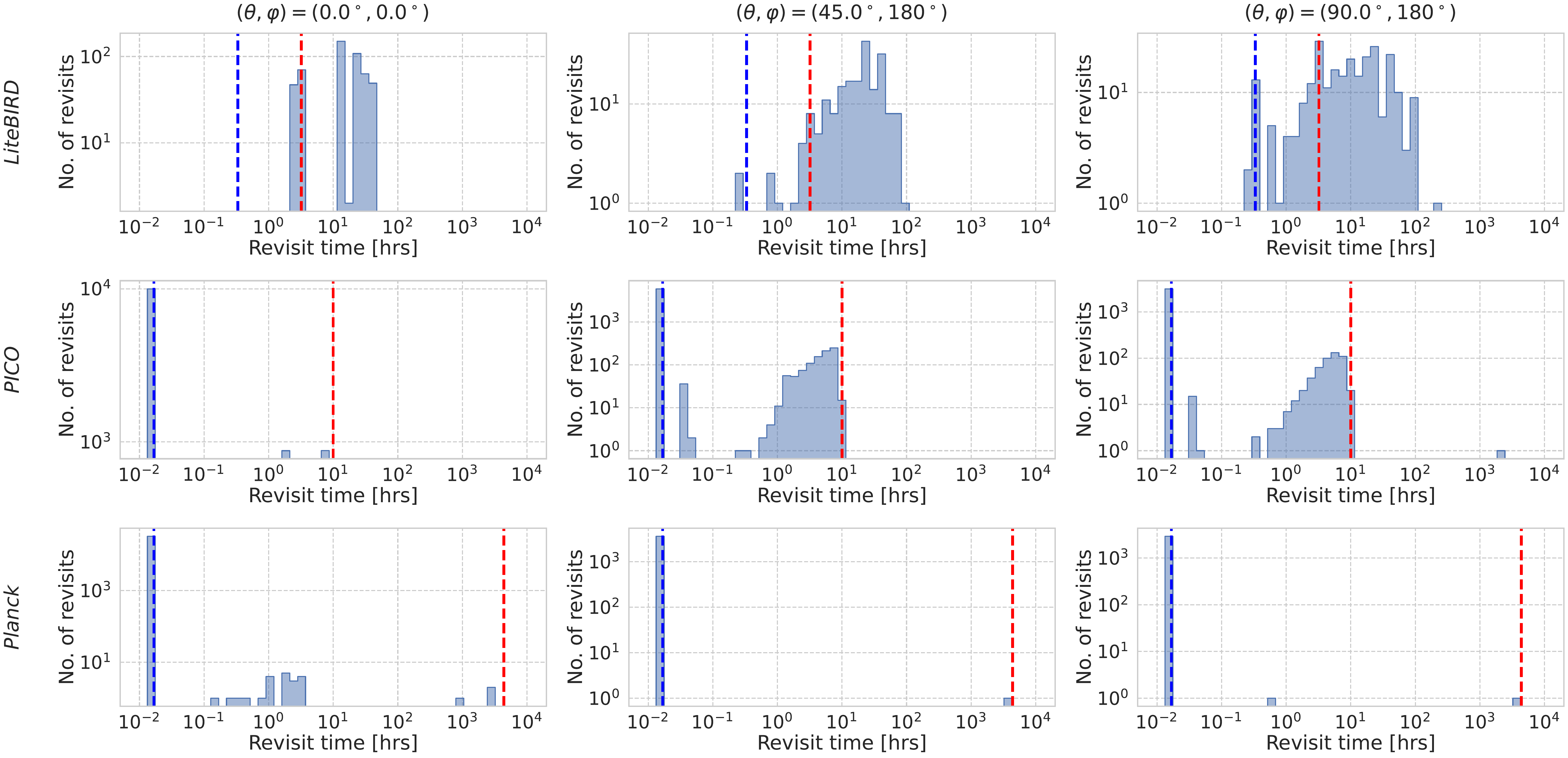}
  \caption{Distribution of revisit times per spacecraft per sky position. The order of the panels corresponds to \cref{fig:visit_hist}. The blue and red dashed lines indicate the spin period and precession period of the corresponding spacecraft, respectively. The reason for the existence of revisits shorter than the spin period is that at the outer edge of the scan pattern, the precession of the sweeping angular velocity reaches $\omega_{\rm{max}}$, which results in a revisit earlier than the spin period at the outer edge.}
  \label{fig:revisit_hist}
\end{figure}

\Cref{fig:revisit_hist} is a histogram of the revisit time against the visit time in \cref{fig:visit_hist}, showing how many revisit events of what time scale occur over the period of visiting the three considered pixels, with the blue and red dashed lines representing the spin period and precession period, respectively.
For the cases of \LiteBIRD and \PICO, the three pixels can be visited for long periods of time and the wide distribution of revisit times has the potential to construct a rich set of null-tests of various time scales, while for the case of \Planck, short timescale revisit event is rich, but the pointing can only visit a specific pixel for a very short period of time. This restricts null-tests, especially those requiring long timescale data sets.

Therefore, it seems that scanning strategies such as those of \LiteBIRD and \PICO, which have small scan pupils and relatively shorter precession periods (of at least $T_\alpha<100$\,hours, in agreement with our discussion in \cref{sec:optimisation}) are likely to be effective for null-tests in the time domain.

\subsection{Planet visit/revisit times} \label{sec:visit}

The long integration time of planet observations discussed in \cref{sec:comp_source_obs} is one important aspect of in-flight calibration, but the length of time that calibration is possible during the mission and how often calibration can be performed are also important.
In this section, visit and revisit times are simulated for target planets to clarify the relationship between the scanning strategy and the visibility of the planet. The simulations use the same setup as in \cref{sec:comp_source_obs}, except that in that section the visit time was integrated over time, whereas here we consider the visit time distribution before integration.

\begin{figure}[h]
  \centering
  \includegraphics[width=0.95\columnwidth]{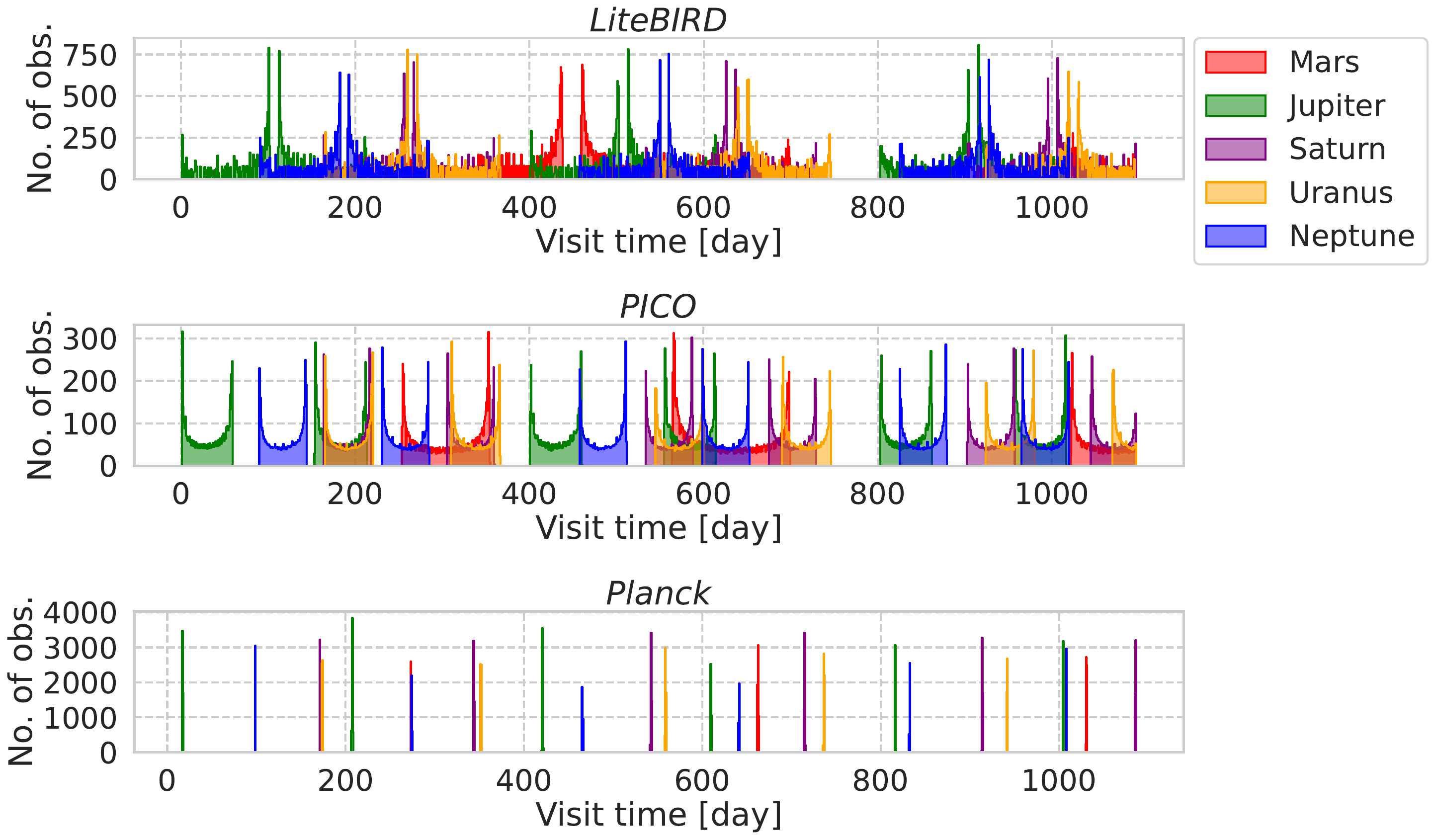}
  \caption{Histogram of the number of observations of the planet over a 3-year mission duration, showing the cases of \LiteBIRD, \PICO and \Planck. The bin width is set to 1\,day, the simulation start time is 2032-04-01T00:00:00, the planet is counted as a hit when scanning is within $0.5^\circ$ of the planet, and the planet is assumed to move every second.}
  \label{fig:planet_visit_hist}
\end{figure}

\Cref{fig:planet_visit_hist} shows a histogram of the visit time for planets for each spacecraft. Because the planets are located near the equator of ecliptic coordinates, the shape of the distribution is similar to the visit time distribution for the sky pixel $(\theta,\varphi) = (90^\circ,180^\circ)$ in \cref{fig:visit_hist}. Thus, here too, a smaller scan pupil leads to a longer period of time during which the planet can be observed. \PICO has a larger scan pupil than \LiteBIRD, so when a planet is spotted within the scan pupil, it is likely to have an unobservable period of about 90\,days. On the other hand, with a rotation period of 1\,minute and a precession period of 10\,hours, a planet also gains a period of observability every minute.
In the case of \Planck, this becomes extreme, and it is possible to observe a planet many times in a very short period. 
However, the visit time distribution creates large gaps, making it difficult to calibrate the gain and pointing drift over long timescales. The integrated visibility time of planets per spacecraft is 3.1\,hours for \LiteBIRD, 1.8\,hours for \PICO, and 1.6\,hours for \Planck, with \LiteBIRD obtaining the longest visibility time. This is related to the smaller scan pupil, but also to the fact that the spin period is 20 times longer than for the other two missions, which increases the time to transit the planet in one visit.

\begin{figure}
  \centering
  \includegraphics[width=0.95\columnwidth]{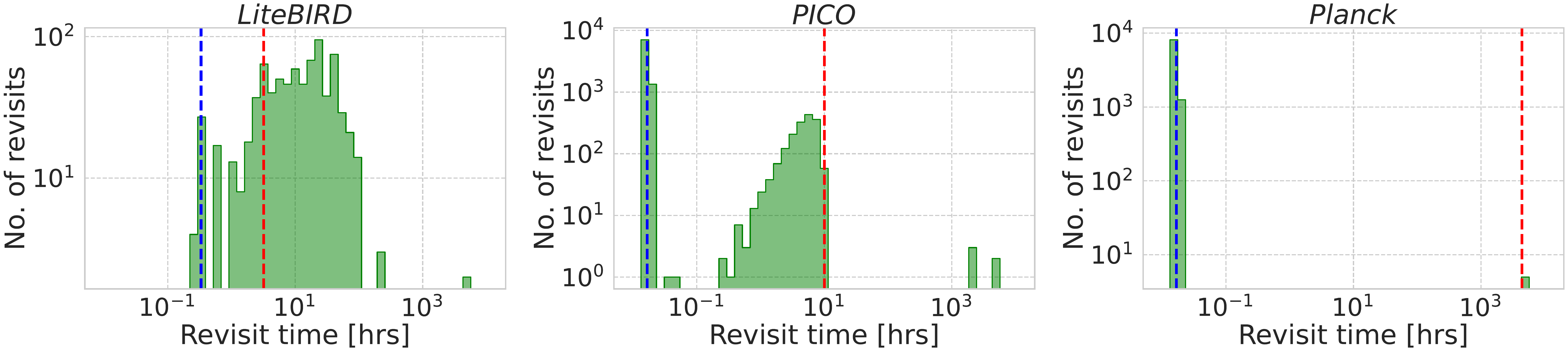}
  \caption{Histogram of Jupiter's revisit times obtained from the visit time distribution in \cref{fig:planet_visit_hist}. The blue and red dashed lines represent the spin period and precession period, respectively. In the three spacecraft cases, the shortest revisit time is given by the spin period, and the longest revisit time is given in about 6\,months, which is the period when the planet is outside the coverage. The second longest revisit time is given by the time between entering and exiting the scan pupil, which is about 10\,days for \LiteBIRD, about 90\,days for \PICO, and about 6\,months for \Planck.}
  \label{fig:planet_revisit_hist}
\end{figure}

\Cref{fig:planet_revisit_hist} shows the distribution of revisit times for Jupiter, which is frequently used for calibration. From this, it is expected that \LiteBIRD and \PICO have revisit events for various timescales and can calibrate a rich data set for each timescale. The revisit events of \Planck occur more often on shorter timescales. 
Therefore, it is important to keep the scan pupil small in order to continue observing the planet as long as possible over the mission duration, which also leads to obtaining revisit events on various timescales. In order to obtain a longer integrated visibility time, it is necessary to increase the spin period.

The shortest revisit event occurs at about the spin period of each spacecraft, while the precession period limits the long-term revisit time for scanning strategies such as \PICO, where the ring trajectory described by the spin overlaps with the next spin cycle, allowing continuous sky mapping. On the other hand, \LiteBIRD does not cause overlapping of the rings at each spin cycle, which results in many revisit times longer than the precession period. In all cases, the longest revisit time will be about 6\,months until the planet is next observed, and the second longest revisit time will be the period of spotting it in the scan pupil.

\section{Discussion and conclusion} \label{sec:conclusion}

In this study, we have explored the parameter space of the scanning strategy of a spacecraft mission equipped with a HWP, using $\{\alpha, T_\alpha, T_\beta\}$ as free parameters. We examined the impact of these parameters on three crucial metrics for $B$-mode observation: visibility time of planets; hit-map uniformity; and cross-link factor. 
In particular, the map-based simulations using the \spin formalism that we introduced, helps us to understand the relationship between the bias of tensor-to-scalar ratio $r$, or $\Delta r$, and cross-link factors, which have been considered important in order to suppress systematic effects.
The coupling between scanning strategies and systematic effects for HWP-equipped spacecraft has been non-trivial, but this method may provide an interface to understand it.

Using the scan simulator \Falcons developed here, we studied the distribution of each metric in the $\{\alpha, T_\alpha, T_\beta\}$ space. As a result, we were able to justify that the configuration used by \LiteBIRD in ref.~\cite{PTEP2023}, namely the \SC, provides an effective trade-off in term of instrumental calibration, suppression of systematic effects, and null-tests for the mission.

From a more general perspective, maintaining the considered metrics in a favorable state can be achieved by setting $\alpha\simeq\beta$ and $T_\alpha < 100$\,hours in the scanning strategy parameter space we defined in \cref{fig:standard_config_and_T_beta} (right). All the metrics considered were shown to be largely independent of $T_\alpha$ in the region $T_\alpha<100$\,hours (see e.g., \cref{fig:planet_and_sigma_hit,fig:cross-links}), such that the geometric parameters play a crucial role in order to optimise the scanning strategy. In \cref{sec:Opt_geometric}, we narrowed down the candidates for scanning strategies worth considering and concluded that $(\alpha,\beta)=(45^\circ,50^\circ)$ provides a good compromise between the considered metrics and the requirements set by the instrumental design of the \LiteBIRD mission.

All the figures presented were displayed in the $\{\alpha, T_\alpha\}$ space, fixing $T_\beta = \tbl$ as the input of the scanning strategy. Such a choice was justified because the structure of the metric distributions in the $\{\alpha, T_\alpha\}$ scales rigidly with the value of $T_\beta$ for $T_\beta<20$\,min, leaving the value of the optimum unchanged. Such behavior is discussed in \cref{apd:T_beta_scaled}.

This rigid scaling can also been observed in \cref{fig:rot_period_opt}, in which the dependence of the cross-link factors values is presented in the $T_\beta$ and $T_\alpha$ space. We extended our discussion in \cref{sec:Opt_kinetic}, finding that the choice of $T_\beta=20$\,min and $T_\alpha=192$\,min provides a good compromise in order to obtain the smallest cross-link factors.
In order to derive these results, we assumed the current rotation rate of the HWP in \LiteBIRD. Generalising this study to another experiment with a high $f_{\rm knee}$ would require a faster HWP rotation in order to obtain an efficient modulation. In such cases, the lower limit of the spin period determined by \cref{eq:T_spin} becomes smaller and a faster spin is allowed. However, it is clear from \cref{fig:rot_period_opt} that there is no benefit to shortening the spin period bellow $T_\beta=20$\,min. 

Longer spin periods are preferable for spacecraft attitude control, and shorter precession periods allow the CMB solar dipole signal used for gain calibration to be shifted to a higher frequency, which can reduce the effects of low frequency gain fluctuations.
As shown in \cref{fig:scanbeam}, a smaller precession period also leads to a wider range of scanning beam angles, which eliminates the degeneracy between beam shape and certain systematic effects. 
To avoid the appearance of a \moire pattern in the hit-map and the cross-link factor map due to resonance between spin and precession (\cref{fig:prec_tuning_maps}), a fine-tuning of the precession period was considered and it was decided to adopt $T_\alpha=192.348$\,min, where the resonance peak with a large standard deviation in these maps was not found in the neighborhood around $T_\alpha=192.348$\,min (\cref{fig:prec_tuning}).

In addition to actively suppressing systematic effects by reducing cross-link factors, it is also the role of scanning strategies to design abundant null-tests to inspect them if there are effective in-flight calibration procedures and unknown systematic effects. 
In \cref{sec:implications}, we compared the sky pixel visit/revisit time and the visit/revisit time to planets -- which are important indicators for designing calibration strategies and effective null-tests -- between the scanning strategies of \Planck, \PICO, and \LiteBIRD's \SC. 
Among these, \LiteBIRD and \PICO were found to have wide coverage per day and could continuously visit specific sky pixels and planets for a long time. 
Moreover, since these missions revisit pixels frequently and on various time intervals, it is suggested that these two spacecraft can perform calibration and null-tests by dividing the data for specific sky pixels and planets into data sets of various timescales. 
However, it is true that \Planck's scanning strategy, which achieves deep sensitivity in a short time and observes stably with many sky pixels, is suitable for temperature observations, although it was not designed with for polarisation observations. 
Furthermore, although not mentioned in the manuscript, the spin direction of the spacecraft and the precession direction are also parameters that can be freely determined. All the results here presented were obtained by imposing counterclockwise spin, precession, and orbital motion on the spacecraft. We simulated all indicators for all four combinations consisting of clockwise or counterclockwise rotations, but no changes due to differences in rotation direction were observed. A detailed discussion on this point can be found in \cref{apd:rotation_direction}.

Our analysis was greatly facilitated by the use of \Falcons, which was useful for avoiding the need for both grid search via supercomputer and thread parallelisation for efficient memory usage. Moreover, the relationship between scanning strategy parameter space and \LB's \SC in our devised constraints offer valuable insights for future polarimetric space missions' scanning strategy design.
This, coupled with our scan simulator \Falcons, should allow for other applications beyond the \LiteBIRD mission.

\acknowledgments
% =====================================================================================
% LiteBIRD Standard Acknowledgements, long format
% =====================================================================================
%
% Japan; point-of-contact: Masashi Hazumi <masashi.hazumi@kek.jp>
This work is supported in Japan by ISAS/JAXA for Pre-Phase A2 studies, by the acceleration program of JAXA research and development directorate, by the World Premier International Research Center Initiative (WPI) of MEXT, by the JSPS Core-to-Core Program of A. Advanced Research Networks, and by JSPS KAKENHI Grant Numbers JP15H05891, JP17H01115, and JP17H01125.
% Canada point-of-contact: Matt Dobbs <Matt.Dobbs@mcgill.ca>
The Canadian contribution is supported by the Canadian Space Agency.
% France; point-of-contact: Ludovic Montier <ludovic.montier@irap.omp.eu>
The French \textit{LiteBIRD} phase A contribution is supported by the Centre National d’Etudes Spatiale (CNES), by the Centre National de la Recherche Scientifique (CNRS), and by the Commissariat à l’Energie Atomique (CEA).
% Germany; point-of-contact: Eiichiro Komatau <komatsu@MPA-Garching.MPG.DE>
The German participation in \textit{LiteBIRD} is supported in part by the Excellence Cluster ORIGINS, which is funded by the Deutsche Forschungsgemeinschaft (DFG, German Research Foundation) under Germany’s Excellence Strategy (Grant No.~EXC-2094 - 390783311).
% Italy; point-of-contact: Paolo Natoli <ntlpla@unife.it>
The Italian \textit{LiteBIRD} phase A contribution is supported by the Italian Space Agency (ASI Grants No.~2020-9-HH.0 and 2016-24-H.1-2018), the National Institute for Nuclear Physics (INFN) and the National Institute for Astrophysics (INAF).
% Norway; point-of-contact: Hans Kristian Eriksen <h.k.k.eriksen@astro.uio.no>
Norwegian participation in \textit{LiteBIRD} is supported by the Research Council of Norway (Grant No.~263011) and has received funding from the European Research Council (ERC) under the Horizon 2020 Research and Innovation Programme (Grant agreement No.~772253 and 819478).
% Spain; point-of-contact: Enrique Martinez-Gonzalez <martinez@ifca.unican.es>
The Spanish \textit{LiteBIRD} phase A contribution is supported by MCIN/AEI/10.13039/501100011033, project refs.~PID2019-110610RB-C21, PID2020-120514GB-I00, PID2022-139223OB-C21 (funded also by European Union NextGenerationEU/PRTR), and by MCIN/CDTI ICTP20210008 (funded also by EU FEDER funds).
% Sweden; point-of-contact: "NAME" "EMAIL ADDRESS"
Funds that support contributions from Sweden come from the Swedish National Space Agency (SNSA/Rymdstyrelsen) and the Swedish Research Council (Reg. no.~2019-03959).
% UK; point-of-contact: Erminia Calabrese <calabrese.erminia@gmail.com>
% To be included.
% USA; point-of-contact: Adrian Lee <Adrian.Lee@berkeley.edu>
The US contribution is supported by NASA grant no.~80NSSC18K0132.
%
% =====================================================================================
%==== YUSUKE DC2/C2C/OU fellow/RECTOR =======
%\yusuke{Yusuke}
This work has also received funding by the European Union’s Horizon 2020 research and innovation program under grant agreement no.~101007633 CMB-Inflate. 
YT acknowledges the support by the Grant-in-Aid for JSPS Fellows, JSPS core-to-core program, JST the establishment of university fellowships towards the creation of science technology innovation, Grant no.~JP23KJ1602, JPJSCCA20200003, and JPMJFS2128.
GP and HI acknowledge the JSPS Invitation Fellowships for Research in Japan. 
%\leocomment{Léo:}
LV acknowledges partial support by the Italian Space Agency \LiteBIRD Project (ASI Grants No.~2020-9-HH.0 and 2016-24-H.1-2018), as well as the RadioForegroundsPlus Project HORIZON-CL4-2023-SPACE-01, GA 101135036.

\bibliographystyle{Class/JHEP}

%--- During a paper writing, it should be on---
%\bibliography{Class/bibliography.bib}
%----------------------------------------------
% When we submit it to arxiv, we must comment out:
% \bibliography{Class/bibliography.bib}
% and generate output.bbl by overleaf's compile which is avilable 
% [Submit]->[arxiv]->[Download...]
% change its name output.bbl to bibliography.bbl
% and put it Class/bibliography.bbl
% then, turn on: 

\providecommand{\href}[2]{#2}\begingroup\raggedright\endgroup
 % for arXiv

\appendix
\section{Additional derivations}

\subsection{Scanning motion of spacecraft} \label{apd:scan_motion}

This section aims at defining the spin and precession motions in detail.
Consider an orthonormal coordinate frame $xyz$ such that the spin axis of the spacecraft is along the $z$-axis. $\beta$ is the angle between the telescope boresight and the spin axis, such that in the $xyz$ frame -- the coordinates of the vector indicating the boresight direction can be expressed as $\mathbf{n}_0=(\sin\beta, 0, \cos\beta)$.
The motion of $\mathbf{n}_0$ around the $z$-axis with angular velocity $\omega_\beta$ is called the `spin' and can be expressed by acting on $\mathbf{n}_0$ with the time dependent rotation matrix $R_z$ as
\begin{align}
    \mathbf{n}_{\rm spin}(t) &= R_z(\omega_\beta t)\mathbf{n}_0 \nonumber\\
    &=
    \mqty(
    \cos\omega_\beta t  &  -\sin\omega_\beta t  & 0 \\
    \sin\omega_\beta t  &   \cos\omega_\beta t  & 0 \\
    0                   &  0                    & 1
    )
    \mqty(
    \sin\beta \\
    0         \\
    \cos\beta
    ),
\end{align}
where $R_j~(j\in\{x,y,z\})$ can be defined similarly as the rotation matrices around each individual axis.

Now, let us account for a possible time dependence of the spin axis direction. Let $z$-axis be the Sun-spacecraft axis, as sketched in \cref{fig:standard_config_and_T_beta} (left). In order to tilt the spin axis by an angle $\alpha$ from the previous configuration, a rotation must be induced on it around the $y$-axis
\begin{align}
    \mathbf{n}_{\rm spin}'(t) &= R_y(\alpha)\mathbf{n}_{\rm spin}(t) \nonumber\\
    &=
    \mqty(
    \cos\alpha   &  0  &  \sin\alpha \\
    0            &  1  &  0          \\
    -\sin\alpha  &  0  &  \cos\alpha
    )
    \mathbf{n}_{\rm spin}(t),
\end{align}
where $\mathbf{n}_{\rm spin}'(t)$ represents the rotation of the telescope's boresight around the spin axis with an angle of $\beta$ with respect to the spin axis inclined by $\alpha$ from the $z$-axis.
We thus consider a rotational motion around the $z$-axis with a rotation axis inclined by $\alpha$ from the $z$-axis with an angular velocity of $\omega_\alpha$.
We refer to this rotation as `precession' and express it by
\begin{align}
    \mathbf{n}(t) &= R_z(\omega_\alpha t)\mathbf{n}_{\rm spin}'(t) \nonumber\\
    &=
    \mqty(
    \cos\omega_\alpha t  &  -\sin\omega_\alpha t  & 0 \\
    \sin\omega_\alpha t  &   \cos\omega_\alpha t  & 0 \\
    0                    &  0                     & 1
    )
    \mathbf{n}_{\rm spin}'(t),
\end{align}
where $\mathbf{n}(t)$ represents observation direction of the boresight after imposing spin and precession motion with specific geometric/kinetic parameters.
Finally, the total rotational motion imposed on the spacecraft is summarised by the combined action of spin and precession with the following matrix chain:
\begin{align}
    \mathbf{n}(t) &= R_z(\omega_\alpha t)R_y(\alpha )R_z(\omega_\beta t)\mathbf{n}_0 \nonumber\\
    &=
    \mqty(
    \cos\omega_\alpha t  &  -\sin\omega_\alpha t  & 0 \\
    \sin\omega_\alpha t  &   \cos\omega_\alpha t  & 0 \\
    0                    &  0                     & 1
    )
    \mqty(
    \cos\alpha   &  0  &  \sin\alpha \\
    0            &  1  &  0          \\
    -\sin\alpha  &  0  &  \cos\alpha
    )
    \mqty(
    \cos\omega_\beta t  &  -\sin\omega_\beta t  & 0 \\
    \sin\omega_\beta t  &   \cos\omega_\beta t  & 0 \\
    0                   &  0                    & 1
    )
    \mqty(
    \sin\beta \\
    0         \\
    \cos\beta
    ). \label{eq:matrix_chain}
\end{align}
The full scanning motion is formulated by imposing such a time dependence of $\mathbf{n}(t)$ on an orbital motion at the Sun-Earth $\rm{L_2}$ point.

\subsubsection{Sweep angular velocity on the sky} \label{apd:sweeping_velocity}

Here we derive the maximum angular velocity at which the telescope's boresight sweeps the sky.\footnote{In fact, a rotation matrix related to the orbital rotation around the Sun can be considered, but it is ignored here because its contribution to the spin and precession is small.}
By using \cref{eq:matrix_chain}, we can define the angle $\Delta \xi$ from which $\mathbf{n}(t)$ moved between time $t$ and $t + \Delta t$. It is defined as
\begin{align}
    \Delta \xi = \abs{\mathbf{n}(t+\Delta t) - \mathbf{n}(t)}.
\end{align}
The angular velocity $\dd{\xi}/\dd{t}$ which corresponds the sweep angular velocity in the sky is
\begin{align}
    \dv{\xi}{t} = \abs{\dv{\mathbf{n}(t)}{t}}.
\end{align}
$\abs{\dd{\xi}/\dd{t}}$ is maximised when $\omega_\beta t = 0$ and we obtain maximum angular velocity at which $\mathbf{n}(t)$ sweeps the sky, $\omega_{\rm max}$ as
\begin{align}
    \omega_{\rm max} = \abs{\dv{\xi}{t}}_{\rm max} = \omega_\alpha\sin(\alpha+\beta) + \omega_\beta\sin\beta. \label{eq:sweeping_velocity}
\end{align}
Note that we assumed here that the precession and the spin motion share a common direction of rotation. The behavior of $\omega_{\rm max}$ when one of the directions of rotation is reversed, as well as its impact on the scanning strategy optimisation, is discussed in \cref{apd:rotation_direction}.

\subsection{Analytical description in \spin space} \label{apd:spin_space_analysis}

\subsubsection{Map-making equation} \label{apd:map-making}
In order to derive the map-making equation, we start from the signal without systematic effects:
\begin{align}
    S(\Omega,\psi,\phi) &= I(\Omega) + Q(\Omega)\cos(4\phi-2\psi) + U(\Omega)\sin(4\phi-2\psi)\nonumber \\
      &= I(\Omega) + \frac{1}{2}P(\Omega)e^{-i(4\phi-2\psi)} + \frac{1}{2}P^*(\Omega)e^{i(4\phi-2\psi)},
\end{align}
where we introduce $P=Q+iU$ and its complex conjugate $P^*$.
Here, we define the signal that is measured by $j^{\rm th}$ observation at the sky pixel $\Omega$ as
\begin{align}
    d_j &= \mqty(1 & \frac{1}{2}e^{-i{(4\phi_j-2\psi_j)}} & \frac{1}{2}e^{i{(4\phi_j-2\psi_j)}}) \mqty(I \\ P \\ P^* ) + n_j\nonumber \\
        &= \mathbf{w}_j \cdot \mathbf{s} + n_j,
\end{align}
where $\mathbf{s}$ is the Stokes vector and $n_j$ represents the noise. 
In order to estimate the Stokes vector from the measured signal, we minimise: 
\begin{align}
    \chi^2 = \sum_{i,j}(d_i -\mathbf{w}_i \cdot \mathbf{s})(N^{-1})_{ij}(d_j -\mathbf{w}_j \cdot \mathbf{s}),
\end{align}
where $N_{ij}$ is the noise covariance matrix. After minimising $\chi^2$, we can obtain the equation to estimate the Stokes vector as
\begin{align}
    \hat{\mathbf{s}} &= \left(\sum_{i,j} \mathbf{w}_i^\dagger (N^{-1})_{ij} \mathbf{w}_j\right)^{-1} \left(\sum_{i,j} \mathbf{w}_i^\dagger (N^{-1})_{ij} d_j\right), \label{eq:map-making_eq_noise}
\end{align}
where $\hat{\mathbf{s}}$ represents the estimated Stokes vector and $\dagger$ represents the Hermitian transpose. 

Now, if we assume that the noise is white noise, i.e., it does not have a correlation between the $i^{\rm th}$ and the $j^{\rm th}$ measurement.
Furthermore, we define the symbol for the average of a quantity $x_j$ as $\langle x_j \rangle = \frac{1}{N}\sum_{j}^{N} x_j$, then, \cref{eq:map-making_eq_noise} can be expressed as
\begin{align}
    \hat{\mathbf{s}} &= \left(\sum_{j} \mathbf{w}_j^\dagger \mathbf{w}_j\right)^{-1} \left(\sum_{j} \mathbf{w}_j^\dagger d_j\right) \nonumber \\
    &=
    \mqty(
    1                   & \frac{1}{2}\h[-2,4] & \frac{1}{2}\h[2,-4] \\
    \frac{1}{2}\h[2,-4] & \frac{1}{4}         & \frac{1}{4}\h[4,-8] \\
    \frac{1}{2}\h[-2,4] & \frac{1}{4}\h[-4,8] & \frac{1}{4}
    )^{-1}
    \mqty(
    \langle d_j \rangle \\ \frac{1}{2}\langle d_j e^{i(4\phi_j-2\psi_j)}\rangle \\ \frac{1}{2}\langle d_j e^{-i(4\phi_j-2\psi_j)}\rangle
    ). \label{eq:map-making_TOD}
\end{align}
This equation is equivalent to the simple binning map-making approach (e.g.\ ref.~\cite{binning_brawn}), and the following relation
\begin{align}
    \mqty(
    \langle d_j \rangle \\ \frac{1}{2}\langle d_j e^{i(4\phi_j-2\psi_j)}\rangle \\ \frac{1}{2}\langle d_j e^{-i(4\phi_j-2\psi_j)}\rangle) =
    \mqty(
    \Sd[0,0]  \\ \frac{1}{2}\Sd[2,-4] \\ \frac{1}{2}\Sd[-2,4]
    ),
\end{align}
allows \cref{eq:map-making_TOD} to be \cref{eq:map-making_spin}.
$\Sd[0,0]$, $\Sd[2,-4]$ and $\Sd[-2,4]$ are obtained from \cref{eq:kSd} and can be expressed as a coupling of the signal field and the orientation function in \spin space.

\subsubsection{Systematic effect of pointing offset} \label{apd:pointing_offset}

We study here the impact of a pointing offset caused by a deviation of the absolute pointing of the spacecraft from the expected direction. This can be caused by a miscalibration of the star tracker on-board the spacecraft or a mounting mismatch between it and the telescope.
For the formulation of the pointing offset we refer to the coordinates and formalism introduced in refs.~\cite{OptimalScan,mapbased} for the study of differential pointing.
We use a flat-sky approximation in the small angular scale limit and span the sky with a Cartesian coordinate system. In this coordinate frame, the $x$- and $y$-axis are chosen to be aligned with east and north, respectively.
Assuming that a first-order Taylor expansion around $(x,y)$ is possible when the pointing offset is small, the pointing offset field, $S$ can be written as
\begin{equation}
    \begin{split}
        S(\psi, \phi) = [1 - (\partial_x\Delta x+ \partial_y\Delta y)]I &+ \frac{1}{2}\left[1-(\partial_x\Delta x+ \partial_y\Delta y)\right]Pe^{-i(4\phi-2\psi)} \\
        &+ \frac{1}{2}\left[1-(\partial_x\Delta x+ \partial_y\Delta y)\right]P^*e^{i(4\phi-2\psi)}.
    \end{split}
\end{equation}
The perturbation term can be defined by using the magnitude of the pointing offset, $\rho$ and direction of the pointing offset, $\chi$ as
\begin{equation}
    \begin{split}
        \partial_x\Delta x+ \partial_y\Delta y &= \partial_x [\rho \sin(\psi+\chi)] + \partial_y [\rho \cos(\psi+\chi)]\\
        &=\frac{\rho}{2}\left[ e^{i(\psi+\chi)} \eth + e^{-i(\psi+\chi)} \overline{\eth} \right],
    \end{split}
\end{equation}
where we introduced the \spin-up (-down) ladder operators, $\eth=\partial_y - i\partial_x$, ($\overline{\eth}=\partial_y + i\partial_x$) as in ref.~\cite{mapbased}.
By using these operators, we can rewrite the pointing offset field as
\begin{equation}
    \begin{split}
        S(\psi, \phi) &= I - \frac{\rho}{2}\left[e^{i(\psi+\chi)}\eth + e^{-i(\psi+\chi)}\overline{\eth}\right]I \\
        &+ \frac{1}{2} \left[e^{-i(4\phi-2\psi)} - \frac{\rho}{2}\left(e^{i(-4\phi+3\psi+\chi)}\eth + e^{-i(4\phi-\psi+\chi)}\overline{\eth}\right) \right]P\\
        &+ \frac{1}{2} \left[e^{i(4\phi-2\psi)} - \frac{\rho}{2}\left(e^{i(4\phi-\psi+\chi)}\eth + e^{-i(-4\phi+3\psi+\chi)}\overline{\eth}\right) \right]P^*.\label{eq:pointing_offset_field}
    \end{split}
\end{equation}
Performing a Fourier transform $(\psi,\phi) \to (n,m)$, we can transfer the signal from real space to \spin space as
\begin{align}
    \St[0,0]  &= I, \\
    \St[2,-4] &= \St[-2,4]^* = \frac{P}{2}, \\
    \St[1,0]  &= \St[-1,0]^* = -\frac{\rho}{2}e^{i\chi}\eth I, \label{eq:temperature_syst}\\
    \St[1,-4] &= \St[-1,4]^* = -\frac{\rho}{4}e^{-i\chi}\overline{\eth} P, \\
    \St[3,-4] &= \St[-3,4]^* = -\frac{\rho}{4}e^{i\chi}\eth P.
\end{align}
The first two rows above are pure signals, with \spin-$(0,0)$ for temperature and \spin-$(\pm2,\mp4)$ for polarisation $P$ or $P^*$, while the latter three rows are systematic effects.
Looking at the systematic signal in \cref{eq:temperature_syst}, it appears that the \spin-$(0,0)$ signal from the original temperature field  is transferred to a \spin-$(\pm1,0)$ signal under the action of the \spin ladder operators. The \spin ladder operators, resulting from a perturbation of the pointing offset, expresses the gradient of the field and produces a spurious odd \spin component that is not present in the expected signal.
By using \cref{eq:kSd}, we can describe the coupling between the scanning strategy and the systematic effects, in the case of the pointing offset systematics:
\begin{align}
    {}_{0,0}\tilde{S}^d(\Omega)
    = {} & \sum_{n'=-\infty}^{\infty} \sum_{m'=-\infty}^{\infty} {}_{0-n',0-m'}\tilde{h}(\Omega) {}_{n',m'}\tilde{S}(\Omega) \\
    \begin{split}
        = {} &  \h[3,-4]\St[-3,4] + \h[2,-4]\St[-2,4] + \h[1,-4]\St[-1,4] \\
             &+ \h[1,0] \St[-1,0] + \h[0,0] \St[0,0]  + \h[-1,0]\St[1,0] \\
             &+ \h[-1,4]\St[1,-4] + \h[-2,4]\St[2,-4] + \h[-3,4]\St[3,-4], \nonumber
    \end{split}\\
    {}_{2,-4}\tilde{S}^d(\Omega) = {} & {}_{-2,4}\tilde{S}^{d*}(\Omega) = \sum_{n'=-\infty}^{\infty} \sum_{m'=-\infty}^{\infty} {}_{2-n',-4-m'}\tilde{h}(\Omega) {}_{n',m'}\tilde{S}(\Omega)\\
    \begin{split}
        = {} &  \h[5,-8]\St[-3,4] + \h[4,-8]\St[-2,4] + \h[3,-8]\St[-1,4] \\
             &+ \h[3,-4]\St[-1,0] + \h[2,-4]\St[0,0]  + \h[1,-4]\St[1,0] \\
             &+ \h[1,0] \St[1,-4] + \h[0,0] \St[2,-4] + \h[-1,0]\St[3,-4]. \nonumber
    \end{split}
\end{align}
These results can be obtained by multiplication of the orientation functions $\h[n,m]$ and the $\St[n,m]$ fields, making explicit how systematic effects with a particular \spin-$(n,m)$ can be suppressed by the orientation function defined by the scanning strategy.

\subsubsection{Systematic effect of HWP non-ideality}\label{apd:HWP_sys}

While HWPs are expected to modulate polarisation and effectively eliminate the $1/f$ noise, reduce the uncertainties of polarisation observations even with a single detector and suppress many systematic effects, there are also systematic effects produced by HWP's imperfections that need to be carefully considered e.g.\ \cite{Giorgio_hwp,Lorenzo_hwp,Essinger_hwp}. In particular, non-ideality of the HWP can create non-diagonal terms in its Mueller matrix, which can induce instrumental polarisation that causes temperature-to-polarisation leakage.

This systematic effect is discussed in ref.~\cite{guillaume_HWPIP} and it has been shown that the CMB solar dipole and the intensity of the Galactic emission produce the main systematic effects.
In order to predict the bias on $r$ induced by the instrumental polarisation to the measured signal, ref.~\cite{guillaume_HWPIP} introduced the Mueller matrix $\Delta M$ as the deviation from the ideal HWP Mueller matrix considering only the instrumental polarisation at modulation frequency $4f_\phi$, taking only the elements that contribute to the temperature-to-polarisation leakage, we can write the systematic signal $\Delta S$ as
\begin{align}
    \Delta S = [\epsilon_1\cos(4\phi-4\psi+\phi_{QI})\cos2\psi_0 + \epsilon_2\cos(4\phi-4\psi+\phi_{UI})\sin2\psi_0]I,
\end{align}
where $\psi_0$ describes the detector's polarisation angle respect to a reference axis in the focal plane coordinates, while $\phi_{QI}$ and $\phi_{UI}$ refer to the phase in ref.~\cite{guillaume_HWPIP}. 
Here $\epsilon_{1}$ and $\epsilon_{2}$ are the amplitude of the Muller matrix elements at the $4f_\phi$, ideally zero but in general they are expected to have finite value due to the HWP non-ideality, which contributes to the magnitude of leakage from temperature to polarisation. 
The signal is given by the combination of two physical effects: first, for the component of the incident polarisation perpendicular to the HWP, the difference in retardation between the two polarisation states of $180^\circ$ produces a spurious $2\phi$-signal (i.e.\ $2f_\phi$-signal); second, the $\rm{s}$- and $\rm{p}$-polarisations of the incident polarisation, which are not perpendicular to the HWP, flip every $180^\circ$ rotation of the HWP, producing a spurious $2\phi$-signal. 
The coupling of the two effects producing $2\phi$-signals will produce spurious $4\phi$-signals, then the modulated temperature signal will mimic a polarisation \cite{imada2018instrumentally}. Since this effect is caused by the HWP rotation, it is not suppressed by its rotation.

By assuming $\epsilon_1=\epsilon_2$ and $\phi_{QI}=\phi_{UI}+\frac{\pi}{2}$, we can simplify the systematic signal as
\begin{align}
    \Delta S = \frac{\epsilon_1}{2}[e^{i(4\phi-4\psi+\phi_{QI}-2\psi_0)} +  e^{-i(4\phi-4\psi+\phi_{QI}+2\psi_0)}]I.\label{eq:HWP_IP_field}
\end{align}
Now consider a coordinate system where $\psi_0=0$ and perform a Fourier transform to obtain the following equation in \spin space
\begin{align}
    {}_{4,-4}\Delta \tilde{S} = {}_{-4,4}\Delta \tilde{S}^* = \frac{\epsilon_1}{2}e^{-i\phi_{QI}}I. \label{eq:HWPIP_syst}
\end{align}
Combining the signals with the scanning strategy according to \cref{eq:kSd} yields
\begin{align}
    {}_{0,0}\Delta \tilde{S}^d  &= \sum_{n'=-\infty}^{\infty} \sum_{m'=-\infty}^{\infty} {}_{0-n',0-m'}\tilde{h} {}_{n',m'}\Delta\tilde{S} \nonumber \\
    &= \h[4,-4]{}_{-4,4}\Delta \tilde{S} + \h[-4,4]{}_{4,-4}\Delta \tilde{S},\\ 
    {}_{2,-4}\Delta \tilde{S}^d &= {}_{-2,4}\Delta \tilde{S}^{d*} =  \sum_{n'=-\infty}^{\infty} \sum_{m'=-\infty}^{\infty} {}_{2-n',-4-m'}\tilde{h} {}_{n',m'}\Delta\tilde{S} \nonumber\\
              &= \h[6,-8]{}_{-4,4}\Delta \tilde{S} + \h[-2,0]{}_{4,-4}\Delta \tilde{S}.
\end{align}
Of these, $\h[\pm4,\mp4]$ and $\h[\pm6,\mp8]$ have a very small value in its real and imaginary parts due to the HWP contribution, so the systematic signal coupled to this term is suppressed. 
On the other hand, $\h[\pm2,0]$ has no HWP contribution, so the signal coupled to this term is only suppressed by the scanning strategy, which produces non-negligible leakage from temperature. 
Although it has been shown that the bias on $r$ caused by this term can be compensated \cite{guillaume_HWPIP}, we emphasise it here as an example of the importance of the scanning strategy itself, even in the presence of a HWP.

\subsubsection{Method to estimate the bias on \texorpdfstring{$r$}{r}}\label{apd:delta_r}

Using the formalism described in this section, we can obtain the temperature and polarisation maps that are injected with specific systematic effects. From these maps, we can make residual maps by subtracting the original input map from the output map with the systematic effect as
\begin{align}
    \Delta I &= {}_{0,0}\tilde{S}^d - \St[0,0],\\
    \Delta P &= {}_{2,-4}\tilde{S}^d - \St[2,-4],
\end{align}
where $\Delta P$ can be decomposed into Stokes parameters as $\Delta Q=\Re[\Delta P]$ and $\Delta U = \Im[\Delta P]$.
If the signal fields have been already described by only systematics signal fields shown as the case of HWP non-ideality in \cref{apd:HWP_sys}, we can obtain $\Delta I$, $\Delta Q$, and $\Delta U$ without taking a residual as
\begin{align}
    \Delta I &= {}_{0,0}\Delta \tilde{S}^d,\\
    \Delta Q &= \Re[{}_{2,-4}\Delta \tilde{S}^d],\\
    \Delta U &= \Im[{}_{2,-4}\Delta \tilde{S}^d].
\end{align}

Now, we define a likelihood function, $L(r)$ to assess the potential systematic bias on $r$ that we call $\Delta r$, which is caused by the systematic effect as
\begin{align}
    \log{L(r)}=\sum^{\ell_{\rm max}}_{\ell=\ell_{\rm min}}\log{P_\ell(r)},
\end{align}
where $\ell_{\rm max}$ ($\ell_{\rm min}$) represents the maximum (minimum) multipole number of the power spectrum \cite{PTEP2023}. In this paper, we apply $\ell_{\rm max}=191$ and $\ell_{\rm min}=2$, as was used in ref.~\cite{PTEP2023}, and define $P_\ell(r)$ as
\begin{align}
    \log{P_\ell(r)}=-f_{\rm sky}\frac{2\ell+1}{2}\left[\frac{\hat{C}_\ell}{C_\ell} +\log{C_\ell} - \frac{2\ell-1}{2\ell+1}\log{\hat{C}_\ell} \right],
\end{align}
where $\hat{C}_\ell$ ($C_\ell$) represents the measured (modeled) $B$-mode power spectrum \cite{likelihood}. For our analysis, we apply $f_{\rm sky}=1$, which means the full-sky coverage because we do not use a Galactic mask in this paper.
The $\hat{C}_\ell$ and $C_\ell$ are defined as
\begin{align}
    \hat{C}_\ell &= C_\ell^{\rm sys} + C_\ell^{\rm lens} +N_\ell,\\
    C_\ell &= r C_\ell^{\rm tens} + C_\ell^{\rm lens} + N_\ell,
\end{align}
where $C_\ell^{\rm sys}$ is the estimated systematic effects power spectrum, which is given by the spherical harmonic expansion of $\Delta Q$ and $\Delta U$, $C_\ell^{\rm lens}$ is the lensing $B$-mode power spectrum, and $C_\ell^{\rm tens}$ is the tensor mode with $r = 1$ \cite{PTEP2023}. The quantity $N_\ell$ is the expected noise power spectrum but we do not consider it in this paper because we do not include noise in the simulations.

The potential systematic bias $\Delta r$ is defined by the value of $r$ where $L(r)$ is maximised:
\begin{align}
    \left. \dv{L(r)}{r} \right|_{r=\Delta r} = 0.
\end{align}
The $\Delta r$ that is shown in \cref{sec:Propagation} follows this definition.

\subsubsection{Comparison with TOD simulation} \label{apd:TOD_comparison}
\begin{figure}
  \centering
  \includegraphics[width=0.49\columnwidth]{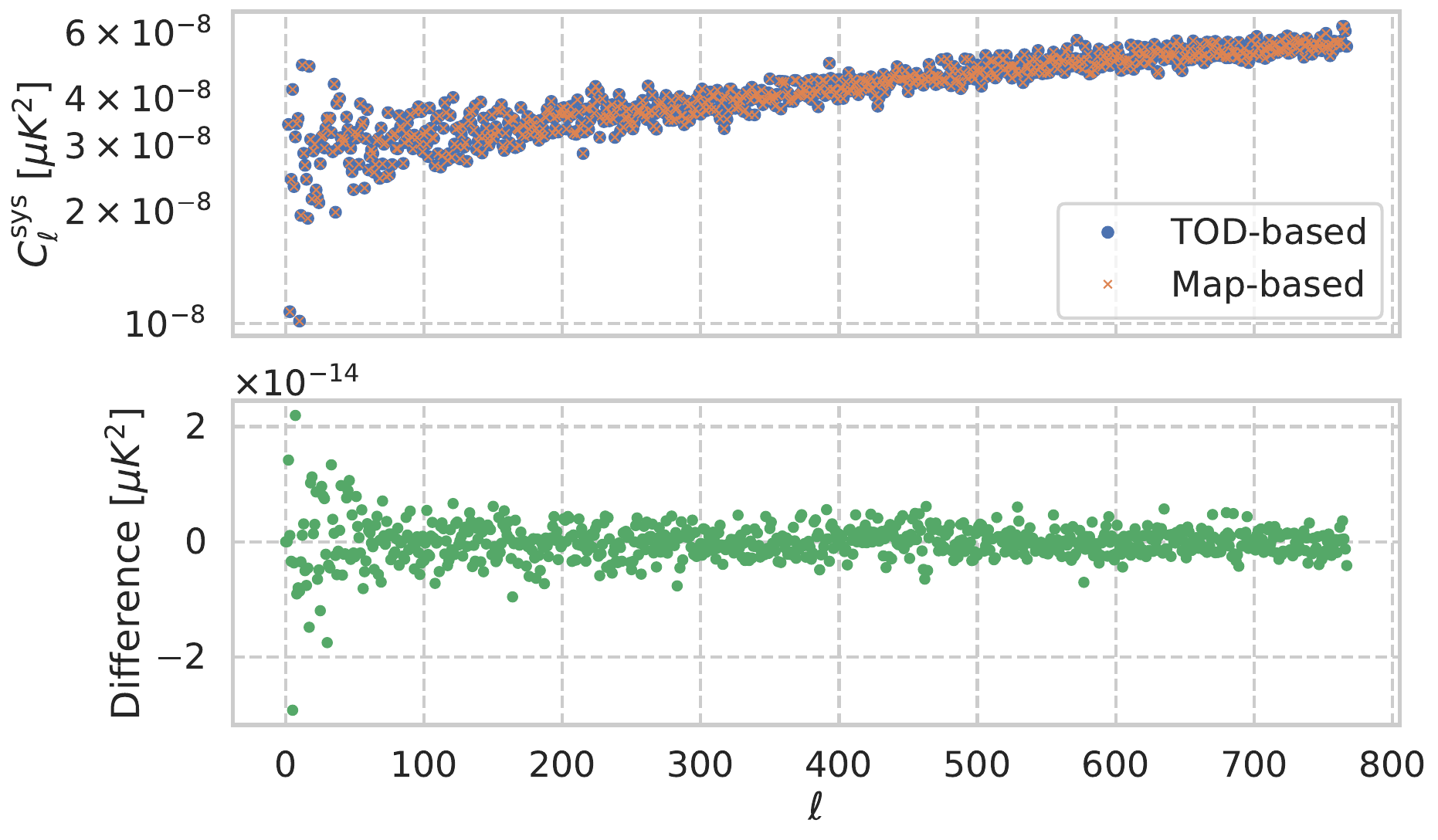}
  \includegraphics[width=0.49\columnwidth]{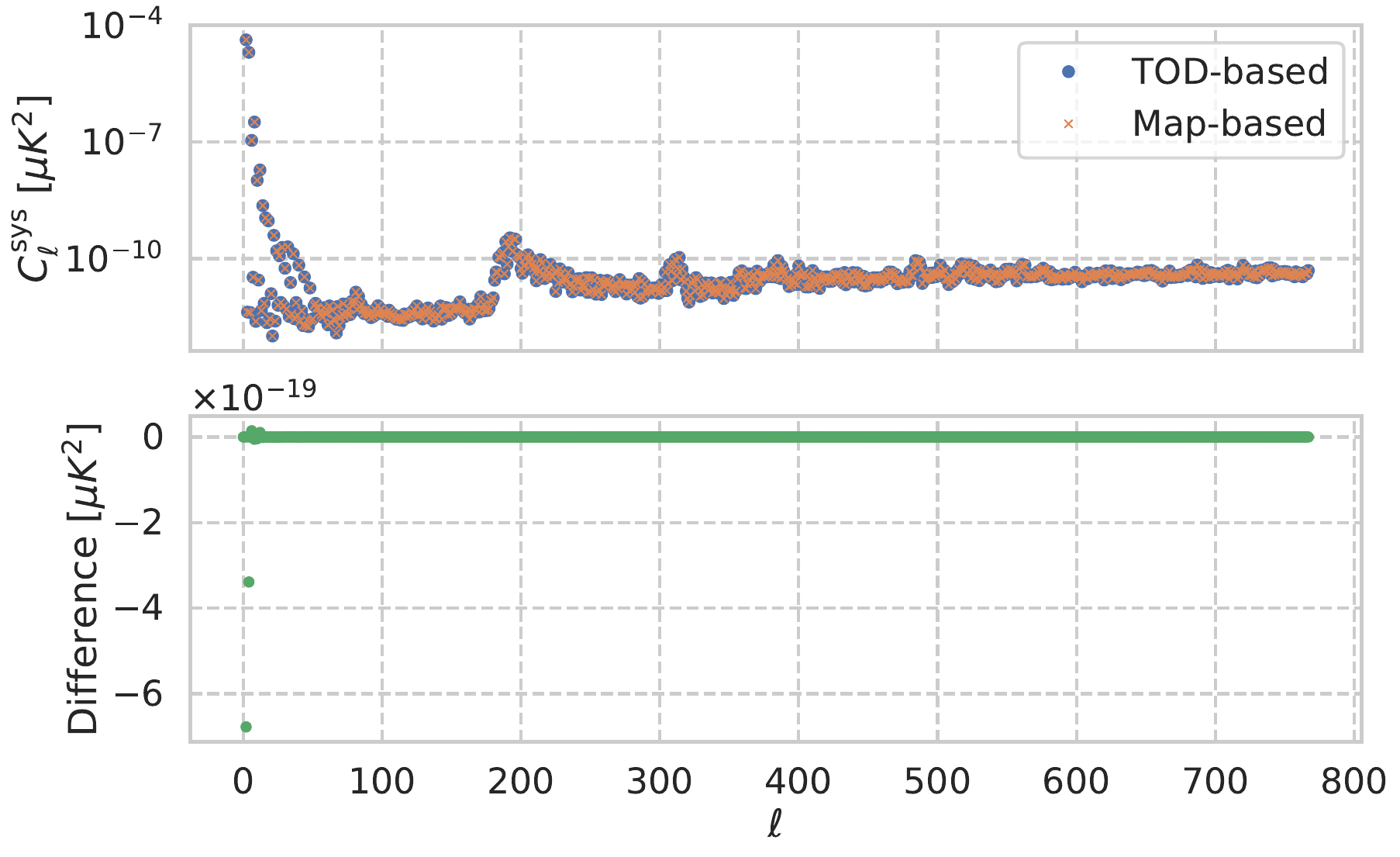}
  \caption{
  (left/right) Systematic effect $B$-mode power spectrum, $C_\ell^{\rm sys}$ due to the pointing offset/HWP non-ideality. Blue dots and orange crosses show $C_\ell^{\rm sys}$ given by the TOD- and map-based simulation, respectively. The bottom plots with green dots show the difference of $C_\ell^{\rm sys}$ between two methods.
  For simulation of pointing offset, we used CMB only as an input map, and $(\rho,\chi)=(1,0)$ arcmin as a systematics parameter. 
  For simulation of HWP non-ideality, we used solar dipole only as an input map, and $(\epsilon_1,\phi_{QI})=(1.0\times10^{-5},0)$ as a systematics parameter.
  }
  \label{fig:cl}
\end{figure}

We confirm in this section whether our new map-based simulation, which can account for HWP using \spin, is consistent with the general TOD-based simulation defined by \cref{eq:map-making_TOD}. 
The input maps and systematics parameters used are the same as for \cref{sec:Propagation}.
We injected the systematic effect, performed map-making, and calculated $\Delta Q$ and $\Delta U$ maps, and systematic $B$-mode power spectrum, $C_\ell^{\rm sys}$ as shown in \cref{apd:delta_r}.
\Cref{fig:cl} (left) shows $C_\ell^{\rm sys}$ for a pointing offset of $(\rho,\chi)=(1,0)$ arcmin, calculated by the TOD-based method (blue dots) and the map-based method (orange crosses). The input map is CMB only.
The plot shown by the green dots at the bottom is the difference in $C_\ell^{\rm sys}$ between two methods, and it is clear that the two methods produce numerically equivalent results, since the order of the difference is $10^{-14}$.
\Cref{fig:cl} (right) shows when the HWP non-ideality is present at $\epsilon_1=1.0\times10^{-5}$ and $\phi_{QI}=0$, and it is also clear that there is no significant difference between the two methods. 
The input map is solar dipole only; the jump at $\ell\simeq200$ originates from the structure on the east-west running line that the hit map has (as seen in middle bottom of \cref{fig:hitmaps}). 

This validation ensures that map-based simulations with spin are scientifically equivalent to TOD simulations, even when HWP is introduced.

\section{Supplementary results and discussions}

\subsection{Impact of the value of the spin period on the metrics} \label{apd:T_beta_scaled}
In this paper, we fixed $T_\beta=\tbl=16.9$\,min and displayed our results in the $\{\alpha,T_\alpha\}$ space. $T_\beta$ is, however, a free parameter in the range $\tbl<T_\beta$, such that the complete parameter space is the three dimensional $\{\alpha,T_\alpha, \tbl<T_\beta\}$ space. However, it can be shown that the $T_\beta$ dimension implies a rescaling of the metrics of interest without changing the values of the optimum solution.

\begin{figure}[h]
  \centering
  \includegraphics[width=0.49\columnwidth]{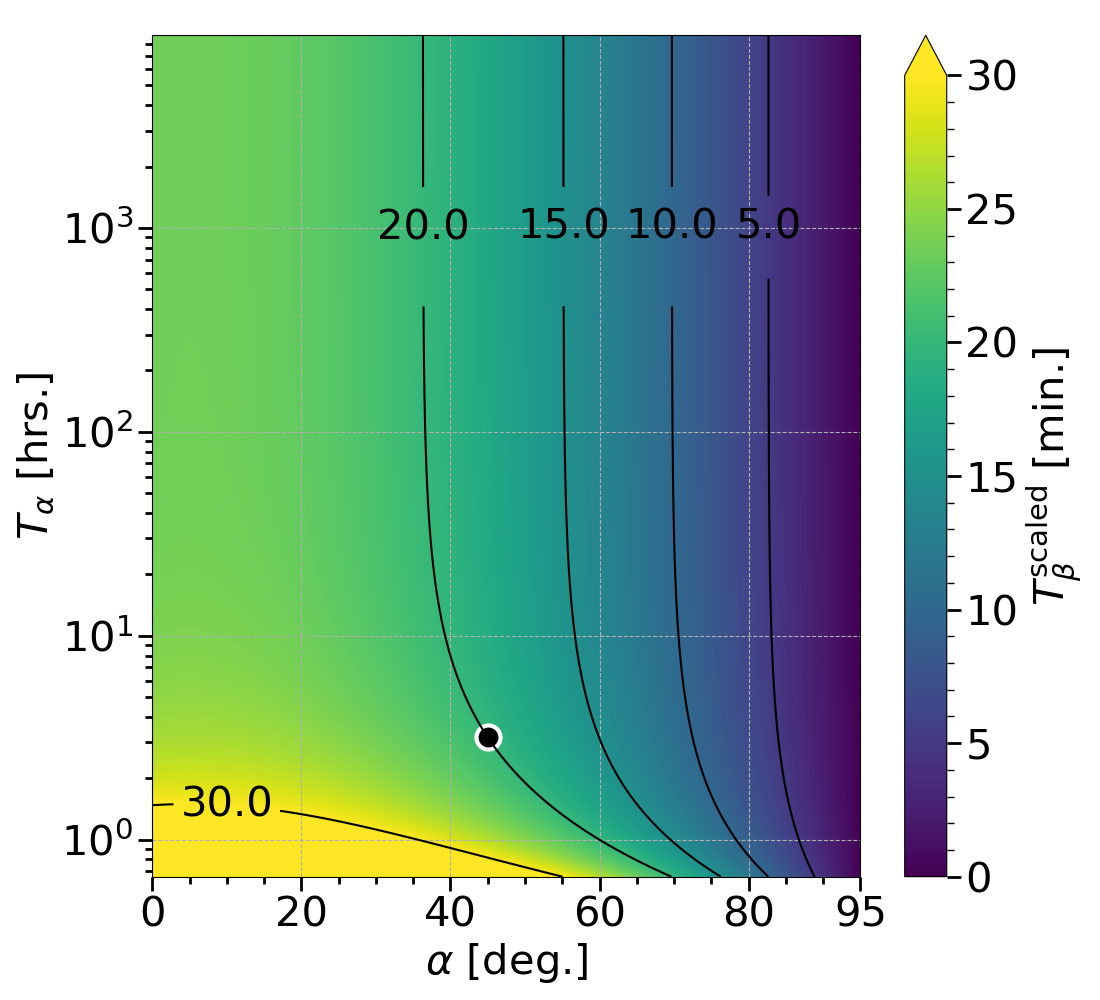}
  \includegraphics[width=0.49\columnwidth]{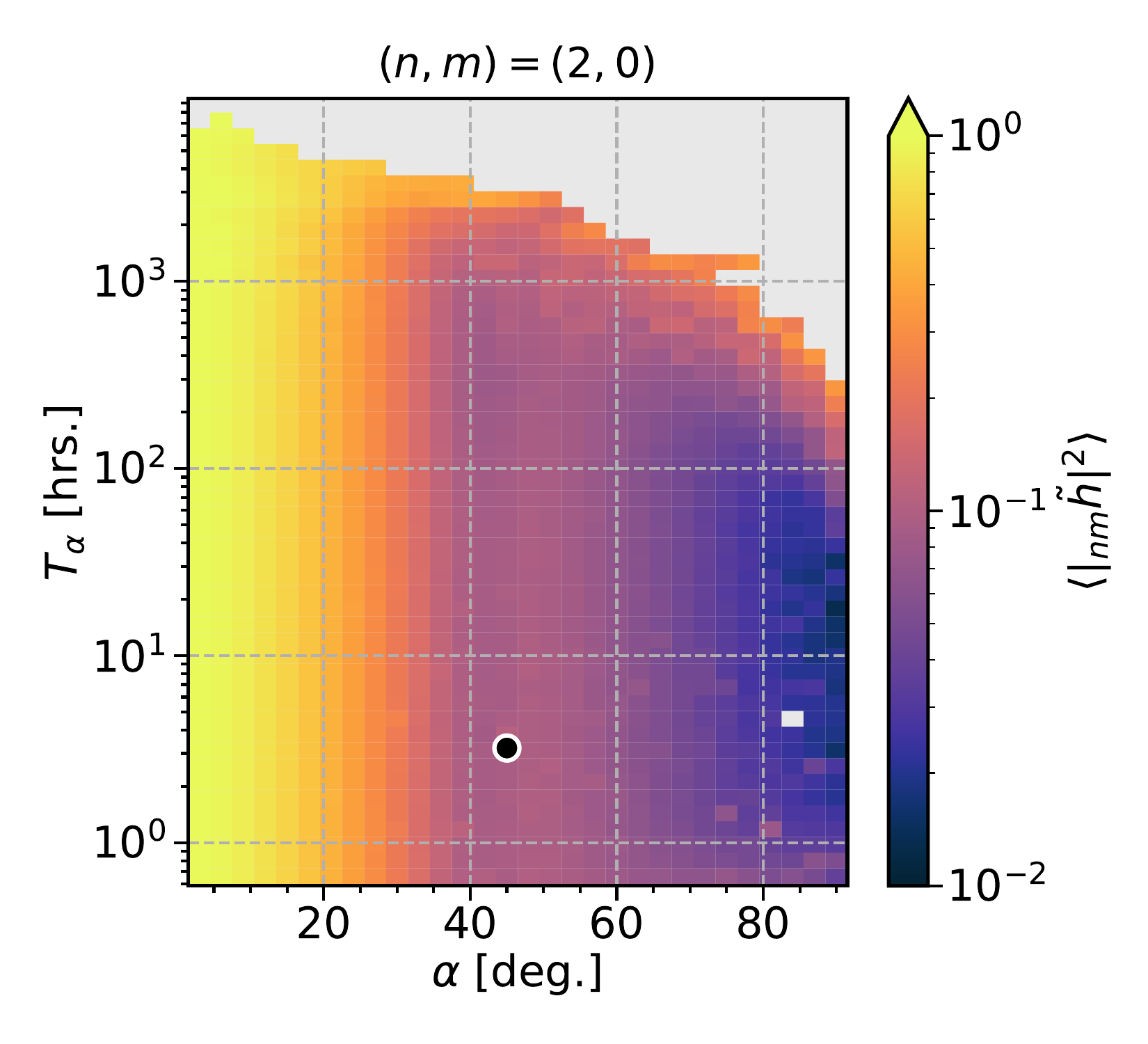}\\
  \includegraphics[width=0.49\columnwidth]{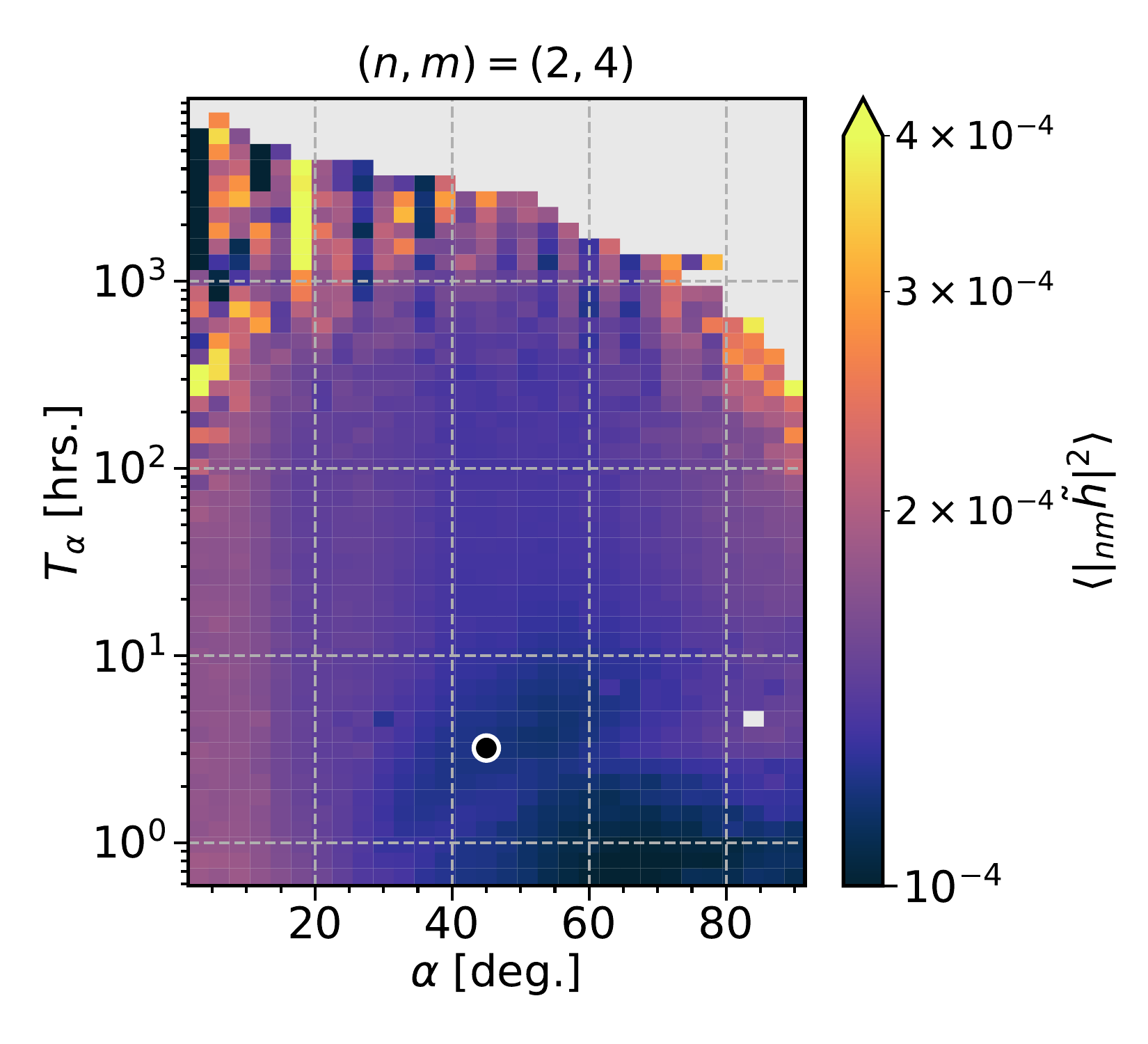}
  \includegraphics[width=0.49\columnwidth]{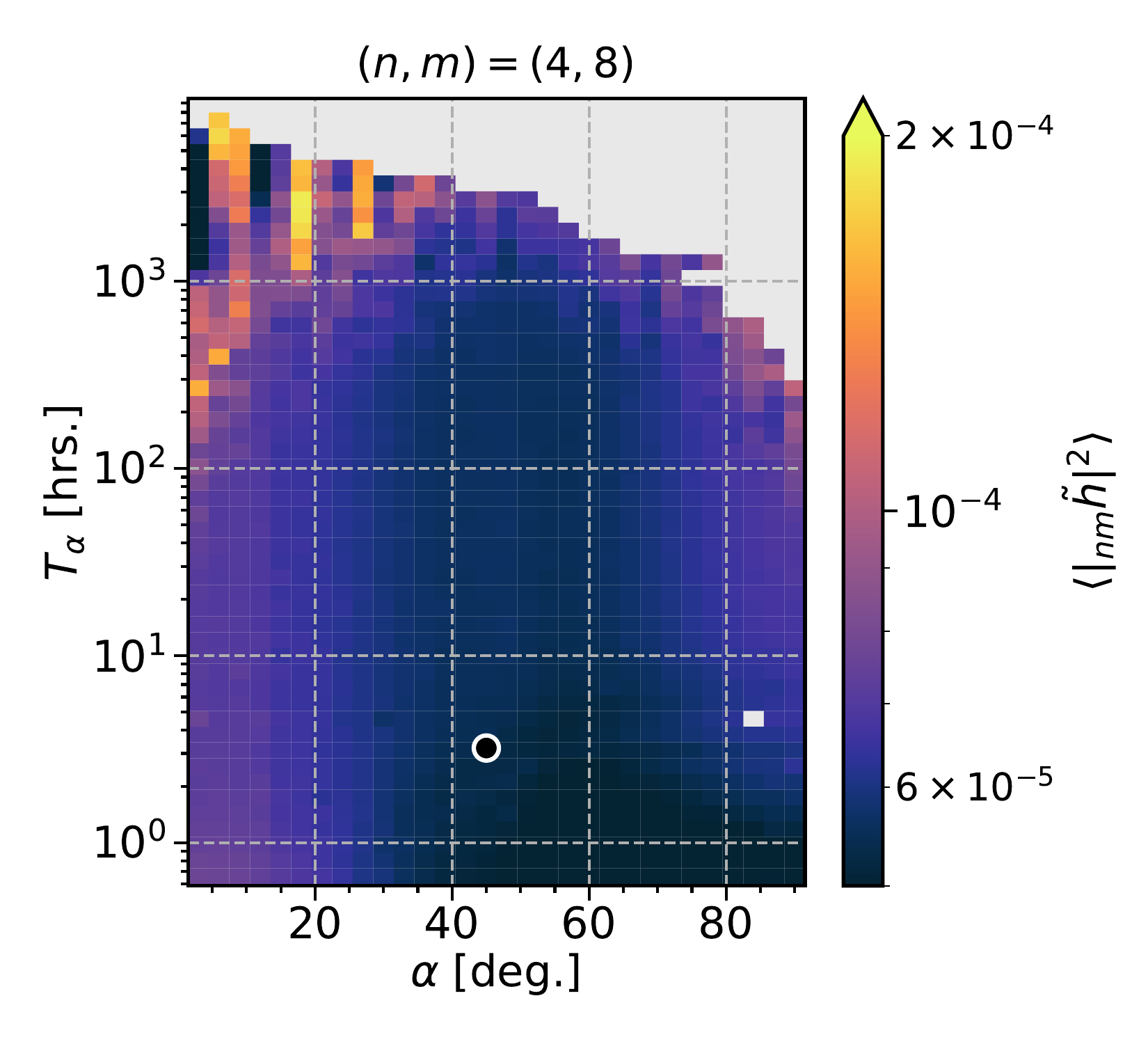}
  \caption{(top left) $\{\alpha,T_\alpha,T_\beta^{\rm scaled}\}$ space created by multiplying $\tbl$ in \cref{fig:standard_config_and_T_beta} (right) by a constant and scaling the entire space so that the \SC has $T_\beta=20$\,min.
  (top right) \spin-$(2,0)$ cross-link factors simulated using $T_\beta^{\rm scaled}$. (bottom left/right) \spin-$(2,4)$/\spin-$(4,8)$ cross-link factors simulated using $T_\beta^{\rm scaled}$. 
  The cross-link factors with non-zero \spin-$m$ has smaller values than those calculated using $\tbl$ due to HWP rotation within the sky pixel; however, the change in $T_\beta$ in all cross-link factors do not cause a change in the extreme value structure. The point with no value in the bottom right-hand corner of the cross-link factor panel is due to a few unobserved sky pixels caused by resonances between spin and precession, which can be eliminated by a small change in the precession period.}
  \label{fig:T_beta_scaled}
\end{figure}

As an illustration, let us consider the results in the $\{\alpha,T_\alpha\}$ space through a rescaling of $\tbl$ and observe the impact on the cross-link factor distribution.
\Cref{fig:T_beta_scaled} (top left) shows the $\{\alpha,T_\alpha\, T_\beta^{\rm scaled}\}$ space, where the point with the \SC, $(\alpha,T_\alpha)=(45^\circ, 192.348\,\rm{min})$, is multiplied by a constant throughout, so that $T_\beta=20$\,min.
This is used as input for the cross-link factor simulation.

\Cref{fig:T_beta_scaled} (top right) shows the \spin-$(2,0)$ cross-link factor, while bottom left/right show the \spin-$(2,4)$/\spin-$(4,8)$ cross-link factors. 
The \spin-$(2,0)$ cross-link factor without the HWP contribution are almost identical to the results calculated with $\tbl$.
The \spin-$(2,4)$ and \spin-$(4,8)$ cross-link factor results also have a flat distribution, and again the structure is not much different from the results presented in \cref{sec:result_crosslink}.
However, it can be seen that the cross-link factors with HWP contributions are generally scaled to smaller values. 
This is because the spin is slower, allowing extra rotation of the HWP while the pointing transits the sky pixel.
Of course, this works well to suppress systematic effects. On the other hand, if the rotation is slowed down any further, the value of the \spin-$(n,0)$ cross-link factors starts to degrade, as discussed in \cref{sec:Opt_kinetic}.
The results on cross-link factors obtained in this supplement emphasise that the global structure does not change with respect to changes in the rotation period, but the values are simply scaled.

A similar argument can be made identically for the other two metrics: hit-map uniformity and visibility time of planets. Indeed, the quality of these two metrics is mostly driven by geometric considerations, such as the shape of the scanning pattern on the sky, and most importantly, the size of the scan pupil.

\subsection{Metrics for detectors located away from the boresight} \label{apd:other_detector}
In this section we investigate how the metrics of detectors located away from the boresight (i.e.\ center of focal plane) behave in our \SC.
Different detectors on the focal plane are often described as having different effective $\beta$ (written as $\beta^{\rm eff}$) than the one of the boresight. This can be expressed by considering the shifted effective spin axes per detector as
\begin{align}
    \beta^{\rm eff}_i = \beta + \beta_i,
\end{align}
where $i$ represents detector index. 
Detectors that observe in a direction that is inclined by $\beta_i$ from the boresight have effectively different geometric parameters, so the scan trajectory is different from that of the boresight and, in general, the metrics also have different values. Conversely, a detector that is not boresight but whose $\beta^{\rm eff}_i$ is close to $\beta$ has a trajectory that is almost redundant with the boresight, so the metrics do not change much.
In the MFT/HFT field of view with $14^\circ$ in radius, we simulated the integration time of planet visibility and cross-link factor by a detector which has minimum/maximum $\beta_i$ ($-14^\circ/14^\circ$).

\begin{figure}[h]
  \centering
  \includegraphics[width=0.32\columnwidth]{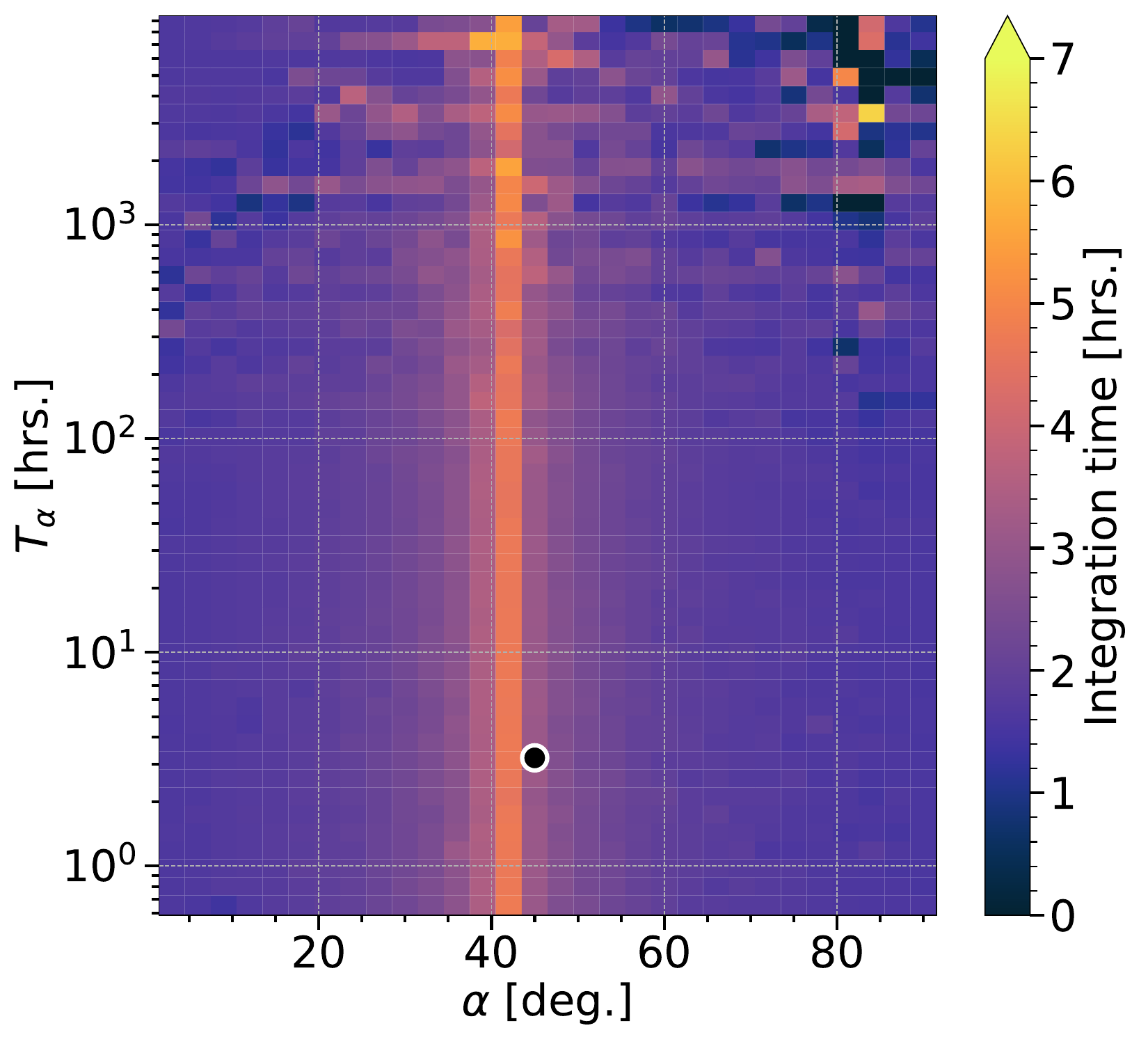}
  \includegraphics[width=0.32\columnwidth]{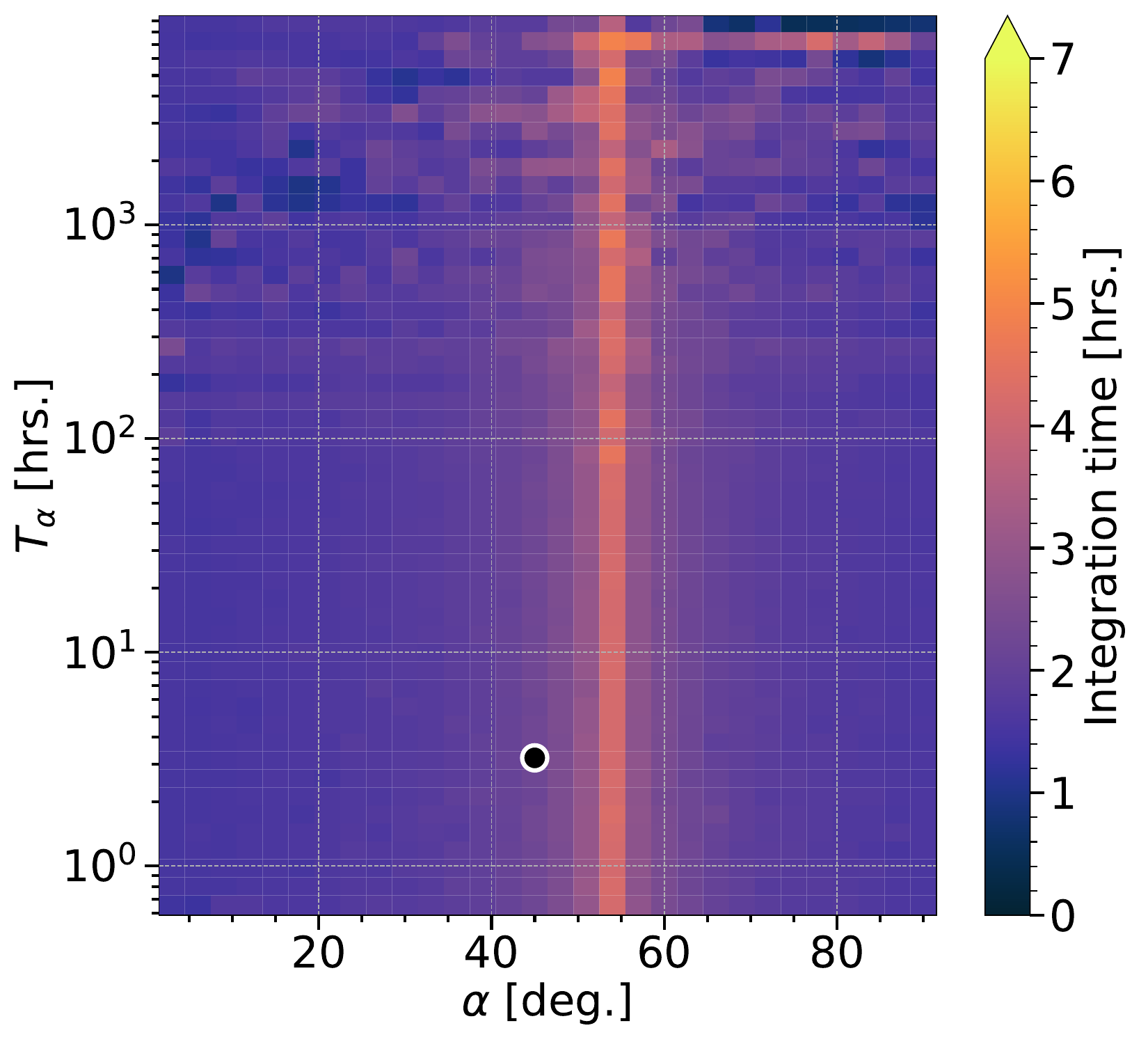}
  \includegraphics[width=0.32\columnwidth]{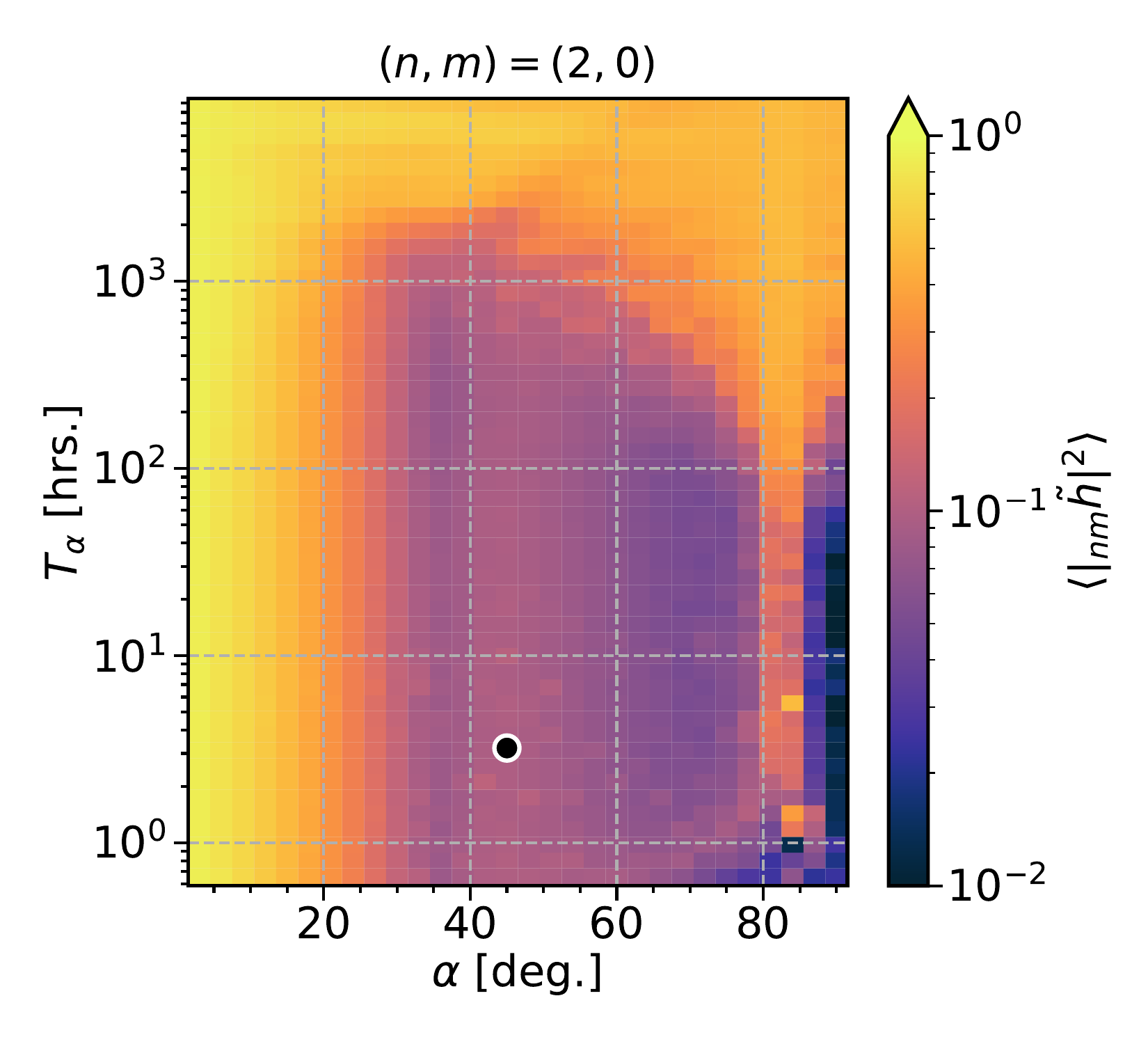}\\
  \includegraphics[width=0.32\columnwidth]{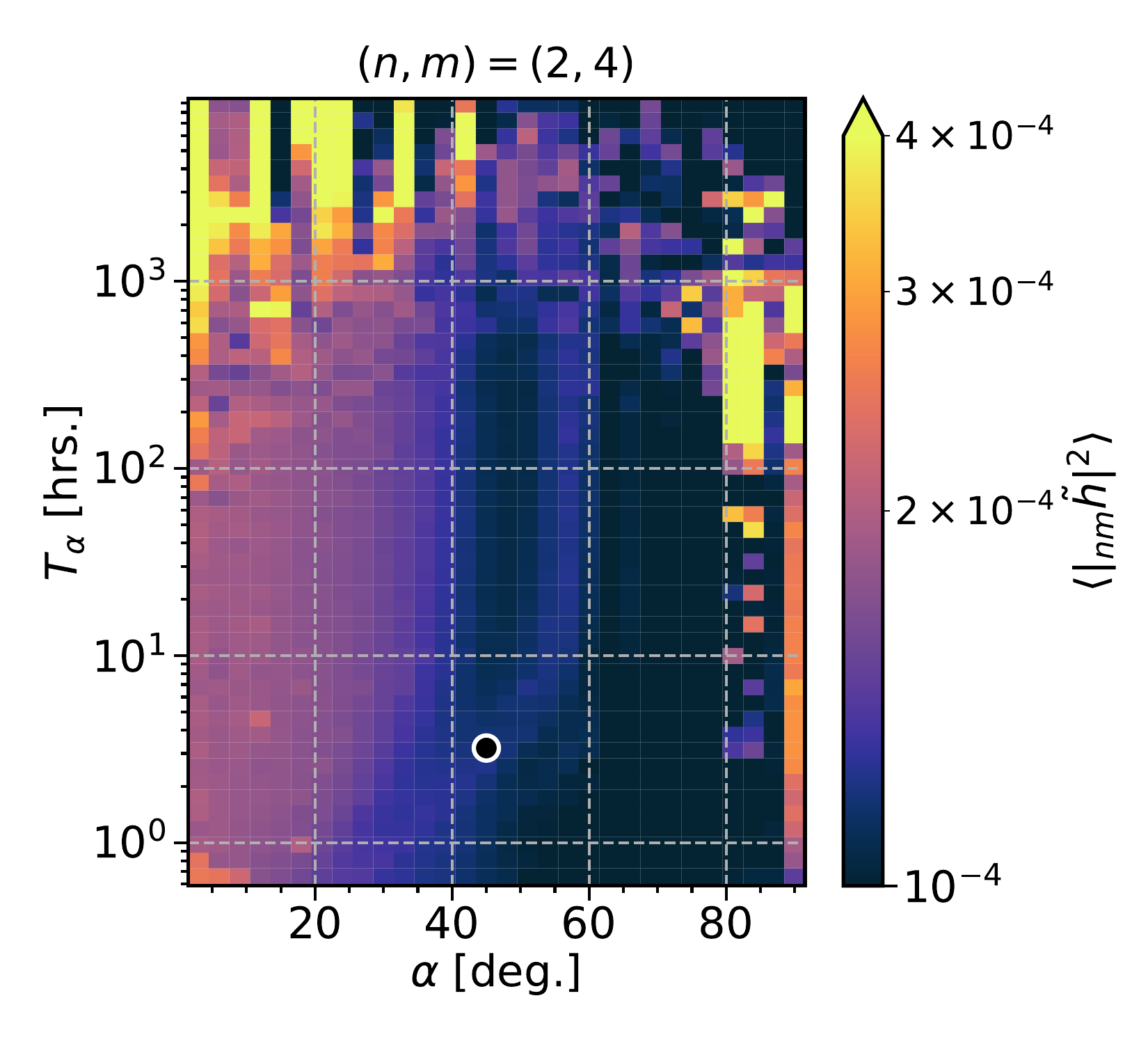}
  \includegraphics[width=0.32\columnwidth]{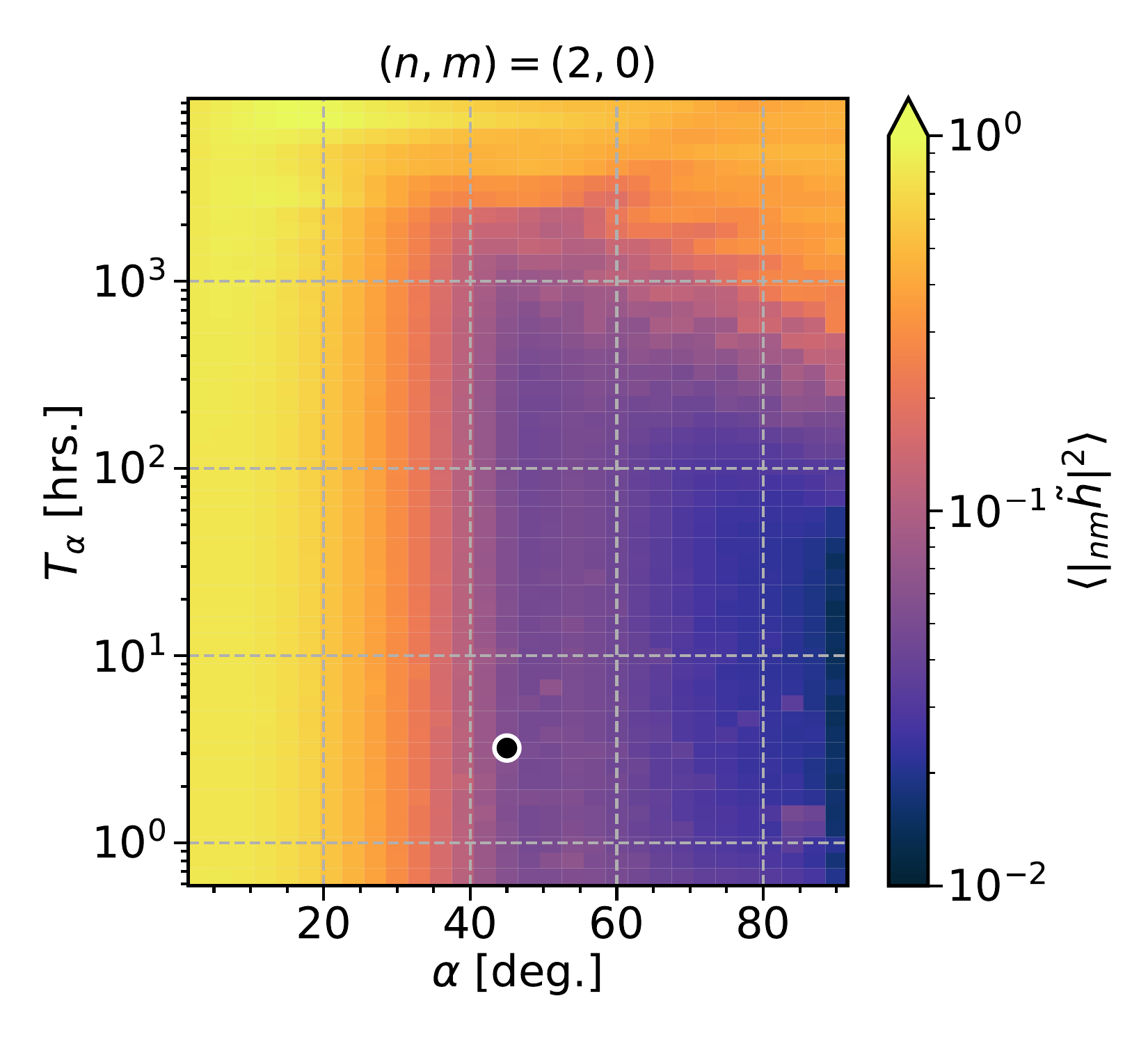}
  \includegraphics[width=0.32\columnwidth]{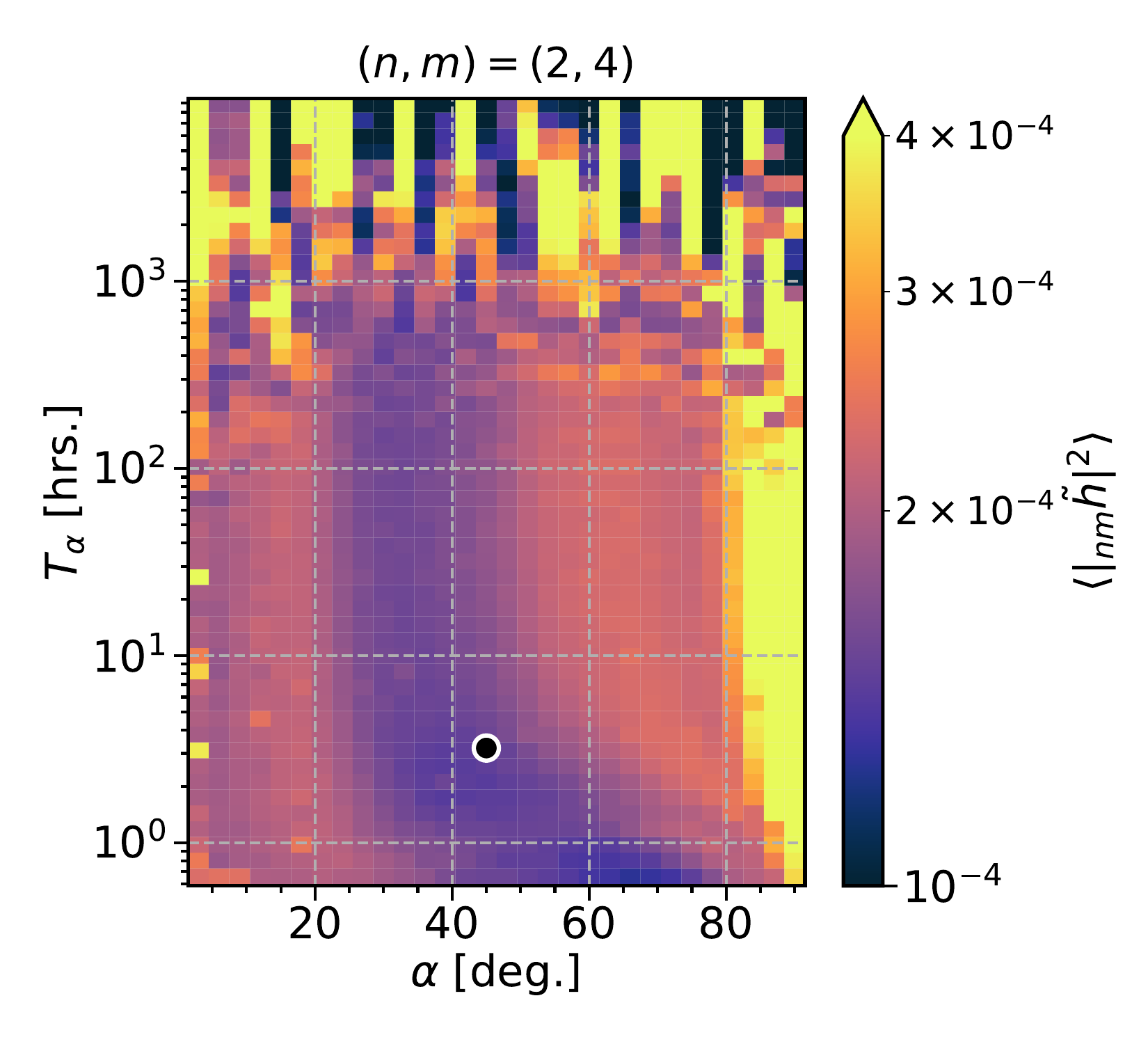}
  \caption{(top left/top middle) Integrated planet visibility time simulated by a detector which has $\beta_i=-14^\circ/14^\circ$. 
  (top right/bottom left) \spin-$(2,0)$ cross-link factor simulated by a detector which has $\beta_i=-14^\circ/14^\circ$. 
  (bottom middle/bottom right) \spin-$(2,4)$ cross-link factor simulated by a detector which has $\beta_i=-14^\circ/14^\circ$. 
  Values for sky pixels with diverging cross-link factors are ignored and averaged over the entire sky.}
  \label{fig:metrics_other_det}
\end{figure}

\Cref{fig:metrics_other_det} shows integration time of planets visibility (top left: $\beta_i=-14^\circ$, top middle: $\beta_i=14^\circ$).
As in \cref{fig:planet_and_sigma_hit} (left), the results for the integration time of planets visibility are independent of $T_\alpha$, with a peak at $\alpha=(\kappa + \beta_i)/2$.
From \cref{fig:metrics_other_det} (top left/middle), it is clear that our \SC is a scanning strategy that provides relatively long integration time of planet visibility not only at the boresight but also at the detector with the minimum or maximum $\beta^{\rm eff}_i$.

\Cref{fig:metrics_other_det} shows \spin-$(2,0)$ cross-link factor (top right: $\beta_i=-14^\circ$, bottom left: $\beta_i=14^\circ$), and \spin-$(2,4)$ cross-link factor (bottom middle: $\beta_i=-14^\circ$, bottom right: $\beta_i=14^\circ$).
When considering the effective $\beta$, certain detectors cannot satisfy \cref{eq:const_geometric}, and the area around the pole of the ecliptic coordinates becomes unobservable, so the cross-links factor in that region diverge. Here we show the results of calculating the full-sky average using the detector which has $\beta_i=\pm14$, ignoring the pixels where the cross-link factor diverge.
First, the $\beta_i=-14^\circ$ detector is not able to observe the full-sky because its trajectory does not go beyond the poles of the ecliptic coordinates. Comparing the cross-link factors for a detector which has $\beta_i=\pm14^\circ$, we can see that the distribution of \spin-$(2,0)$ cross-link factor is similar in structure, but the overall value is larger for $\beta_i=-14^\circ$. This is due to the fact that the uniformity of the crossing angle for each pixel has decreased due to the narrower area which can be accessed by the scanning.
The \spin-$(2,4)$ cross-link factor also shows a change in structure, and the values are generally smaller for $\beta_i=-14^\circ$. This is due to the fact that a smaller $\beta_i$ slows down the scan speed defined by \cref{eq:sweeping_velocity}, causing the HWP to rotate extra within the sky pixel.
It is clear that although the structure and value of the distribution changed, the \SC still provides small cross-link factors for detectors other than the boresight.

\subsection{Discussion on the rotation direction of the spacecraft} \label{apd:rotation_direction}

Although we defined the rotation matrix in \cref{eq:matrix_chain} as forward rotation, i.e., counterclockwise, there are in principle four possible combinations of spin and precession with respect to the direction of rotation.
Since these four combinations are independently positioned with respect to the orbital direction and will have different trajectories, it is necessary to check whether our conclusions are independent of the direction of rotation.

First, we distinguish the four combinations by the sign of the rotation matrix and define the precession-spin coherence from the product of the signs, as shown in \cref{tab:rotation_convention}.

\begin{table}[ht]
\centering
\begin{tabular}{lllll}
\hline
Sign of precession         & $+$ & $+$ & $-$ & $-$ \\ \hline
Sign of spin               & $+$ & $-$ & $+$ & $-$ \\ \hline
Precession-spin coherence  & $+$ & $-$ & $-$ & $+$ \\ \hline
\end{tabular} 
\caption{Convention for rotations and theirs coherence}
\label{tab:rotation_convention}
\end{table}
When the coherence is positive, $\omega_{\rm max}$ is consistent with \cref{eq:sweeping_velocity}, but when it is negative, one rotation has the effect of pulling back the other, effectively slowing down the sweep angular velocity. To obtain it, the rotation matrix $R_z$ in \cref{eq:matrix_chain} can be reversed, i.e., transposed, and rewritten as
\begin{align}
    \mathbf{n}(t) &= R_z(\omega_\alpha t)R_y(\alpha )R_z^\top(\omega_\beta t)\mathbf{n}_0,
\end{align}
where $\top$ represents the transposition of the matrix.
Here the sweep angular velocity is defined as $\omega_{\rm max}^+$ when the coherence is positive and $\omega_{\rm max}^-$ when the sign is negative:
\begin{align}
    \omega_{\rm max}^\pm = \omega_\alpha\sin(\alpha\pm\beta) + \omega_\beta\sin\beta.
\end{align}
This modification propagates to $T_\beta^{\rm lower}$ and $T_\alpha^{\rm lower}$ in our constraints of \cref{eq:T_spin,eq:T_alpha}, and they can be written with the precession-spin coherence as
\begin{align}
    T_\beta^{\rm{lower},\pm}  &= \frac{2\pi \Nmod T_\alpha \sin\beta}{\Delta \theta f_\phi T_\alpha - 2\pi \Nmod \sin(\alpha\pm\beta)},\label{eq:T_spin_minus} \\
    T_\alpha^{\rm{lower},\pm} &= \frac{2\pi \Nmod (\sin\beta + \sin(\alpha\pm\beta))}{\Delta \theta f_\phi}.
\end{align}

\begin{figure}
  \centering
  \includegraphics[width=0.49\columnwidth]{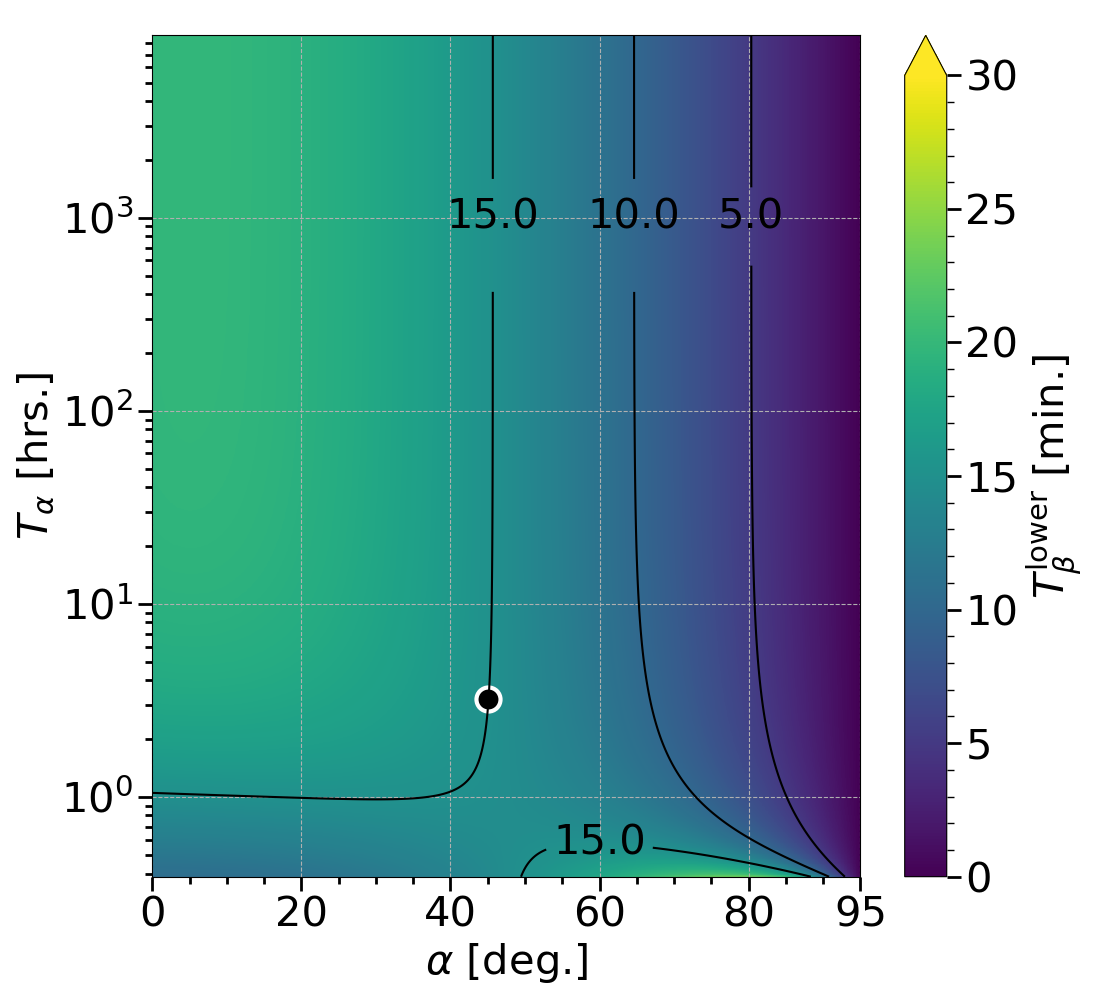}
  \includegraphics[width=0.49\columnwidth]{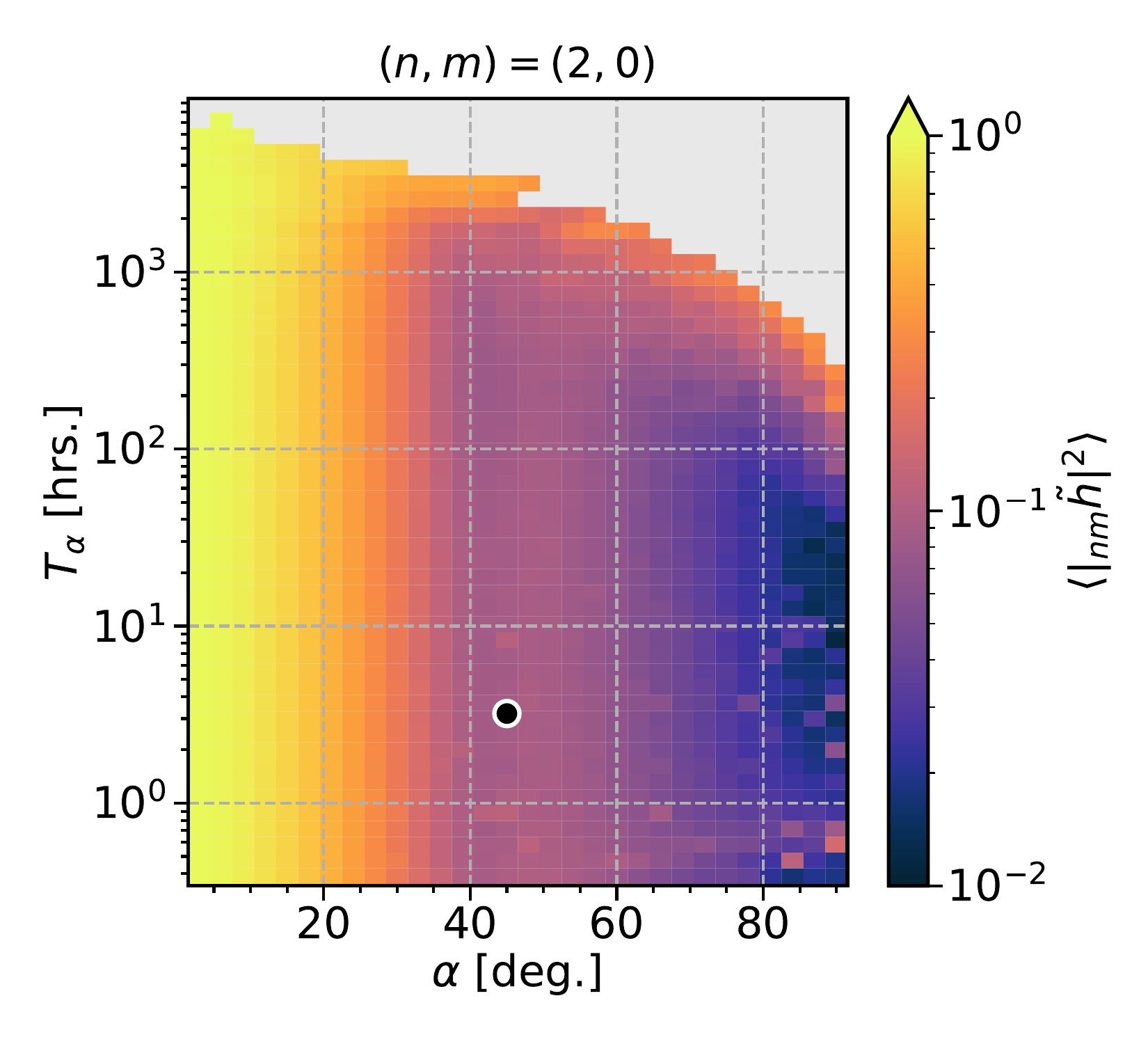}\\
  \includegraphics[width=0.49\columnwidth]{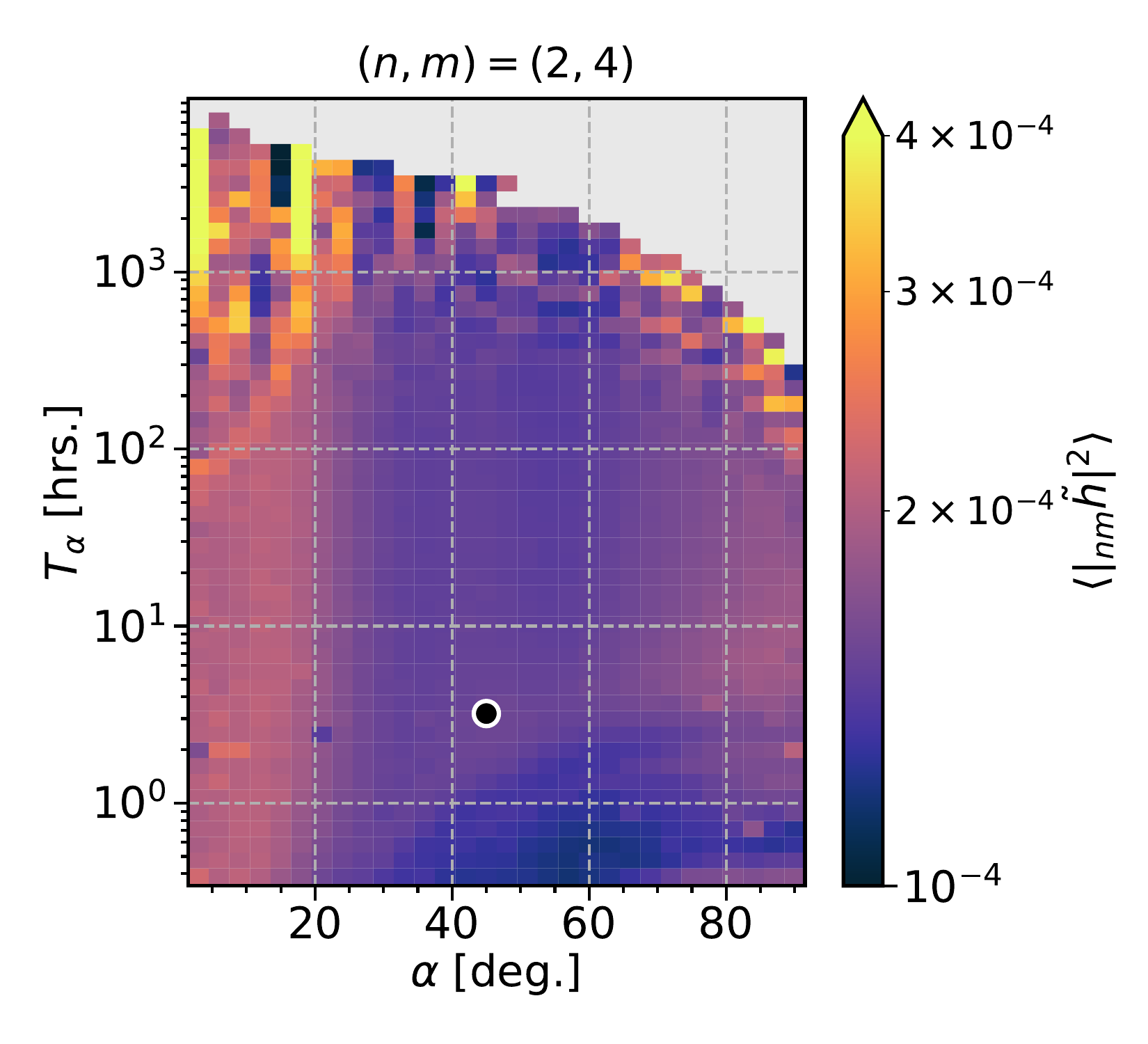}
  \includegraphics[width=0.49\columnwidth]{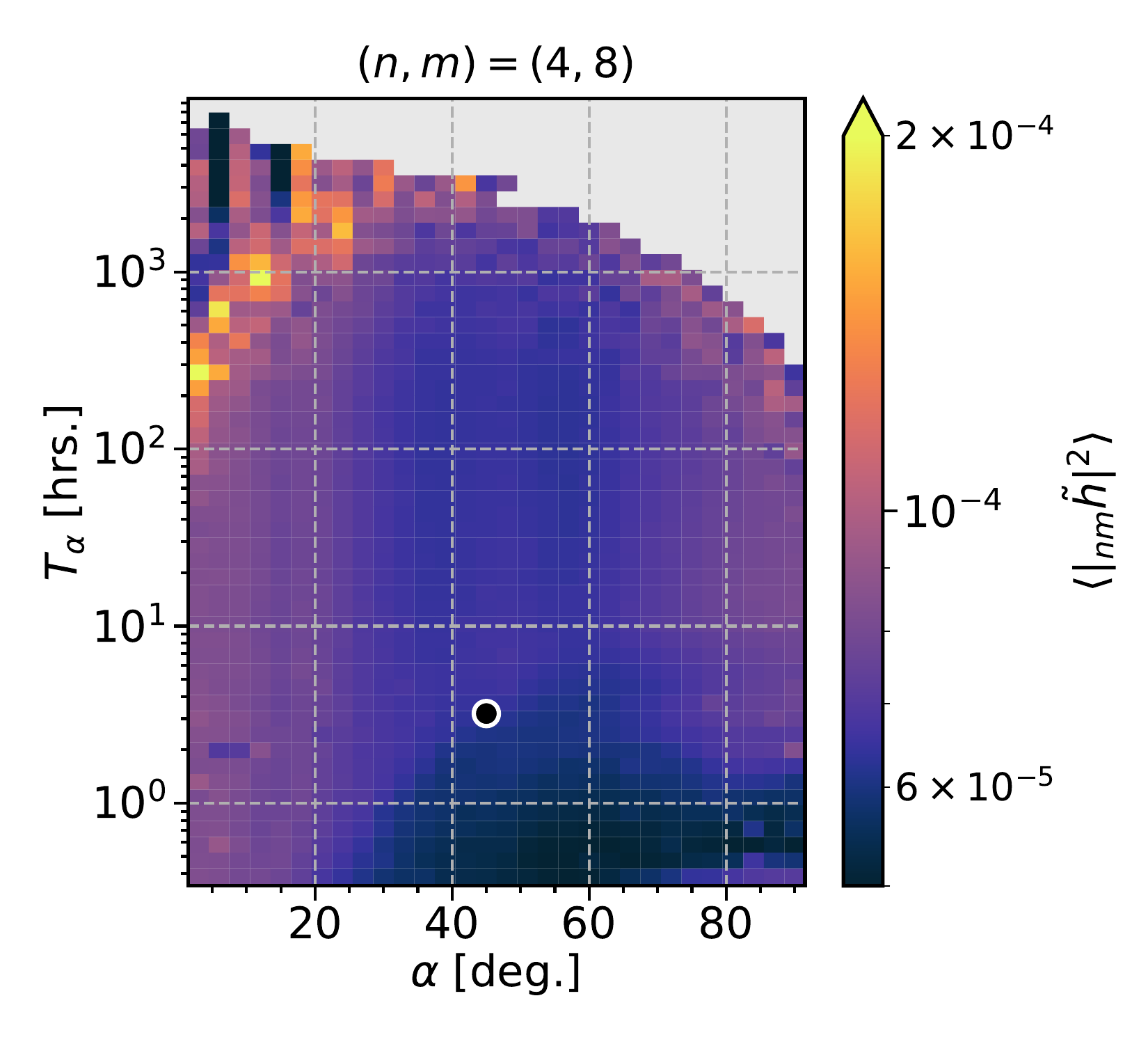}
  \caption{(top left) $\{\alpha, T_\alpha, T_\beta^{\rm lower,-}\}$ space given by \cref{eq:T_spin_minus}. (top right) \spin-$(2,0)$ cross-link factors simulated using $T_\beta^{\rm lower,-}$. (bottom left/right) \spin-$(2,4)$/\spin-$(4,8)$ cross-link factors simulated using $T_\beta^{\rm lower,-}$. Each map is simulated using the $(\rm{prec.},\rm{spin})=(+,-)$ configuration. The change from $\omega_{\rm max}^+ \rightarrow \omega_{\rm max}^-$ effectively slows down the sweep speed and increases the number of HWP rotations per visit for a sky pixel. This results in a smaller cross-link factor with a non-zero \spin-$m$ which has a larger HWP contribution than in the case simulated with $T_\beta^{\rm lower,+}$ shown in \cref{sec:result_crosslink}.}
  \label{fig:CCW_figures}
\end{figure}

If the coherence is assumed to be negative, a different $T_\beta^{\rm lower}(\alpha,T_\alpha)$ space is constructed than in \cref{fig:standard_config_and_T_beta} (right), as shown in \cref{fig:CCW_figures} (top left), and a shorter $T_\beta^{\rm lower}$ can be obtained due to changing $\omega_{\rm max}^+ \to \omega_{\rm max}^-$.
We re-examined all metrics with $T_\beta^{\rm lower}$ defined by the four combinations and the corresponding coherence, and the advantages of the \SC were still seen in all metrics.
The remaining panels of \cref{fig:CCW_figures} show the cross-link factors simulated with the configuration $(\rm{prec.},\rm{spin})=(+,-)$, which shows no significant change from \cref{fig:cross-links}.
This alienates our particular concern about the degradation of \SC's cross-link factors due to trajectory changes and the emergence of different best solutions for the parameters.

On the other hand, the interesting thing about negative coherence combinations such as $(\rm{prec.},\rm{spin})=(+,-)$ is that only the maximum sweep angular velocity can be slowed down without changing the scan parameter values.
For example, the \SC has $\omega_{\rm max}^+=0.26$\,[deg/s], while the spin direction switch $(\rm{prec.},\rm{spin})=(+,-)$ slows it down to $\omega_{\rm max}^-=0.23$\,[deg/s].
This would slightly increase the number of data samples per unit time for the sky pixels and improve the number of rotations of the HWP per sky pixel.

Whether this subtle slowdown changes the physical results in a favorable direction will only become clear with some kind of end-to-end simulation that includes time-correlated noise and systematic effects that are synchronised with the HWP rotation, so we leave that for a future study.
However, we would like to emphasise that regardless of the direction of spin and precession of the \SC that we have presented, it certainly retains the capabilities required for a scanning strategy, and it contains sufficient information for the basic design of the spacecraft.

\end{document}